\documentclass[12pt]{article}
\usepackage{amsmath}
\usepackage{graphicx}
\usepackage{enumerate}
\usepackage{natbib}
\usepackage{url} 
\usepackage{amssymb}
\usepackage{float}
\usepackage{mathtools}
\usepackage{multirow}
\usepackage{caption}
\usepackage{xcolor}
\usepackage{bm}
\usepackage{xr}
\usepackage[ruled,vlined]{algorithm2e}

\DeclareMathOperator*{\argmax}{argmax} 
\newcommand{\blind}{0}

\begin{document}

\def\spacingset#1{\renewcommand{\baselinestretch}%
{#1}\small\normalsize} \spacingset{1}


\if0\blind
{
  \title{\bf Bayesian spline-based hidden Markov models with applications to actimetry data and sleep analysis}
  \author{
  	Sida Chen \\
  	Department of Statistics, University of Warwick \\
  	MRC Biostatistics Unit, University of Cambridge 
  	\and
  	B{\"a}rbel Finkenst{\"a}dt \\
  	Department of Statistics, University of Warwick
  }
  \date{Last modified: May 26, 2023}
  \maketitle
} \fi

\if1\blind
{
  \bigskip
  \bigskip
  \bigskip
  \begin{center}
    {\LARGE\bf Bayesian spline-based hidden Markov models with applications to actimetry data and sleep analysis}
\end{center}
  \medskip
} \fi

\bigskip
\begin{abstract}
B-spline-based hidden Markov models employ B-splines to specify the emission distributions, offering a more flexible modelling approach to data than conventional parametric HMMs. We introduce a  Bayesian framework for inference, enabling the simultaneous estimation of all unknown model parameters including the number of states. A parsimonious knot configuration of the B-splines is identified by the use of a trans-dimensional Markov chain sampling algorithm, while model selection regarding the number of states can be performed based on the marginal likelihood within a parallel sampling framework. Using extensive simulation studies, we demonstrate the superiority of our methodology over alternative approaches as well as its robustness and scalability. We illustrate the explorative use of our methods for data on activity in animals, i.e. whitetip-sharks. The flexibility of our Bayesian approach also facilitates the incorporation of  more realistic assumptions and we demonstrate this by developing  a  novel hierarchical  conditional HMM to analyse human activity for circadian and sleep modelling. 
\end{abstract}

\noindent%
\vfill

\newpage
\spacingset{1.9} 
\section{Introduction}
\label{sec:introduction}
\addtolength{\textheight}{.5in}
The class of hidden Markov models (HMMs) offers a powerful approach for extracting information from sequential data \citep{rabiner1989tutorial}. A basic N-state HMM consists of a discrete-time stochastic process $(x_{t},y_{t})$ where $x_{t}$ is an unobserved N-state time-homogeneous Markov chain, and $y_{t}|x_{t} \sim f_{x_{t}}(y_{t})$ with the \textit{emission} distributions $f_{x_{t}}$ belonging to some parametric family such as normal or gamma. However, a parametric HMM is often too restrictive for complex real data \citep{zucchini2016hidden}. 
It is recognized that simple parametric choices for the emission distributions are not always  justified, and moreover, their misspecification can lead to seriously erroneous inference on the number and classification of  hidden states \citep{yau2011bayesian,pohle2017selecting}. Semi- and nonparametric modelling of emission distributions offer more flexibility and/or may serve as exploratory tools to investigate the suitability of a parametric family; see \citet{piccardi2007hidden} for activity recognition in videos, \citet{yau2011bayesian} for the analysis of genomic copy number variation, \citet{langrock2015nonparametric,langrock2018spline} for modelling animal movement data and \citet{kang2019bayesian} for delineating the pathology of Alzheimer's disease, among many others. Theoretical guarantees for inference in such models have been studied in a number of recent papers. Notably \citet{alexandrovich2016nonparametric} proved that model parameters and the order of the Markov chain are identifiable (up to permutations of the hidden states labels) if the transition probability matrix of $\{x_{t}\}$ has full rank and is ergodic, and if the emission distributions are distinct. These conditions are fairly generic and in practice will usually be satisfied. We also refer to \citet{gassiat2016inference,gassiat2016nonparametric} for further identifiability results, and to \citet{vernet2015posterior}, \citet{de2017consistent}, and references therein, for further theoretical results on inference under  nonparametric settings. The increased flexibility and modelling accuracy obtained by  non-parametric  emission distributions  comes at  a higher computational cost. For instance, the cost of the standard HMM algorithms (e.g. the forward-backward algorithm of  \citet{rabiner1989tutorial}) for kernel-based HMMs \citep{piccardi2007hidden} is subject to a quadratic growth with  data size $n$ and thus can be prohibitive for long time series data. Bayesian nonparametric HMMs \citep{yau2011bayesian}  built on Dirichlet process mixture models pose challenges to the existing sampling methods due to the increased complexity of the model space \citep{hastie2015sampling}.

Splines have good approximation properties for a rich class of functions \citep{de1978practical,schumaker2007spline}.
A spline function of order $O$ is a piecewise polynomial function of degree $O-1$ where the polynomial pieces are connected at the so-called knot points. Provided  these are distinct, the derivatives of piecewise polynomials are ($O-2$)-times continuously differentiable at the knots. B-splines (short for basis splines) of order $O$ provide basis functions for representing spline functions of the same order defined over the same set of knots \citep{de1978practical}. 
The great flexibility and nice computational properties of the B-splines make them a popular tool in semi-/nonparametric statistical modelling, especially in nonlinear regression analysis \citep{denison2002bayesian,zanini2020flexible} and density estimation \citep{koo1996bivariate,edwards2019bayesian}.
Incorporating B-splines into HMMs is attractive for real applications as  two powerful aspects can be exploited,  the forward-backward algorithm for efficient HMM inference, and the flexibility  for estimating the emission densities. A frequentist estimation approach for HMM based on penalized B-splines (P-splines) was  introduced by \citet{langrock2015nonparametric,langrock2018spline}. It requires pre-specifying the number and positions of knots where, in practice, a large number of knots are needed to ensure  flexibility, leading to  computational challenges (e.g. convergence to suboptimal local extrema of the likelihood) and cost. Also to date the selection of the state-specific smoothing parameters and the quantification of parameter uncertainty  remain  challenging inferential tasks in the frequentist framework. Current methods rely on cross-validation and parametric bootstrap techniques \citep{langrock2015nonparametric}, which are extremely computationally intensive and can be numerically unstable especially for increasing cardinality $N$. Hence their approach is so far only feasible for models with a small $N$ which may severely limit its applicability.

As far as we are aware spline-based methods have not yet been considered for emission density estimation in a Bayesian formulation of  HMMs. The aim of this article is to propose and develop a methodology that achieves exactly this by means of  an almost ``tuning-free" reversible jump Markov chain Monte Carlo (RJMCMC) algorithm \citep{green1995reversible}, which exploits (i) the  forward filtering backward sampling (FFBS) procedure for efficient simulation of the hidden state process, (ii) a stochastic approximation based adaptive MCMC scheme for automatic tuning, (iii) a reparameterization scheme for enhancing the sampling efficiency and (iv) an adaptive knot selection scheme that modifies and extends ideas considered in other scenarios such as \citet{dimatteo2001bayesian} and  \cite{sharef2010bayesian} for flexible emission modelling. We report results demonstrating significant advantages of our proposed adaptive spline based algorithm. Compared to current alternative spline-based approaches, namely the frequentist P-spline approach of \citet{langrock2015nonparametric}, and a Bayesian adaptive P-spline \citep{lang2004bayesian} approach which is newly adapted here for density estimation in HMMs, our method generally achieves higher estimation accuracy and efficiency while maintaining a much lower model complexity. It also performs favourably over the Gaussian mixture based HMM in more challenging data-generating scenarios.

Estimating $N$ is often a question of scientific interest in itself and introduces an additional level of complexity to HMM inference. While order estimation has been extensively studied  for parametric HMMs, such as in \citet{celeux2008selecting}, \citet{pohle2017selecting} and \citet{fruhwirth2006finite}, few theoretical or practical results have been obtained for the semi- or nonparametric case, which often requires the number of states to be known or fixed in advance \citep{piccardi2007hidden,yau2011bayesian,lehericy2018state}.
Recently, \citet{lehericy2019consistent} proposed two estimators for $N$, which are theoretically attractive but suffer from implementation difficulties such as non-convex optimization problems and heuristic tuning. In this paper, we address this issue with a fully Bayesian approach to the selection of $N$ through a parallel sampling scheme that is both easy to implement and computationally efficient. Quantities such as the marginal likelihood for each model can be easily estimated.

Among the numerous application fields of HMMs, there is a growing interest  within the context of e-Health to gain insight into an individual's health status based on relevant biomarker data. Physical activity (PA) is receiving much  attention  as an important biomarker of the sleep-wake cycle and circadian timing system, which is closely associated with our physical and mental health \citep{roenneberg2016circadian}.  PA can be easily and objectively measured  in a non-obtrusive way under normal living conditions using accelerometry or actigraphy through wearable sensing devices. \citet{huang2018hidden}  investigated  PA  using a HMM with circadian-clock driven  transition probabilities and, amongst various circadian parameters of interest, they  proposed a novel model-derived circadian parameter for monitoring and quantifying a subject's circadian rhythm. An advantage of a further Bayesian formulation is that the modularity of its components can be used to perform inference rigorously even in a more complex hierarchical HMM model. As a further contribution  based on our proposed Ansatz, we develop a hierarchical conditional HMM that may be applied to (i) characterize the sleep-wake patterns in the overall PA data of an individual  and (ii) analyze sleep patterns of an individual in a refined way through a ``sub-HMM" that is conditional on the ``rest" state inferred from (i).

The manuscript is structured as follows: Section 2 provides details of a Bayesian formulation of the spline-based HMM, Section 3 details the structure of our proposed inference algorithms including model selection on the number of states and summarizes the performance of our methods in comparison to other related methods under various simulation settings. Section 4 illustrates our methods on animal activity data and introduces the conditional HMM approach that is applied to human PA data from the Multi-Ethnic Study of Atherosclerosis
(MESA)\footnote{We refer to \citet{chen2015racial} and \citet{zhang2018national}) for more background information on the MESA dataset}. Section 5 provides a discussion and possible directions of further work.

\section{A Bayesian HMM with spline-based emissions}
\label{sec:The model}
We approximate the emission densities $f_{1},\ldots,f_{N}$, focusing on univariate emissions, using mixtures of standardized cubic B-spline basis functions of order $O=4$ \citep{langrock2015nonparametric}. The knots are located between boundary knots $a$ and $b$ (assumed fixed), and we use $R_{K}=(r_{1},\ldots,r_{K})$ to denote the interior knot configuration shared across states, with the left and right external knots set to $a$ and $b$, respectively \citep{friedman2001elements}. Note that $K=k$ corresponds to the case of $k+4$ B-spline basis functions, and we assume $K\geq 2$ for identifiability. Under these settings, $f_{i}$ is formulated as
\begin{equation}\label{eq:emission}
	f_{i}(y_{t})=\sum_{k=1}^{K+4}a_{i,k}B_{k}(y_{t}), \quad\quad  i=1,\ldots,N,
\end{equation}
where $B_{k}(y)$, $k=1,\ldots,K+4$, denotes the $k$-th normalized (such that it integrates to one) B-spline basis function of degree 3, and the $a_{i,k}$ are the corresponding coefficients such that $\sum_{k=1}^{K+4}a_{i,k}=1$ and $a_{i,k}\ge 0$ for all $k=1,\ldots,K+4$. 
In the time-homogeneous case, i.e. where the transition probabilities of the Markov chain are constant over time, the resulting class of HMMs is fully specified by the initial state distribution, $\bm{\delta}=(\delta_{1},\ldots,\delta_{N})$, with $\delta_{i}=P(x_{1}=i)$, the transition probability matrix, $\Gamma=(\gamma_{i,j})_{i,j=1,\ldots,N}$, with $\gamma_{i,j}=P(x_{t}=j|x_{t-1}=i)$, and the emission densities defined in \eqref{eq:emission}. The joint (complete) likelihood of observations $\mathbf{y}^{(n)}=(y_{1},\ldots,y_{n})$ and the hidden states $\mathbf{x}^{(n)}=(x_{1},\ldots,x_{n})$ is
\begin{equation}\label{eq:completellk}
	f(\mathbf{y}^{(n)},\mathbf{x}^{(n)}|K,R_{K},\bm{\delta},A_{K},\Gamma)=\delta_{x_{1}}\prod_{t=2}^{n}f(x_{t}|x_{t-1},\Gamma)\prod_{t=1}^{n}f_{x_{t}}(y_{t}),
\end{equation}
where here, and throughout this paper, we  use $f(\cdot|\cdot)$ as a generic notation to represent conditional densities as specified by their arguments and $A_{K}$ denotes the set of spline coefficients $a_{i,k}$, $i=1,\ldots,N, k=1,\ldots,K+4$. Integrating out the hidden states the marginal likelihood can be evaluated in $O(N^{2}n)$ steps using the forward algorithm (in the form of \citet{zucchini2016hidden}), via the matrix product expression 
\begin{equation} \label{eq:obsllk}
	\begin{split}
		f(\mathbf{y}^{(n)}|K,R_{K},\bm{\delta},A_{K},\Gamma) & = \int f(\mathbf{y}^{(n)},\mathbf{x}^{(n)}|K,R_{K},\bm{\delta},A_{K},\Gamma)d\mathbf{x}^{(n)} \\
		& = \bm{\delta} P(y_{1})\Gamma P(y_{2})\cdots\Gamma P(y_{n})\mathbf{1},
	\end{split}
\end{equation}
where $P(y_{t})$ is a diagonal matrix with $i$-th diagonal entry given by $f_{i}(y_{t})$ and $\mathbf{1}$ is a column vector  of dimension $N$ of ones.

To complete the Bayesian formulation of the model, we assume the following factorization of the complete joint density
\begin{displaymath}
	f(K,R_{K},\bm{\delta},A_{K},\Gamma,\mathbf{y}^{(n)},\mathbf{x}^{(n)})=f(K)f(\bm{\delta})f(\Gamma)f(R_{K}|K)f(A_{K}|K).
\end{displaymath}
\begin{equation*}
	\times f(\mathbf{y}^{(n)},\mathbf{x}^{(n)}|K,R_{K},\bm{\delta},A_{K},\Gamma).
\end{equation*}
The assumption that the parameters associated with the observed and hidden process are a-priori independent is commonly adopted in Bayesian HMMs. We use a uniform prior on $\{2,\ldots,K_{max}\}$ for $K$, with $K_{max}$ fixed to $50$ in our examples\footnote{Clearly larger default values, including the sample size $n$, may be used instead and the estimation results do not appear to be sensitive to its choice as long as $K_{max}$ is large enough.} where a preliminary study suggested that this was large enough to cover the support of $K$. For the knot positions, we propose that the $r_{k}$ are taken to be the $k$-th order statistics of $K$ independent uniform random variables on $[a,b]$, i.e. $f(R_{K}|K)=K!/(b-a)^{K}$. The state-specific spline coefficients $(a_{i,1},\ldots,a_{i,K+4})$, $i=1,\ldots,N$, are reparameterized as $a_{i,j}=\exp(\Tilde{a}_{i,j})/\sum_{l=1}^{K+4}\exp(\Tilde{a}_{i,l})$, $\Tilde{a}_{i,j}\in \mathbb{R}$, so that the positivity and unit sum constraints will not hinder the design of our RJ moves. The fact that the $\Tilde{a}_{i,j}$ are not identifiable is not a concern as we are only interested in the $a_{i,j}$, which are identifiable, and in this way the mixing of the MCMC may be improved \citep{cappe2003reversible}. We choose to use a log-gamma prior with shape parameter $\zeta$ and rate parameter $1$ on the $\Tilde{a}_{i,j}$, i.e. $\exp(\Tilde{a}_{i,j})\sim$ Gamma$(\zeta,1)$, giving a symmetric Dirichlet, i.e. Dir($\zeta,\ldots,\zeta$) distribution on the corresponding $(a_{i,1},\ldots,a_{i,K+4})$. We choose a vague Gamma$(1,1)$ hyperprior\footnote{In our experiments we did not find the values of these hyperparameters to be very influential, and other reasonable values may be used.} on $\zeta$ to reflect our uncertainty on $\zeta$ and our prior belief of sparse distributions on the spline coefficients (when $\zeta <1$). For the transition probability matrix we followed the literature (see, e.g. \citet{ryden2008versus}) assuming that the rows are a-priori independent, each of which has a vague Dirichlet prior $(\gamma_{i,1},\ldots,\gamma_{i,N})\sim Dir(1,\ldots,1)$, $i=1,\ldots,N$, where we assume that the initial distribution\footnote{Note that it is not possible to estimate it consistently as there is only one unobserved variable associated with it.}
is fixed and uniform on $\{1,\ldots,N\}$.
Thus the complete joint density incorporating the reparametrization can be rewritten as 
\begin{displaymath}
	f(\zeta,K,R_{K},\Tilde{A}_{K},\Gamma,\mathbf{y}^{(n)},\mathbf{x}^{(n)})=f(\zeta)f(K)f(\Gamma)f(R_{K}|K)f(\Tilde{A}_{K}|K,\zeta)
\end{displaymath}
\begin{equation}\label{eq:target}
	\times f(\mathbf{y}^{(n)},\mathbf{x}^{(n)}|K,R_{K},\Tilde{A}_{K},\Gamma), 
\end{equation}
where $\Tilde{A}_{K}$ represents the set of $\Tilde{a}_{i,k}$ ($i=1,\ldots,N, k=1,\ldots,K+4$). We remark that in cases where N is large, using state-specific knot configurations for emissions may be preferred over a shared knot configuration across states, and our Bayesian model can be readily adapted. See supplementary Section \ref{sec:AP_statespec_adsp} for more details.

\section{Inference}
\label{sec:The algorithm}
Our aim is to obtain realisations from the posterior distribution of $(K,R_{K},A_{K},\Gamma,\zeta,\mathbf{x}^{(n)}|\mathbf{y}^{(n)})$, which can be achieved by simulating from the joint posterior density defined via \eqref{eq:target}. To allow for model searches between parameter subspaces of different dimensionality, we develop a RJMCMC algorithm which combines a Metropolis-within-Gibbs sampler with trans-dimensional moves generated by  births and deaths of knot points. The structure of our algorithm is listed in Algorithm 1, where $b_{K}=\mathbf{I}(K=2)+0.5\times \mathbf{I}(3\leq K<K_{max})$ and $\mathbf{I}(\cdot)$ is the indicator function (therefore $b_K=0.5$ for $3\leq K<K_{max}$). The RJMCMC algorithm is conditioned on the cardinality $N$ noting that model selection will be addressed in Section \ref{sec:model selection}. Steps (a)-(e) propose moves within a dimension while the last step proposes a birth or death of a knot point which changes the model dimension. We now outline the rules for each of the updating steps while further details of this algorithm together with an extension assuming state-specific knots are provided in the supplementary Section \ref{sec:AP_statespec_adsp}. 
\begin{algorithm}[h!]
	\SetAlgoLined
	Initialize $K$, $R_{K}$, $\zeta$, $\Tilde{A}_{K}$, $\Gamma$  \;
	\For{i=1, \ldots, T}{
		
		(a) update the hidden state sequence $\mathbf{x}^{(n)}$\;
		(b) update the transition probability matrix $\Gamma$\;
		(c) update the knot location vector $R_{K}$\;
		(d) update the set of B-spline coefficients $A_{K}$ (via $\Tilde{A}_{K}$)\; 
		(e) update the hyperparameter $\zeta$\;	
		draw $U\sim U(0,1)$\;
		\eIf{$U<b_{K}$}{
			consider the birth of a knot point in the B-spline representation in \eqref{eq:emission}\;
		}{
			consider the death of a knot point in the B-spline representation in \eqref{eq:emission}\;
		}
	}
	\caption{Reversible jump MCMC algorithm for spline-based HMMs}
\end{algorithm}
\subsection{Within-model moves}
The moves in steps (a) and (b) are of Gibbs type whereas those in steps (c) to (e) are of Metropolis-Hastings (MH) type, all of which are conditioned on the current number of knot points $K$. In step (a), $\mathbf{x}^{(n)}$ can be simulated exactly and efficiently from its full conditional distribution, $f(\mathbf{x}^{(n)}|\mathbf{y}^{(n)}, K, R_{K}, A_{K}, \Gamma)$, via a standard FFBS procedure with transition matrix $\Gamma$ and emission densities $f_{i}(y_{t})$ given in \eqref{eq:emission} (see e.g. \citet{cappe2007inference}). In step (b), the rows of $\Gamma$ are conditionally independent and are updated from their conjugate Dirichlet posterior
\begin{displaymath}
	(\gamma_{i,1},\ldots,\gamma_{i,N})\sim Dir(1+n_{i,1},\ldots,1+n_{i,N}),\quad i=1,\ldots,N,
\end{displaymath}
where $n_{i,j}$ denotes the number of transitions from state $i$ to $j$ in $\mathbf{x}^{(n)}$. In step (c) a knot $r_{k^{*}}$ is chosen uniformly from the set of existing knots $\{r_{1},\ldots,r_{K}\}$ and proposed to be moved to a candidate point, $r_{c}$, which is generated from a normal distribution with mean $r_{k^{*}}$ and standard deviation $\tau_{1}$, truncated to $[a,b]$ \citep{dimatteo2001bayesian}. The proposal in step (d) is generated by a random walk on the reparameterized spline coefficients $\Tilde{a}_{i,j}$ $(i=1,\ldots,N; j=1,\ldots,K+4)$, i.e. $\Tilde{a}_{i,j}^{'}=\Tilde{a}_{i,j}+\eta_{i,j}$, where $\eta_{i,j}\sim \mathcal{N}(0,\tau_{2}^{2})$. In step (e), we update $\zeta$ via a log-normal random walk $\log(\zeta^{'})=\log(\zeta)+\nu$, where $\nu\sim \mathcal{N}(0,\tau_{3}^{2})$. 
To allow for automatic tuning of the variance parameters $\tau_{1}$, $\tau_{2}$ and $\tau_{3}$, we adopt a simple well-used adaptive MCMC scheme based on a stochastic approximation procedure \citep{atchade2011adaptive}, without incurring additional computational burden. More details are given in supplementary Section \ref{sec:AP_RJMCMC}.

\subsection{Birth and death moves}
The birth and death moves allow for increasing or decreasing the number of knots, or equivalently, the number of B-spline basis elements. Our design extends the ideas of \citet{dimatteo2001bayesian} and \citet{sharef2010bayesian} to the framework of HMMs. Suppose that the current model has knot configuration $(K,R_{K})$, we  make a random choice between birth or death with probabilities $b_{K}$ and $d_{K}=1-b_{K}$, respectively.  In the birth move, we select a knot, $r_{b^{*}}$, at random from the existing knots and create a candidate new knot, $r_{c}$, by drawing from a normal distribution (truncated to $[a,b]$) with mean $r_{b^{*}}$ and standard deviation $\tau(R_{K},b^{*})$, where $\tau$ is chosen as a function having the form $(r_{b^{*}+1}-r_{b^{*}-1})^{\alpha}$ and $\alpha$ is a positive real constant. The intuition here is that a new knot is more likely to be needed in locations where existing knots are relatively ``dense". To complete the birth step we update the corresponding spline coefficients, which now has dimension $K+5$ for each state. Here, our design is guided by the deterministic knot insertion rule described in \citet{de2001practical} which allows a new knot to be inserted without changing the shape of the overall B-spline curve, noting that  this exact relationship becomes approximate in our context as we are working with normalized basis functions. We extend the scheme by adding more degrees of freedom in order to meet the dimension matching condition required for the validity of the RJMCMC algorithm. More specifically, for the birth of a candidate knot point $r_{c}\in (r_{n^{*}},r_{n^{*}+1})$, the associated spline parameters $\Tilde{a}^{'}_{i,j}$, for $i=1,\ldots,N$, are created as 
\begin{equation}\label{eq:birth}
	\Tilde{a}^{'}_{i,j} =
	\begin{cases}
		\Tilde{a}_{i,j} & 1\leq j\leq n^{*}+1 \\
		c_{j}\Tilde{a}_{i,j}+(1-c_{j})\Tilde{a}_{i,j-1} & n^{*}+1<j<n^{*}+4 \\
		u_{i}\Tilde{a}_{i,j}+(1-u_{i})\Tilde{a}_{i,j-1} & j=n^{*}+4\\
		\Tilde{a}_{i,j-1} & n^{*}+4<j\leq K+5
	\end{cases}
\end{equation}
where $c_{j}=(r_{c}-r_{j-4})/(r_{j-1}-r_{j-4})$ and $u_{i}\stackrel{iid}{\sim} U(0,1)$. Here the $\Tilde{a}^{'}_{i,j}$ are generated using the deterministic rule in \citet{de2001practical}, except for $\Tilde{a}^{'}_{i,n^{*}+4} $ where we introduce one degree of freedom through $u_{i}$. This way of updating allows us to effectively use knowledge from current spline parameters, while also allowing for a possible improvement on the fit resulting from the introduction of a new knot point. This design can also be related to the idea of ``centering" reversible jump proposals proposed in \citet{brooks2003efficient} where current and proposed parameters produce similar likelihoods. 

Next, consider the death of a knot point from the current knot configuration $(K,R_{K})$. A knot, $r_{d^{*}}$, is chosen at random from the set of existing knots $\{r_{1},\ldots,r_{K}\}$ and then deleted. The spline parameters associated with this move are updated according to the inverse transformation of \eqref{eq:birth}: 
\begin{displaymath}
	\Tilde{a}^{'}_{i,j} =
	\begin{cases}
		\Tilde{a}_{i,j} & 1\leq j\leq d^{*} \\
		\frac{\Tilde{a}_{i,j}-(1-c_{j})\Tilde{a}^{'}_{i,j-1}}{c_{j}} & d^{*}<j<d^{*}+3 \\
		\Tilde{a}_{i,j+1} &  d^{*}+3\leq j\leq K+3
	\end{cases} 
\end{displaymath}
where $c_{j}=(r_{d^{*}}-r_{j-4})/(r_{j-1}-r_{j-4})$. The parameters for the state process remain unaltered in either birth or death move\footnote{We note the difference between our birth and death proposals to those in \citet{sharef2010bayesian} (see equation 3.1 therein), who propose a parameterization where the transformation acts on the exponentials of the spline coefficients (restricted to be positive). Such a scheme may be problematic as the proposed parameters from the death step based on the deterministic rules are not guaranteed to be positive.}.  

\subsection{Bayesian model selection: Estimating the cardinality $N$}
\label{sec:model selection}
While it is theoretically possible to extend Algorithm 1 by introducing an additional reversible jump step on the number of states, or by using a product space search algorithm to sample from the joint posterior of parameters from all competing models (e.g. \citet{carlin1995bayesian}), it is challenging to design computationally practical trans-dimensional algorithms in the HMM setting due to the large and complex parameter space. Another potential strategy is to use Dirichlet process based priors for the transition matrix, which allows for a potentially infinitely large state space (see e.g. \citet{fox2011sticky}). However, combining such a framework with the spline-based emission model would pose significant computational challenges.
Instead, we propose to perform  model selection based on the marginal likelihood, also known as ``evidence"
\begin{equation}\label{evidence}
	f(\mathbf{y}^{(n)}|N=j)=\int f(\mathbf{y}^{(n)}|\bm{\theta}_{j},N=j)f(\bm{\theta}_{j}|N=j)d\bm{\theta}_{j}, \quad j=1,\ldots,M,
\end{equation}
where $\bm{\theta}_{j}$ is the parameter set (excluding $\mathbf{x}^{(n)}$) associated with the $j$-state model, $f(\mathbf{y}^{(n)}|\bm{\theta}_{j},
\\N=j)$ is the observed likelihood given in \eqref{eq:obsllk} and $M$ denotes some maximum number of states that we want to consider. Given prior model probabilities $P(N=j)$ and evidences, the posterior model probabilities can be computed using Bayes' theorem. Following Bayesian decision theory we can pick the model that gives the highest posterior probability, i.e. $N^{*}=\argmax_{k=1,\ldots,M}P(N=k|\mathbf{y}^{(n)})$. For most models of interest (including HMMs), however, the integral in \eqref{evidence} has no closed-form expression and needs to be approximated. Various Monte Carlo based approximation schemes have been proposed, see \citet{friel2012estimating} and \citet{llorente2020marginal} for  recent reviews. 

We propose to approximate the evidence of a spline-based HMM by using a harmonic mean estimator  \citep{gelfand1994bayesian}, which allows direct estimation of the evidence using the simulation output and thus is straightforward to implement. The estimator relies on the simple fact that for any proper density function $h$, we have for the expectation
\begin{displaymath}
	\mathbf{E}_{\bm{\theta}_{j}|\mathbf{y}^{(n)}}\bigg[\frac{h(\bm{\theta}_{j})}{f(\bm{\theta}_{j})f(\mathbf{y}^{(n)}|\bm{\theta}_{j})}\bigg]=\int \frac{h(\bm{\theta}_{j})}{f(\bm{\theta}_{j})f(\mathbf{y}^{(n)}|\bm{\theta}_{j})}f(\bm{\theta}_{j}|\mathbf{y}^{(n)})d\bm{\theta}_{j}=\frac{1}{\mathbf{M}_{j}},
\end{displaymath}
where $\mathbf{M}_{j}=\int f(\bm{\theta}_{j})f(\mathbf{y}^{(n)}|\bm{\theta}_{j})d\bm{\theta}_{j}$. A Monte Carlo approximation of the evidence is thus obtained as 
\begin{displaymath}
	\mathbf{\hat{M}}_{j}=\bigg\{\frac{1}{T}\sum_{i=1}^{T}\frac{h(\bm{\theta}_{j}^{(i)})}{f(\bm{\theta}_{j}^{(i)})f(\mathbf{y}^{(n)}|\bm{\theta}_{j}^{(i)})}\bigg\}^{-1},
\end{displaymath}
where $\bm{\theta}_{j}^{(i)}$ is the $i$-th sample simulated from the posterior $f(\bm{\theta}_{j}|\mathbf{y}^{(n)})$. This estimator enjoys a finite variance if $\int h^{2}(\bm{\theta})/(f(\bm{\theta})f(\mathbf{y}^{(n)}|\bm{\theta}))d\bm{\theta}<\infty$, i.e. $h(\bm{\theta})$ must have lighter tails than $f(\bm{\theta})f(\mathbf{y}^{(n)}|\bm{\theta})$ \citep{diciccio1997computing}. To ensure this we follow \citet{robert2009computational} and \citet{marin2009importance} to construct an appropriate density $h$ based on truncated highest posterior density (HPD) regions derived from the MCMC samples. The resulting estimator is known as a truncated harmonic mean estimator and has been successfully used in various other model settings, see for instance \citet{durmus2018efficient} and \citet{acerbi2018bayesian}. More specifically, we define a sample-based $100\beta\%$ HPD region as (omitting the dependence on the index of state j for clarity) $\Tilde{\mathbf{H}}_{\beta}=\{\bm{\theta}^{(i)}:\ f(\bm{\theta}^{(i)})f(\mathbf{y}^{(n)}|\bm{\theta}^{(i)})>\Tilde{q}_{\beta}\}$, where $\Tilde{q}_{\beta}$ is the empirical upper $\beta$ quantile of the $(f(\bm{\theta}^{(i)})f(\mathbf{y}^{(n)}|\bm{\theta}^{(i)}))$ produced in the output of the MCMC. Here we propose to construct the density $h$ as
\begin{equation*}
	h(\bm{\theta})=\frac{1}{V(\xi)\beta T}\sum_{j:\bm{\theta}^{(j)}\in \Tilde{\mathbf{H}}_{\beta},dim(\bm{\theta}^{(j)})=dim(\bm{\theta})}\mathbf{I}(d(\bm{\theta}^{(j)},\bm{\theta})<\xi),
\end{equation*}
where $V(\xi)$ is the volume of a ball centered at $\bm{\theta}$ with radius $\xi$ (small), $dim(\cdot)$ is the dimensionality of the argument and $d(\cdot,\cdot)$ is a suitable distance measure. It is easy to check that $h$ is a proper density function and has a finite support, noting that in our context the parameter space of $\bm{\theta}=(K,\zeta,R_{K},A_{K},\Gamma)$ is a union of subspaces of varying dimension. Our proposal $h$ can be viewed as a histogram-like nonparametric estimator of the posterior $f(\bm{\theta}|\mathbf{y}^{(n)})$ based solely on samples in the HPD regions. Note that $V(\xi)$ does not need to be computed as it cancels out when computing posterior model probabilities, provided that $\xi$ is fixed across models.   

 \subsection{Performance in Simulations}
 \label{sec:simulation}
To thoroughly evaluate the empirical performance of our proposed adaptive spline (adSP) Bayesian methodology, we conducted simulation studies in various hypothetical and realistic settings. Here we briefly summarize the design and results, with additional details provided in supplementary sections \ref{sec:appsimulation} and \ref{sec:AP_simulation2}.

We first compared our adSP with three other relevant candidate methods: (i) the frequentist P-spline (fpSP) approach of \cite{langrock2015nonparametric}, (ii) a Bayesian adaptive P-spline approach (bpSP) that, to the best of our knowledge, represents the first implementation within HMMs, and (iii) a frequentist Gaussian mixture-based HMM (GMM) motivated by \citet{volant2014hidden}.
Our comparison is based on estimation accuracy, using two criteria, namely the average  Kullback-Leibler divergence (KLD) from the true emission distributions, and the decoding accuracy/error,  quantified via the proportion of correctly/incorrectly classified states, as well as computational cost. We generated artificial data from four simulation models with emissions exhibiting  features such as multi-modality, skewness, heavy-tailedness and excess kurtosis. Model 1 is a 2-state HMM  with a normal and a normal mixture emission as considered in \citet{langrock2015nonparametric}. Model 2 is a 3-state HMM with a unimodal positively skewed emission distribution in state 1, a bimodal distribution in state 2, and a unimodal negatively skewed distribution in state 3. Model 3 is motivated by the \textit{bimod} model considered in \citet{yau2011bayesian} with emissions using a mixture of a Laplace and a generalized Student's t distribution.  Model 4 corresponds to the \textit{trimod} case of \citet{yau2011bayesian}, a 2-state HMM  with emissions specified as a mixture of three well-separated normal distributions. Models 3 and 4 pose the most serious  challenges, even when the correct number of states is assumed to be known as in \citet{yau2011bayesian}. 

First, we verified the performance of our model selection method, which is based on the marginal likelihood, for the two Bayesian approaches, adSP and bpSP.  We found that the adSP method identified the correct number of states  in all 60 replicates for each of the four simulation models, with averaged posterior probability of the correct model equal to one. In contrast, bpSP had a lower accuracy and underestimated  the number of states in some repetitions. 
Furthermore, comparing the performance of the four methods across the  simulation models for fixed cardinality $N$, we conclude that the proposed adSP method is the only method that performs consistently well in all scenarios, with a particular advantage in decoding. Within the spline-based methods, adSP and bpSP had roughly comparable accuracy in the easier settings (Models 1 and 2), while the advantages of using adSP became apparent in the more challenging scenarios (Models 3 and 4) where bpSP suffered from poor mixing or convergence issues. Although fpSP was found to have better convergence than bpSP, it yielded lower accuracy than adSP in almost all cases. The GMM method performs well when the true emissions were close to Gaussian or Gaussian mixtures, but its performance was weak when emissions possessed non-Gaussian properties (e.g. skewness or heavy tails). The choices of its associated hyperparameters are influential to the results (overfitting or over-penalization), which is in agreement with previous findings \citep{baudry2015mixtures,fan2021simultaneous}. It also suffered from serious convergence issues in challenging scenarios like Model 4. In terms of computational efficiency, the adSP was generally the second most efficient approach after GMM, while bpSP was the most computationally costly. It is important to note that when calculating the computational time for the frequentist approaches, i.e. fpSP and GMM, the additional time required for performing uncertainty quantification of the parameters was not included. In practice this poses a significant computational burden, especially for fpSP, while it is provided by the Bayesian methods without extra cost.

We considered two additional simulation scenarios to further assess the applicability of our proposed methods (see supplementary Section \ref{sec:AP_simulation2}). We examined the robustness of adSP in the presence of misspecified transition dynamics, where data were generated using either a semi-Markov or non-homogeneous state process. We also tested the feasibility and scalability of the algorithm based on state-specific knot configurations in handling systems with a larger cardinality $N$. In both cases, our proposed methods showed strong performance in estimation accuracy and efficiency.


\section{Applications}
\label{sec:application}
\subsection{Analysis of oceanic whitetip shark acceleration data}
\begin{figure}[h!]
	\centering
	\includegraphics[width=0.33\textwidth]{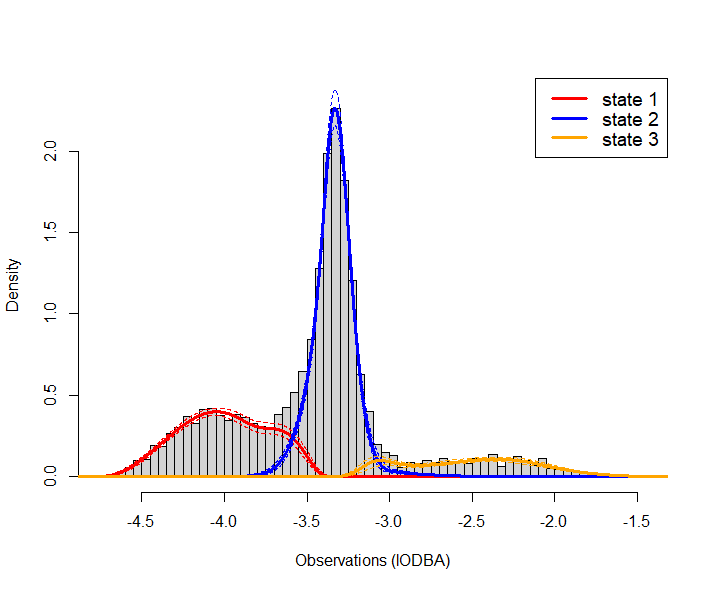}\includegraphics[width=0.33\textwidth]{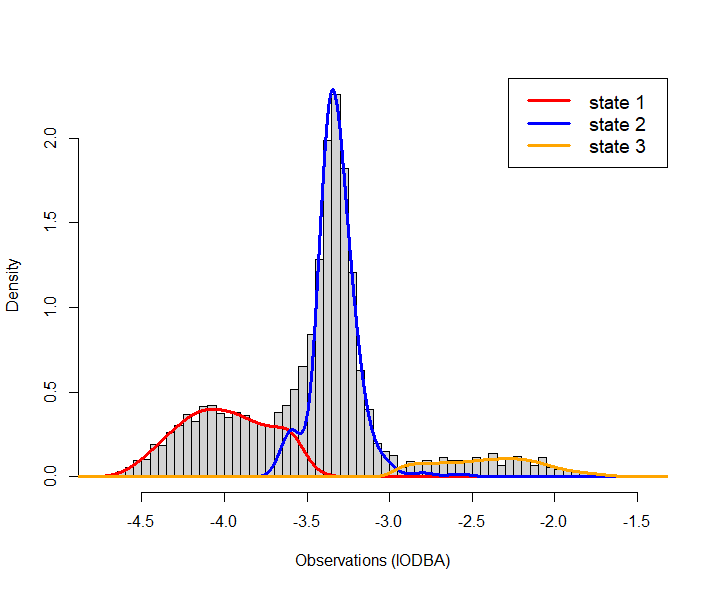}
	\includegraphics[width=0.33\textwidth]{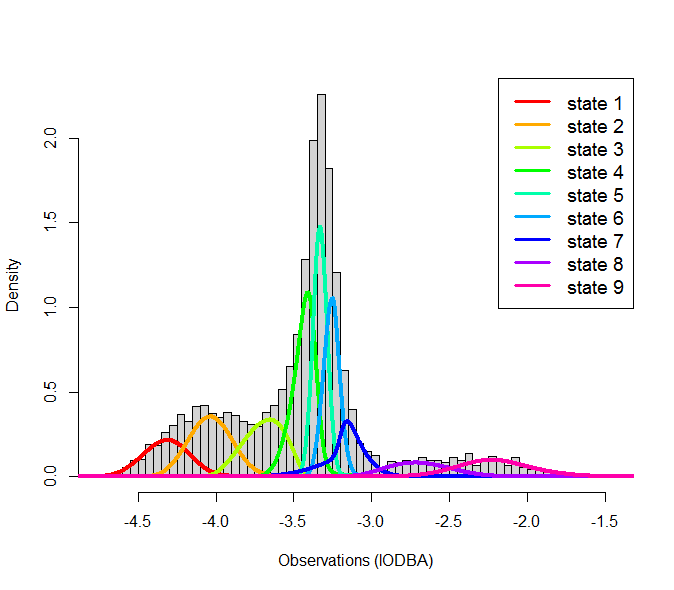}
	\caption[Estimation results for shark data]{\label{fig:sharkN3} Left, middle and right panels show the histogram of 15s-averaged lODBA values along with the estimated emission densities (weighted according to their proportion in the stationary distribution of the estimated Markov chain) obtained from our method ($N=3$), Langrock et al.'s method ($N=3$) and the 9-state model, respectively. Here the state labels are sorted according to their mean lODBA levels.}
\end{figure}

HMMs provide a useful tool for modelling animal movement metrics to study the dynamic patterns of an animal's behavioural states (e.g., resting, foraging or migrating) in ecology \citep{langrock2012flexible,langrock2018spline}. Here we consider a time series of the overall dynamic body acceleration (ODBA) collected from an oceanic whitetip shark at a rate of $16$ Hz over a time span of $24$ hours. A larger replicate data set was analyzed in \citet{langrock2018spline}. For our analysis, the raw ODBA values are averaged over non-overlapping windows of length 15 seconds and log transformed (lODBA), resulting in a total of 5760 observations. The marginal distribution of the transformed data is shown in Figure \ref{fig:sharkN3}. We  modelled the lODBA values using our  adSP  with  cardinality set to $N=3$  as in \citet{langrock2018spline}, 
who present biological reasons for this assumption. Implementational details of the MCMC algorithm are provided in supplementary Section \ref{sec:AP_shark}. Figure \ref{fig:sharkN3} (left panel) shows the estimated emission densities (obtained as in supplementary Section B.1) along with pointwise $95\%$ credible intervals. The posterior modal number of knots is $13$, with $\hat{P}(K=13|data)=0.689$. For comparison, we also fitted a B-spline HMM using the  method of 
\citet{langrock2018spline}, where we have set $K=39$ to ensure enough flexibility and selected 
$\lambda=(300,1,400)$ for the smoothing parameters based on our experiments. 
While the resulting transition probability estimates and the density fits seem to be comparable 
between the two approaches  (see Figure \ref{fig:sharkN3} middle panel), their method
uses approximately three times the number of parameters as our method for estimating the emissions, for which we experienced numerical stability issues during  estimation\footnote{This finding is consistent with our experience in the simulation studies.}. Therefore, the bootstrap-based uncertainty quantification approach would be challenging and costly to implement, whereas we obtained posterior uncertainties for the parameters at no extra cost. 

The computational feasibility and stability of the proposed algorithm allowed us to increase the value of $N$ and perform model selection. Preliminary runs indicated that $N>5$ was likely to be favoured, so we used adSP with state-specific knots along with the marginal likelihood approach described above. Using a discrete uniform prior over $\{2,\ldots,10\}$, the posterior modal number of states was estimated to be $N=9$, with a posterior probability of 1, thus the data strongly support a considerably larger number of states than was originally assumed in \citet{langrock2018spline}. In addition, individual emissions of the 9-state model now appear unimodal (see right panel of Figure \ref{fig:sharkN3}). This is interesting: apart from investigating the biological interpretation of these states, our results suggest that one might consider fitting a fully parametric HMM with standard unimodal forms of emission densities for $N=9$. Further detailed results of the application to the shark data can be found in  supplementary Section \ref{sec:AP_shark}.

\subsection{A conditional HMM for analysing circadian and sleep patterns in human PA data}
\label{sec:condHMM}
\subsubsection{The  model}
The use of PA data obtained from wearable sensors  for monitoring circadian rhythm and sleep pattern  outside laboratory settings is well justified \citep{ancoli2015sbsm,quante2018actigraphy}.  We next use our proposed Ansatz to introduce a conditional hidden Markov model in its general form and illustrate the method using publicly available human PA data  from MESA. The conditional HMM consists of a ``main model" to characterize the general pattern of the overall time series, and a ``sub-HMM" that is invoked based on a specific state $i$ of the main model with $N_{S}$ possible sub-states. Without loss of generality, we set $i=1$ and for simplicity omit this subscript in what follows. The posterior distribution of the parameters in the sub-HMM can be expressed as
\begin{equation}\label{subhmmAS}
	f(\bm{\theta}^{S}|\mathbf{y}^{(n)})=\int f(\bm{\theta}^{S}|\mathbf{x}^{(n)},\mathbf{y}^{(n)})f(\mathbf{x}^{(n)}|\mathbf{y}^{(n)})d\mathbf{x}^{(n)},
\end{equation}
where $\bm{\theta}^{S}$ is the parameter set for a $N_{S}$-state sub-HMM, $\mathbf{x}^{(n)}$ is the hidden state sequence associated with the main-HMM and we assume that
\begin{equation}\label{subhmmcondpost}
	f(\bm{\theta}^{S}|\mathbf{x}^{(n)},\mathbf{y}^{(n)})\propto f(\bm{\theta}^{S})f(\mathbf{y}^{(n)}|\bm{\theta}^{S},\mathbf{x}^{(n)}).
\end{equation}
We refer to the second term in \eqref{subhmmcondpost} as the ``conditional likelihood" for the sub-HMM. The key idea here is that by conditioning on state $1$ of the main-HMM, we want to restrict the observations that contribute to the likelihood to only those associated with the time points where $x_t=1$, while also maintaining the temporal dependence of these observations. To achieve this, we could simply treat observations $\{y_{t}: x_{t}\neq 1\}$ as ``missing data". This strategy offers the advantage that the resulting conditional likelihood can be easily evaluated in the HMM framework. More specifically, let $(t_{1},\ldots,t_{T_{1}})$ be the collection of time points in ascending order such that $x_{t_{j}}=1$, $j=1,\ldots,T_{1}$. Using the notation of Section 2, we have
\begin{equation}\label{subhmmcondllk}
	\begin{split}
		f(\mathbf{y}^{(n)}|\bm{\theta}^{S},\mathbf{x}^{(n)}) & = f(y_{t_{1}},\ldots,y_{t_{T_{1}}}|\bm{\theta}^{S}) \\
		& = \sum_{x_{t_{1}},\ldots,x_{t_{T_{1}}}}f(y_{t_{1}},\ldots,y_{t_{T_{1}}},x_{t_{1}},\ldots,x_{t_{T_{1}}}|\bm{\theta}^{S}) \\
		& =\bm{\delta} P(y_{1})\Gamma P(y_{2})\cdots\Gamma P(y_{n})\mathbf{1}
	\end{split},
\end{equation}
where $P(y_{t})=I_{N_{S}}$, the identity matrix of dimension $N_{S}$, for $t\neq t_{1},\ldots,t_{T_{1}}$. The last row of \eqref{subhmmcondllk} takes the same form as the marginal likelihood of a standard HMM, and therefore, the standard forward algorithm can be used to efficiently evaluate the conditional likelihood. 
Note that the uncertainty regarding the state classification is properly taken into account, as the state sequence will be integrated out to obtain the marginal posterior as defined in \eqref{subhmmAS}, on which our inference for the sub-HMM will be based.

It should be pointed out that our conditional HMM approach is different from what is called the hierarchical HMM (see e.g. \citet{adam2019joint}) in that for the latter, a joint model is formulated for multiple observed processes at different temporal resolutions, each of which is modelled via a hidden Markov process and the process at the coarser level determines the onset of a specific finer level process for each epoch. In contrast, our method may operate on a single time scale and allows us to refine our analysis  of a chosen state. It is also important to note that fitting the main HMM with $N+N_S-1$ states will not necessarily split state 1 of the original model into $N_S$ sub-states as desired, whereas in our framework we  control this directly  through the  conditioning.

In our application we are interested in studying sleep from accelerometer data, where state 1 corresponds to the lowest activity state that contains the sleep bouts. However, accelerometer data often contain a large number of zeros during a rest or sleep state \citep{ae2018missing}, which could cause issues for the spline-based model.  To address this,  we assume a ``zero inflation" of the  emissions at both HMM levels\footnote{The  model can  be easily adapted to  applications where further discrete mass points  for low observations  are needed.}
\begin{displaymath}
	f_{x_{t}}(y_{t})=w_{x_{t},1}\delta_{0}+w_{x_{t},2}f^{B}_{x_{t}}(y_{t}),
\end{displaymath}
where $x_{t}$ indicates the underlying state at time t, $w_{x_{t},1}$ represents the state-specific zero weight 
such that $0\leq w_{x_{t},1}\leq 1$ and $w_{x_{t},1}+w_{x_{t},2}=1$, $\delta_{0}$ is the Dirac delta distribution and $f^{B}_{x_{t}}(y_{t})$ is a spline-based emission density as defined in Section 2. 
Following \citet{gassiat2016inference} we can establish identifiability of the resulting HMM provided 
that at most one $w_{x_{t},1}$ is equal to one and that $\{\delta_{0},f^{B}_{1},\ldots,f^{B}_{N}\}$ are linearly independent. In our analysis these conditions are always satisfied. 

\subsubsection{Application to the MESA dataset}
To illustrate our proposed method, we consider two example subjects, A  and B, corresponding  to  subjects 921 and 3439 in the  MESA dataset, respectively,  who both have no diagnosed sleep related diseases. 
Subjects wore an actigraph (Actiwatch Spectrum) on the non-dominant wrist for one week and activity was measured in each 30-s epoch by counting the number of times movement intensity crossed a threshold. The resulting values reflect the overall activity intensity in each epoch.
Additionally, each subject undertook a polysomnography (PSG) session for one night during the monitoring period.
PSG is a multi-sensor approach that collects multiple physiological signals from the body and is considered as the gold standard of measuring sleep \citep{berry2012aasm}. Wake and four sleep stages (N1, N2, N3 and REM) were identified for every 30-s epoch using the criteria set out by the American Academy of Sleep Medicine. Among these, N1 and N3 correspond to light and deep sleep, respectively, while N2 is an intermediate stage. N1, N2 and N3 are collectively referred to as non-REM stages  \citep{berry2012aasm}. The REM stage is physiologically distinct from the other stages and associated with dreaming \citep{stein2012heart}. Typically, individuals go through the four stages several times during a night's sleep. For the main-HMM, we used 5-min averaged PA and set $N=3$ as in \citet{huang2018hidden}\footnote{One could  first perform model selection with the marginal method introduced earlier. However, for comparison and since model selection is not the main theme of this application we use the settings of \citet{huang2018hidden}. They found that $N=3$ tended to be the optimal choice in parametric HMM modelling of this kind of data, which, furthermore,  consistently over many individuals, assigned the rest/sleep periods to  night times}.  For the sub-HMM  we assume 2 sub-states,  1.1 and 1.2,  to potentially capture the {\it ultradian} oscillations between higher and lower intensity of movement during sleep. Such were found in accelerometer data by \citet{winnebeck2018dynamics} who concluded that they   correspond  to the circa 120-min periodic transitions between the Non-REM and REM stages of sleep. However, their Ansatz could not account for a stochastic oscillating pattern. We based inference for the sub-HMM  on the finest possible time resolution of the raw 30-s PA counts to focus on the detail of activity during sleep. The implementational details of the MCMC algorithm are provided in  supplementary Section \ref{sec:AP_condHMM}.

\begin{figure}[h!]
	\centering
	\includegraphics[width=0.496\textwidth,height=4.2cm]{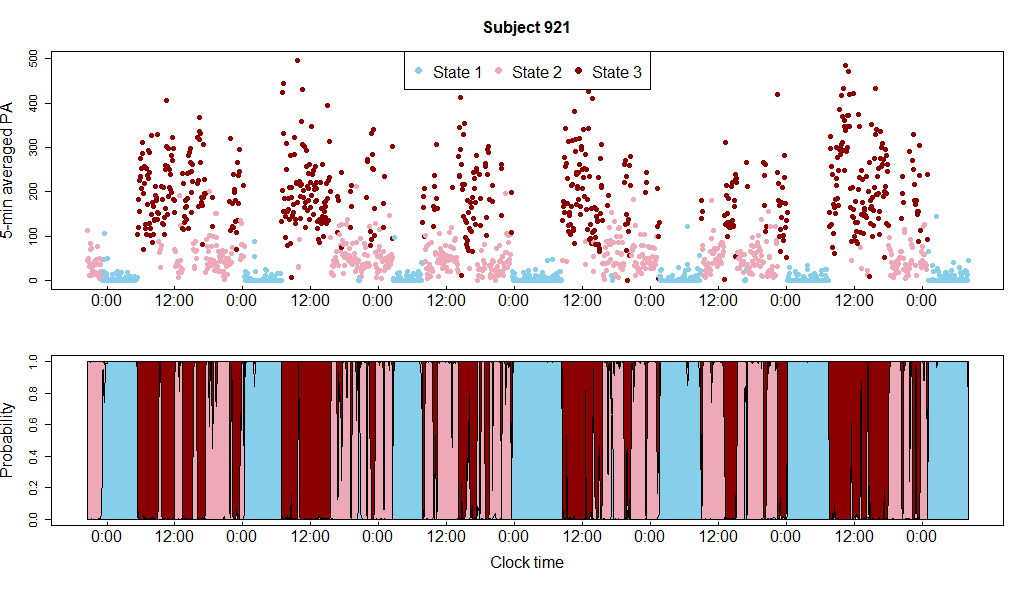}
	\includegraphics[width=0.496\textwidth]{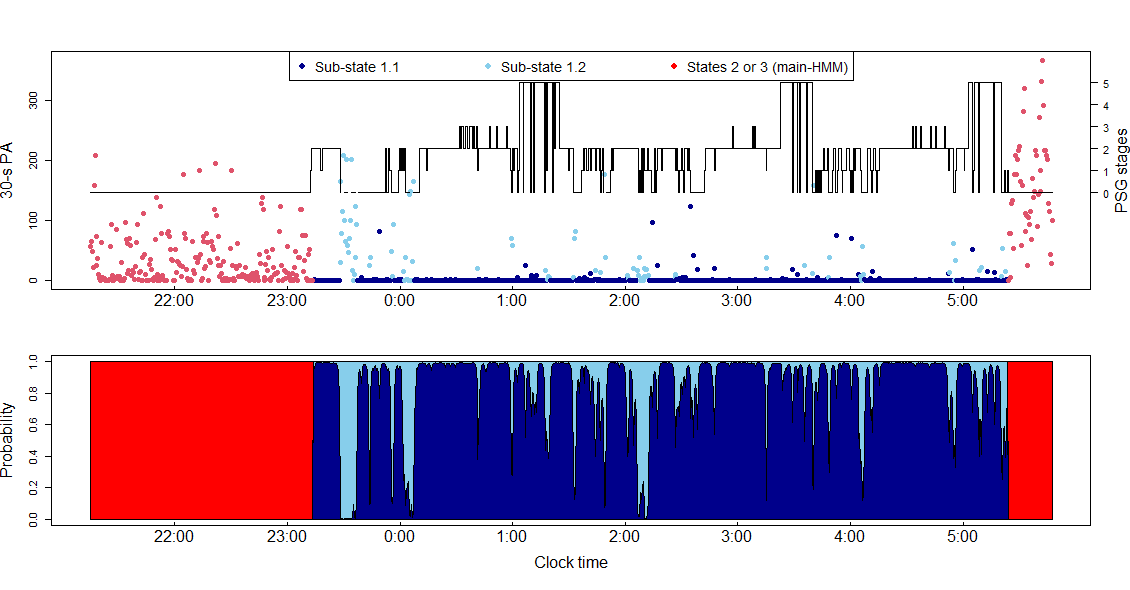}
	\includegraphics[width=0.496\textwidth]{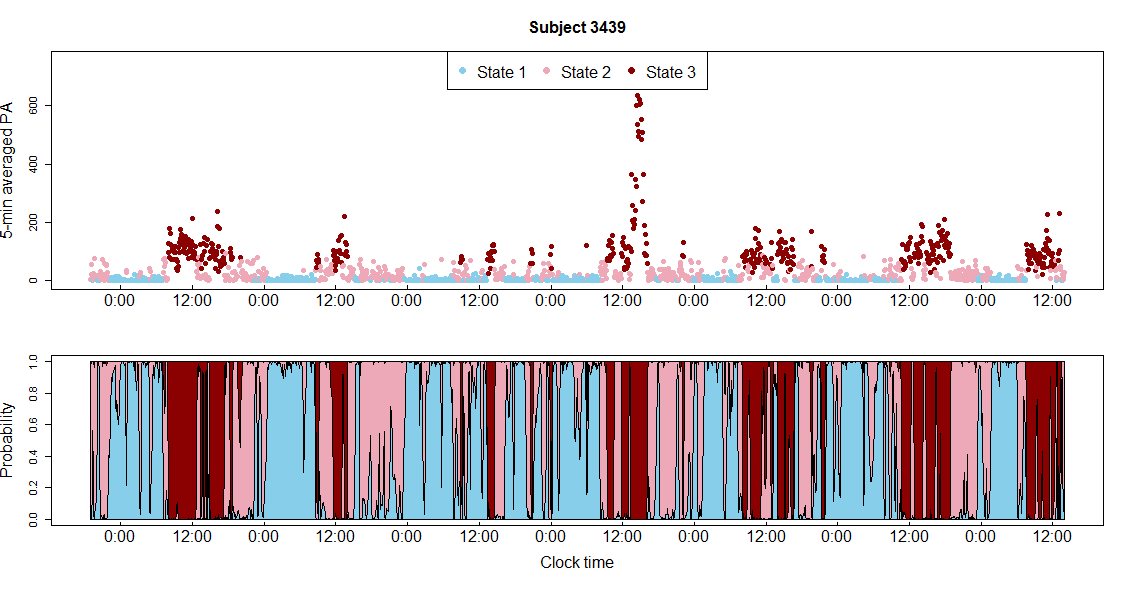}
	\includegraphics[width=0.496\textwidth]{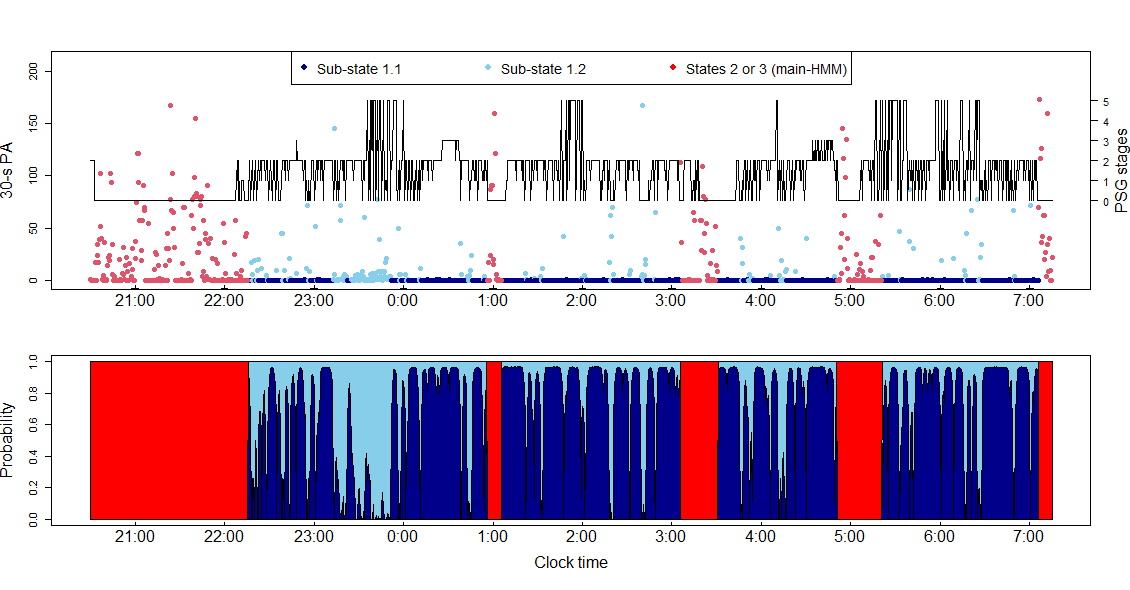}
	\caption[Conditional HMM results for MESA subjects A (top row) and B (bottom row).]{\label{fig:CHMM} 
 Left: Results from main-HMM fitted to 5-min averaged PA data over a monitoring period of 7 days 
 (Top panel: data with colours indicating the locally decoded state at each time; bottom panel:  cumulative posterior probability of the state at each time, i.e. $P(x_{t}\leq i|\bm{\theta},\mathbf{y}^{(n)}); i=1,2,3$). Right:  
 Results  of sub-HMM  fitted to the 30-s PA data, focusing on the one-night PSG monitoring period
 (Top panel: locally decoded state at each time; bottom panel: corresponding cumulative probability of each sub-state at each time. Data in red represent those assigned to states outside of state 1 by the local decoding result for the main-HMM).
	}
\end{figure}

The left panels of Figure \ref{fig:CHMM} depict the 5-min averaged PA data for subjects A and B, along with the locally decoded states and the cumulative posterior probabilities  of the three states at each time point (i.e. $P(x_{t}\leq i|\bm{\hat{\theta}},\mathbf{y}^{(n)}); i=1,2,3$) under the fitted main-HMMs.  State 1 (in blue) is characterized by  periods of immobility, usually occurring at night time. Other states (in pink and red shades) usually correspond to day-time activities of varying intensity, which depend on the subject's lifestyle and may be interrupted by daytime naps, as seen for subject B.
The estimated main-HMM suggests that subject A has a more active lifestyle and a more regular sleep-wake routine, with no significant sleep disruptions during the monitoring period. In contrast, subject B appears to suffer from a more disturbed circadian rhythm. These visual impressions are supported by estimating additional HMM-derived parameters that can be used to quantify an individual's  circadian rhythm, such as  the  dichotomy $I<O$ and rhythm indices (computed as in \citet{huang2018hidden}), where lower values indicate more disrupted circadian rhythms. For  B, these values were $96.4\%$ and $0.553$ while  A had  higher values of $99.4\%$ and $0.774$,  respectively. 
These findings are consistent with the sleep questionnaires completed by the subjects, where  B reported having generally restless sleep and sometimes having trouble falling asleep. 
We also evaluated the performance of our main-HMM in classifying sleep (state 1) versus wake (states 2 and 3) by comparing its decoding output to PSG-derived sleep/wake labels (available for one night) on an epoch-by-epoch basis, where the PSG stage for each 5-min epoch are determined by the most frequent stage of the corresponding ten 30-s bins in the raw PSG labels. 
Our main-HMM achieved state-of-the-art performance in terms of overall accuracy, sensitivity for sleep (proportion of true sleep epochs identified correctly) and specificity for wake (proportion of true wake epochs identified correctly), with values of $88.2\%$, $100\%$, $70.7\%$ for subject A and $89.9\%$, $94.4\%$, and $79.5\%$ for B, respectively. The relatively lower accuracy for detecting wake is expected as there are usually in-bed times before falling asleep or gentle sleep interruptions that are characterized by low or no activity. 
Our main-HMM provides useful quantitative summaries of an individual's rest-activity profile, as proposed in a parametric approach  in \citet{huang2018hidden}, but with the added benefit of a more flexible modelling of the emissions. The parameter $\hat{\gamma}_{1,1}$ is of particular interest for circadian and sleep analysis as low values suggest interrupted or fragmented sleep and thus a low quality of sleep, which is another indicator of circadian disruption. For subject B, the posterior mean for $\gamma_{1,1}$ was $0.916$, which is lower than $0.98$ obtained for subject A.

The right panels of Figure \ref{fig:CHMM} show the locally decoded time series of the 30-s PA data during the PSG monitoring period along with the cumulative probability of the two sub-states. As observed, for both subjects, the transitions between 
and the times spent in the  two sub-states are highly stochastic. 
State 1.1 has a high probability of observing zero, with a posterior mean of zero weight $\hat{w}_{1.1,1}$ of $0.908$ and $0.963$ for subjects A and B, while state 1.2  captures a moderately higher level of activity where the posterior mean $\hat{w}_{1.2,1}$ for A and B are $0.325$ and $0.463$, respectively. To investigate the link between the sub-states in our HMM and the true sleep stages from PSG,  we computed the proportion of the five PSG stages (wake, N1, N2, N3, REM) contained in each of the two HMM sub-states (see Table \ref{tab:examplesubstcomp})\footnote{Here we present only results for our two example subjects. Such an analysis could be extended  to include all MESA subjects but this is beyond the remit of this article. }. We can see that while both sub-states contain a mix of all the PSG stages, only state 1.1 contained any deep sleep  stages (N3) and had smaller proportions  of wake and light sleep stages (N1) compared to state 1.2.  This can also be seen by looking at the percentage of the PSG stages decoded as State 1.1, which are $(23.8\%, 82.1\%, 96.1\%, 100\%, 92.9\%)$ for subject A and $(26\%, 67.4\%, 77\%, 100\%, 67.9\%)$ for subject B for (wake, N1, N2, N3, REM). In contrast, State 1.2 tends to be associated with lighter sleep stages as well as disruptions into wake which were not identified by the main-HMM. We therefore anticipate that state 1.2 provides additional useful information regarding the sleep quality of a subject.

The estimated  transition probabilities of the fitted sub-HMM provide a systematic quantitative  summary which could be used, for example, to compare sleep behaviour between subjects. 
For subject A, the diagonal entries of $\Gamma$ have posterior means ($\pm1$ standard deviation) of $\hat{\gamma}_{1.1,1.1}=0.966$ ($\pm 0.01$) and $\hat{\gamma}_{1.2,1.2}=0.665$ ($\pm 0.056$), and those for subject B are $\hat{\gamma}_{1.1,1.1}=0.908$ ($\pm 0.009$) and $\hat{\gamma}_{1.2,1.2}=0.727$ ($\pm 0.047$).
B has a lower $\hat{\gamma}_{1.1,1.1}$ and higher $\hat{\gamma}_{1.2,1.2}$, indicating a higher probability of leaving state 1.1 and a longer expected staying time in state 1.2, which may be associated with poorer sleep quality during the monitoring period. 
Indeed, according to the MESA database, subject B has a lower sleep efficiency $63.15\%$ (computed from PSG) compared to $66.37\%$ for A. Our results are also consistent with Table \ref{tab:examplesubslpcomp}, which shows that subject B spent a larger proportion of sleep time in wake and N1 stages while having a lower proportion of time in the deeper N2 and N3 stages. The decoding and state probabilities of the sub-HMM also allow us to investigate the dynamic variation within and between  bouts of sleep. For instance, the  fragmentation of the blue region in the state probability plots of Figure \ref{fig:CHMM} suggests that subject A seems to experience more interruptions and lighter sleep during the earlier phase of the sleep bout, whereas subject B suffers from  sleep interruptions and transitions to lighter sleep throughout the entire bout. These observations are consistent with their own reports in the sleep questionnaire and the PSG recordings during the single night.

\begin{table}
	\caption{Composition of the states of the sub-HMM with respect to PSG stages  
		\label{tab:examplesubstcomp}}
	\centering
	\renewcommand{\arraystretch}{0.85}
	\begin{tabular}{lcccccc}
		\hline
		Subject   & sub-state & Wake & N1 & N2 & N3 & REM \\ \hline
		Subject A    & 1.1 & 0.155 & 0.098 & 0.569 & 0.038 & 0.14  \\ 
	   	             & 1.2 & 0.568 & 0.162 & 0.176 & 0 & 0.095 \\ 
		Subject B    & 1.1 & 0.174 & 0.167 & 0.529 & 0.051 & 0.078  \\ 
		             & 1.2 & 0.264 & 0.224 & 0.398 & 0 & 0.114  \\  \hline
	\end{tabular}
\end{table}

\begin{table}
	\caption{Proportions of time spent in different PSG stages during sleep for the example subjects   
		\label{tab:examplesubslpcomp}}
	\centering
	\renewcommand{\arraystretch}{0.85}
	\begin{tabular}{lccccc}
		\hline
		Subject &  Wake & N1 & N2 & N3 & REM \\ \hline
		Subject A   & 0.21 & 0.105 & 0.519 & 0.035 & 0.132  \\ 
		Subject B & 0.369 & 0.143 & 0.394 & 0.03 & 0.066  \\  \hline
	\end{tabular}
\end{table}

\section{Summary and further Discussion}
\label{sec:discussion}
In this paper, we propose and develop a Bayesian methodology for inference in spline-based HMMs, which offer attractive properties compared to alternative nonparametric HMMs in terms of simplicity in model interpretation and flexibility in modelling. Our method allows for the number of  states, $N$, to be unknown along with all other model parameters including the spline knot configuration(s). Compared with a P-spline-based construction, we achieve a parsimonious and efficient positioning of the spline knots via a RJMCMC algorithm, where the knots can either be shared across states or be state-specific. Model selection on $N$ is based on the marginal likelihood, which can be effectively estimated via a truncated harmonic mean estimator under an easy-to-implement parallel sampling scheme. Through extensive simulation studies, we demonstrated the effectiveness and superiority of our proposed methods over alternative comparators, including the Gaussian mixture based HMM, the frequentist P-spline-based approach of \citet{langrock2015nonparametric}, and a Bayesian adaptive P-spline approach which is investigated here for the first time. Importantly, the computational efficiency and flexibility of our algorithm allows us to deal with more states, which is a challenging problem even for parametric approaches due to convergence problems with increasing $N$. We highlight this advantage in the application to the animal movement data and illustrate the use of our method as a nonparametric approach for explorative data analysis.

The application to human PA data highlights the flexibility of our Bayesian modelling approach that can be extended in a relatively straightforward way to hierarchical scenarios such as the conditional HMM. The extension to a hierarchical framework of a sub-HMM within an overall HMM here allows us to  estimate many important parameters that characterize an individual's circadian rhythm, and  to model the individual stochastic dynamics of the rest state activity where the sub-states may be associated with deeper and lighter or interrupted sleep stages. Another feature of our method is that the  algorithm operates in an unsupervised manner, i.e. it does not require PSG labels for learning the model, which is desirable in applied settings as these labels are very costly or even impossible to acquire \citep{li2020novel}. The method developed here is thus of imminent interest to sleep and circadian biology researchers using data from wearable sensors.

Our modelling framework opens up several possible extensions in further research.
For instance, the homogeneous assumption on the hidden Markov chain can be relaxed by reparameterizing $\Gamma$ in terms of the covariates via multinomial logistic link functions \citep{zucchini2016hidden}. Efficient MCMC inference can be achieved by incorporating the Polya-Gamma data augmentation scheme of \citet{polson2013bayesian}, which was successfully applied to parametric nonhomogeneous HMMs in \citet{holsclaw2017bayesian}, into the present modelling framework. Our methodology can also be extended in a relatively straightforward manner to Markov switching (generalized) additive models as studied in \citet{langrock2017markov,langrock2018spline}  using frequentist approaches, where the splines can be used to model the functional effects of the covariates instead of the emissions. Without the density constraints on the spline parameters, the design of the RJMCMC algorithm can be simplified, and the efficiency of the resulting algorithm may be further improved. We believe that the advantages of using a Bayesian approach over a frequentist penalized approach as observed in this paper would carry over to this context.
Additionally, it would be interesting to explore the combination of the modern deep-learning-based methods, which excel at handling highly complex temporal dependence in the series and utilizing historical information sets, with the conventional HMM probabilistic framework for achieving better predictive ability while maintaining model interpretability.

\bigskip
\begin{center}
	{\large\bf ACKNOWLEDGEMENT}
\end{center}

We wish to thank Prof. Roland Langrock, Dr. Yannis Papastamatiou and Dr. Yuuki Watanabe for providing the Oceanic Whitetip shark data. We wish to acknowledge Dr. Qi Huang for her support on the analysis of human accelerometer data. We would also like to acknowledge the Warwick Statistics Department and MRC Biostatistics Unit, University of Cambridge for the support of Sida Chen's research. We are grateful to the editors and the  reviewers for their constructive comments  that have significantly improved the manuscript.

\bigskip
\begin{center}
	{\large\bf SUPPLEMENTARY MATERIAL}
\end{center}

\begin{description}
	
	\item[Supplementary document:] Provides additional details on the RJMCMC algorithm, simulation studies, and the case studies. (.pdf file)

\end{description}

\bibliographystyle{apalike}

\bibliography{Bibliography-MM-MC}
\end{document}


\def\spacingset#1{\renewcommand{\baselinestretch}%
	{#1}\small\normalsize} \spacingset{1}

\if0\blind
{
	\title{Supplementary material to the paper \\
		\bf Bayesian spline-based hidden Markov models with applications to actimetry data and sleep analysis}
	\author{
		Sida Chen \\
		Department of Statistics, University of Warwick \\
		MRC Biostatistics Unit, University of Cambridge 
		\and
		B{\"a}rbel Finkenst{\"a}dt \\
		Department of Statistics, University of Warwick
	}
	\maketitle
} \fi

\if1\blind
{
	\bigskip
	\bigskip
	\bigskip
	\begin{center}
		{\LARGE\bf  Bayesian spline-based hidden Markov models with applications to actimetry data and sleep analysis}
	\end{center}
	\medskip
} \fi

\bigskip

\spacingset{1.9} 
\appendix
\counterwithin{table}{section}
\counterwithin{figure}{section}
\section{Further details of the reversible jump MCMC algorithm}
\label{sec:AP_RJMCMC}
In this section, we give further computational and implementational details related to the reversible jump MCMC algorithm presented in Section \ref{sec:The algorithm} of the main paper and establish the validity of the algorithm. 
\subsection{Adaptive MCMC set-up}
The scaling parameter $\tau_{i}$ is adapted from iteration $t-1$ to $t$ as
\begin{displaymath}
	\tau_{i}^{(t)}=\max\bigg(\tau_{i}^{(t-1)}+\epsilon(t)\text{sgn}\big(\frac{1}{T_{a}}\sum_{j=t-T_{a}+1}^{t}\rho_{i}^{(j)}-\rho_{i}^{*}\big),\ \epsilon_{i}\bigg),
\end{displaymath}
where $\epsilon(t)=\min(0.01,1/\sqrt{t})$ following suggestions in \citet{roberts2009examples} and \citet{rosenthal2007amcmc}, sgn$(\cdot)$ is the sign function, $\rho_{i}^{(j)}$ is the MH acceptance rate in iteration j, $\rho_{i}^{*}$ is the targeted acceptance rate and $\epsilon_{i}$ is a sufficiently small positive number. That is, we adjust the scaling parameter at every iteration by adding or subtracting a factor $\epsilon(t)$ (whose magnitude is diminishing) if the averaged acceptance rate over the past $T_{a}$ iterations is below or above the target $\rho_{i}^{*}$, see \citet{green2015bayesian} and references therein for more details on the related theory and methods. In our context we set $\rho_{1}^{*}=\rho_{3}^{*}=0.4$ and $\rho_{2}^{*}=0.24$ based on the optimal scaling results for MH algorithms (see e.g. \citet{gelman1997weak} and \citet{roberts2001optimal}). Furthermore, we set $T_{a}=10$ (inspired by results in \citet{marshall2012adaptive}) and $\epsilon_{i}=10^{-6}$. In our examples, the lower bound $\epsilon_{i}$ is usually not reached as the algorithm usually stabilizes well.
\subsection{Acceptance probabilities for the Metropolis-Hastings moves}
Moves (c) and (d) in Algorithm 1 are standard Metropolis-Hastings updates. For (c), the acceptance probability for relocating the knot $r_{k^{*}}$ to the candidate point $r_{c}$ is:
\begin{displaymath}
	\min\big(1,\frac{f(\mathbf{y}^{(n)},\mathbf{x}^{(n)}|K,R_{K}^{'},\Tilde{A}_{K},\Gamma)f_{N,[a,b]}(r_{k^{*}}|r_{c}, \tau_{1}^{2})}{f(\mathbf{y}^{(n)},\mathbf{x}^{(n)}|K,R_{K},\Tilde{A}_{K},\Gamma)f_{N,[a,b]}(r_{c}|r_{k^{*}}, \tau_{1}^{2})}\big),
\end{displaymath}
where $R_{K}^{'}$ differs from $R_{K}$ only in the replacement of $r_{k^{*}}$ by $r_{c}$, and $f_{N,[a,b]}(\cdot|\mu,\sigma^{2})$ denotes the density of the truncated normal distribution with mean $\mu$, standard deviation $\sigma$ and bounded on $[a,b]$.
For move (d), since the proposal is symmetric, the acceptance probability is:
\begin{displaymath}
	\min\big(1,\frac{f(\mathbf{y}^{(n)},\mathbf{x}^{(n)}|K,R_{K},\Tilde{A}_{K}^{'},\Gamma)f(\Tilde{A}_{K}^{'}|K,\zeta)}{f(\mathbf{y}^{(n)},\mathbf{x}^{(n)}|K,R_{K},\Tilde{A}_{K},\Gamma)f(\Tilde{A}_{K}|K,\zeta)}\big),
\end{displaymath}
where the set of $\Tilde{a}_{i,j}^{'}$ is denoted by $\Tilde{A}_{K}^{'}$. 
In step (e) we used a log-normal random walk to update the parameter due to the positivity constraint, and the corresponding acceptance probability after adjusting for the log-transformation is
\begin{displaymath}
	\min\big(1,\frac{f(\Tilde{A}_{K}|K,\zeta^{'})f(\zeta^{'})\zeta^{'}}{f(\Tilde{A}_{K}|K,\zeta)f(\zeta)\zeta}\big).
\end{displaymath}

\subsection{Acceptance probabilities for the birth and death moves}
Using the notation of \citet{green1995reversible}, the birth move for the spline parameters is accepted with probability $\min(1,A)$, where A could be expressed in the form
\begin{displaymath}
	\text{likelihood ratio} \times \text{prior ratio} \times \text{proposal ratio} \times \text{Jacobian}.
\end{displaymath}
In our context the likelihood ratio is:
\begin{displaymath}
	\frac{f(\mathbf{y}^{(n)},\mathbf{x}^{(n)}|K+1,R_{K+1},\Tilde{A}^{'}_{K+1},\Gamma)}{f(\mathbf{y}^{(n)},\mathbf{x}^{(n)}|K,R_{K},\Tilde{A}_{K},\Gamma)},
\end{displaymath}
where $\Tilde{A}^{'}_{K+1}$ stands for the set of proposed $\Tilde{a}^{'}_{i,j}$ and the complete data likelihood $f(\mathbf{y}^{(n)},\mathbf{x}^{(n)}|\cdot)$ is given by equation \eqref{eq:completellk} of the main paper. The prior ratio is given by the product of the ratio of the priors on each block of parameters that are involved in this update:
\begin{displaymath}
	\frac{f_{U}(K+1)}{f_{U}(K)}\dfrac{\frac{(K+1)!}{(b-a)^{K+1}}}{\frac{K!}{(b-a)^{K}}}\frac{\prod_{i=1}^{N}\prod_{j=1}^{K+5}f_{LG}(\Tilde{a}^{'}_{i,j}|\zeta,1)}{\prod_{i=1}^{N}\prod_{j=1}^{K+4}f_{LG}(\Tilde{a}_{i,j}|\zeta,1)}, 
\end{displaymath}
where $f_{U}(\cdot)$ denotes the probability mass function of a uniformly distributed random variable on $\{2,\ldots,K_{\max}\}$, and $f_{LG}(\cdot|\zeta,1)$ stands for the log-gamma density with shape parameter $\zeta$ and rate parameter $1$. The proposal ratio is given by
\begin{displaymath}
	\frac{d_{K+1}}{(K+1)}\Big\{b_{K}\frac{\sum_{i=1}^{K}f_{N,[a,b]}(r_{c}|r_{i},\tau(R_{K},i)^{2})}{K}\Big\}^{-1},
\end{displaymath}
where $f_{N,[a,b]}(\cdot|\mu,\sigma^{2})$ denotes the density of a truncated normal distribution with mean $\mu$, standard deviation $\sigma$ and bounded on $[a,b]$. Lastly, the Jacobian corresponding to the transformation from $(R_{K},\Tilde{A}_{K},\Gamma,r_{c},u_{1},\ldots,u_{N})$ to $(R_{K+1},\Tilde{A}^{'}_{K+1},\Gamma)$ is 
\begin{displaymath}
	|(r_{n^{*}+2}r_{n^{*}+3})^{N}\prod_{i=1}^{N}(\Tilde{a}_{i,n^{*}+4}-\Tilde{a}_{i,n^{*}+3})|,
\end{displaymath}
where $n^{*}$ is defined as in Section 3.2 of the main paper such that $r_{c}\in (r_{n^{*}},r_{n^{*}+1})$. After simplification A is thus given by
\begin{displaymath}
	A=\frac{f(\mathbf{y}^{(n)},\mathbf{x}^{(n)}|K+1,R_{K+1},\Tilde{A}^{'}_{K+1},\Gamma)}{f(\mathbf{y}^{(n)},\mathbf{x}^{(n)}|K,R_{K},\Tilde{A}_{K},\Gamma)}\frac{1}{(b-a)}\frac{\prod_{i=1}^{N}\prod_{j=n^{*}+2}^{n^{*}+4}f_{LG}(\Tilde{a}^{'}_{i,j}|\zeta,1)}{\prod_{i=1}^{N}\prod_{j=n^{*}+2}^{n^{*}+3}f_{LG}(\Tilde{a}_{i,j}|\zeta,1)} 
\end{displaymath}
\begin{displaymath} 
	\times \frac{d_{K+1}}{b_{K}}\Big\{\frac{\sum_{i=1}^{K}f_{N,[a,b]}(r_{c}|r_{i},\tau(R_{K},i)^{2})}{K}\Big\}^{-1}|(r_{n^{*}+2}r_{n^{*}+3})^{N}\prod_{i=1}^{N}(\Tilde{a}_{i,n^{*}+4}-\Tilde{a}_{i,n^{*}+3})|.
\end{displaymath}
Since the birth and death moves are defined in a symmetric way, the acceptance probability for this death move is $\min(1,A^{-1})$, where $A$ is computed by considering the reverse birth move starting with the $K-1$ interior knots and with $n^{*}=d^{*}-1$.

\subsection{Tackling label switching}
A practical consequence of the properties of the model and its prior is that samples generated by the reversible jump MCMC algorithm are subject to the label switching problem, i.e. the state labels could permute during the MCMC iterations without changing the posterior density \citep{scott2002bayesian}. As a result, the MCMC output cannot be directly used for inference about the state specific parameters. To tackle this issue we choose to use the Kullback–Leibler relabelling algorithm developed in \citet{stephens2000dealing}, which has been successfully applied for both parametric HMMs \citep{rodriguez2014label} and nonparametric HMMs \citep{hadj2020identifying}. The basic ideas are as follows. For each of the $T$ collected MCMC samples (discarding those collected during the burn-in), we first construct a $n\times N$ dimensional ``classification probability matrix" whose $(i,k)$-th entry in our context is given by
\begin{displaymath}
	P_{i,k}^{(t)}=\frac{\pi_{k}^{(t)}f_{k}^{(t)}(y_{i})}{\sum_{j=1}^{N}\pi_{j}^{(t)}f_{j}^{(t)}(y_{i})}, \quad t=1,\ldots,T,
\end{displaymath}
where $f_{k}^{(t)}(\cdot)$ is the emission density for state $k$ constructed from the $t$-th MCMC sample, and $(\pi_{1}^{(t)},\ldots,\pi_{N}^{(t)})$ is the stationary distribution associated with the transition matrix $\Gamma^{(t)}$. The algorithm then involves iteratively searching a specific permutation of state labels to minimize the KLD between classification probabilities averaged over the MCMC iterations, $q_{i,k}=(\sum_{t=1}^{T}P_{i,k}^{(t)})/T$, and the classification probabilities obtained in each MCMC iteration. In other words, we make the state labels associated with each MCMC draw agree on the classification probabilities $[P_{i,k}^{(t)}]$. The ``optimized" permutation searched for each MCMC sample can then be used to relabel the samples to achieve a consistent ordering of the labels. We refer to \citet{stephens2000dealing}, and \citet{rodriguez2014label}, for more details of the algorithm. In our implementations, we use the R package \textit{Label.switching} of \citet{ls2016R} to perform this minimization procedure.

\subsection{Validity of the algorithm}
When the adaptive tuning scheme ceases after a period of burn in, the validity of the proposed reversible jump MCMC algorithm can be established following standard Markov chain theory as presented in \citet{tierney1994markov} and \citet{robert2013monte}. First note that the Markov transition kernel for each of the move steps admits the target posterior distribution, $f$ (defined in equation \eqref{eq:target}), as invariant distribution, so a concatenation of these kernels also admits $f$ as invariant distribution. Irreducibility of the constructed chain can be deduced as the chain can move from one value of $K$ to any other possible value by increasing or decreasing its value by one at a time, with positive probability. In step (a) all possible state allocations have positive probability. In steps (b), (d) and (e) the full conditional distribution/proposal density is positive on the natural parameter space and the same holds true for step (c) if we consider several consecutive sweeps. The chain is also aperiodic as there is a strictly positive probability that the chain remains in a neighbourhood of the current state after one sweep of the MCMC procedure. With the above properties the chain is guaranteed to converge to the posterior distribution from almost all initial states (except for a set of posterior probability zero). To replace ``almost all" by ``all" we require a stronger condition called Harris recurrence,  which is generally difficult to verify in the trans-dimensional MCMC set-up \citep{roberts2006harris,hastie2012model}. However in practice we could tackle this issue by drawing the initial state using a continuous distribution centered around the posterior mode (or other approximations to that such as the maximum likelihood estimate). This strategy is employed in our initialization process. To accelerate the convergence of the chain we initialize the knot points at the empirical quantiles of the data so that more knots are initially placed at data-rich regions. For the remaining parameters the initial values are drawn from appropriate truncated normal distributions centered at their respective maximum likelihood estimates computed given the initial knot configuration.

\subsection{State-specific knot configurations for emission estimation for larger N}
\label{sec:AP_statespec_adsp}
A shared knot configuration or basis functions provides a parsimonious representation of the emission densities, but this approach can become inefficient as the number of states, $N$, grows large. In such scenarios, the usage of basis functions for a particular state can be very low, and a knot point will be more difficult to add or delete due to the shared nature (i.e. in general, a new knot is more likely to be accepted if it contributes to the fit of all the emission densities), resulting in slower mixing of the chain. 
To address this issue, we propose an extension of the basic algorithm allowing different knot configurations, $(K_{i},R_{i,K_{i}})$, for each state $i$.
Priors for state-specific spline parameters are constructed from the same distributions used in Section \ref{sec:The model} for the shared knot setting. 
Building on the Bayesian model presented in Section \ref{sec:The model} of the main paper, the joint complete density for the extended model is
\begin{displaymath}	
	f(\{K_{i}\}_{i=1}^{N}, \{R_{i,K_{i}}\}_{i=1}^{N}, \{\zeta_{i}\}_{i=1}^{N},\{\tilde{A}_{i,K_{i}}\}_{i=1}^{N},\Gamma,\mathbf{y}^{(n)},\mathbf{x}^{(n)})=
\end{displaymath}
\begin{equation*}
	f(\Gamma)\prod_{i=1}^{N}f(K_{i})\prod_{i=1}^{N}f(\zeta_{i})\prod_{i=1}^{N}f(R_{i,K_{i}}|K_{i} \prod_{i=1}^{N}f(\tilde{A}_{i,K_{i}}|K_{i},\zeta_{i})f(\mathbf{y}^{(n)},\mathbf{x}^{(n)}|\cdot),
\end{equation*}
where $\tilde{A}_{i,K_{i}}$ represents the spline coefficient vector for state $i$ (with dimensionality dependent on $K_{i}$), $\zeta_{i}$ denotes the state-specific shape parameter for the log-gamma prior for $\tilde{A}_{i,K_{i}}$. Slightly different from the shared knot setting, we use a Gamma$(1,1)$ hyperprior for the shape parameter $\zeta_{i}$, but with an additional truncation at 0.01, to improve numerical stability.
Posterior sampling for the extended model can be performed in a similar manner as described in Algorithm 1. The state sequence and transition matrix are updated in exactly the same way as before. For $R_{i,K_{i}}$, $\tilde{A}_{i,K_{i}}$ and $\zeta_{i}$, the same types of Metropolis-Hastings updates (i.e. steps (c), (d) and (e) in Algorithm 1) can be used, now scanning through each $i$. Birth or death of a knot point is considered separately for the knot configuration associated with each state, via the reversible jump move as described in Section \ref{sec:The algorithm} of the main paper. 
The resulting algorithm is thus expected to be more computationally intensive than Algorithm 1 due to the increased number of updating steps. However, note that this increase is partially counterbalanced by the potentially reduced number of basis functions and spline coefficients needed for each state, as well as the improved mixing of the Markov chain.  Experimental results indicate that, say for $N \geq 5$, the algorithm with state-specific knots is more efficient overall than the shared knot version, and its advantage becomes more pronounced as the value of $N$ increases. Furthermore, the algorithm demonstrates scalability, remaining computationally feasible for up to 10 states.

\section{Details for the main simulation study}
\label{sec:appsimulation}
In this section, we provide details on the setups and results of our main simulation study, where we evaluate the performance of our proposed adSP method and compare it against fpSP, bpSP and GMM methods.
\subsection{Evaluation metrics for estimation methods}
To compare the estimation accuracy of different approaches, we consider two primary metrics
\begin{enumerate}
	\item \textbf{Ability to recover the true emission distributions}: This is quantified by the average Kullback-Leibler divergence (KLD):
	$
	(\mathbf{KLD}(\hat{f}_{i}||f_{i})+\mathbf{KLD}(f_{i}||\hat{f}_{i}))/2$, $i=1,\ldots,N$, where $\mathbf{KLD}(\hat{f}_{i}||f_{i})=\int \hat{f}_{i}(y)\log(\hat{f}_{i}(y)/f_{i}(y))dy$ and $\hat{f}_{i}$ is the estimated emission density for state $i$. In Bayesian MCMC, used for adSP and bpSP methods, we estimate the unknown emission densities (pointwise) through the posterior expectation $\mathbf{E}[f_{i}(y)|\mathbf{y}^{(n)}]$, which is the Bayes estimator of $f_{i}(y)$ under posterior mean squared error loss and can be approximated by the Monte Carlo average $\hat{f}_{i}(y)=\sum_{j=1}^{T}f^{(j)}_{i}(y)/T$, $i=1,\ldots,N$, where $f^{(j)}_{i}(y)$ is the emission density arising from the $j$-th MCMC sample (as a function of the knot configuration and spline coefficients) and $y$ is a fixed point in the domain of the observed data. Density estimates for the fpSP method are constructed as in \citet{langrock2015nonparametric} where the penalized maximum likelihood estimates of the spline coefficients are plugged into equation \eqref{eq:emission}, and those for the GMM method are constructed in a similar fashion. In our implementations we used the $KLD$ function in the R package \textit{LaplacesDemon} \citep{laplacedemonR} for approximating the average KLD.
	
	\item \textbf{Decoding accuracy}: Decoding is to infer the hidden state process $\mathbf{x}^{(n)}$ based on the observed $\mathbf{y}^{(n)}$ and is one of the key inference tasks in HMMs. We assess the agreement/disagreement between the estimated and true state sequences by calculating the proportion of correctly/incorrectly classified states, where the latter is also known as normalized Hamming distance \citep{fox2011sticky}. For the adSP and bpSP methods we estimate the states by first estimating the marginal state probability based on the MCMC samples of $\mathbf{x}^{(n)}$ as $\hat{P}(x_{t}=k|\mathbf{y}^{(n)})=\sum_{j=1}^{T}\mathbf{I}(x_{t}^{(j)}=k)/T$, $t=1,\ldots,n$, $k=1,\ldots,N$, where $x_{t}^{(j)}$ is the value of $x_{t}$ from the $j$-th MCMC drawing. We then determine the value of $x_{t}$ such that its posterior state probability is maximized. For the fpSP and GMM methods our results are based on global decoding via the Viterbi algorithm, conditional on the point estimates of the model parameters \citep{zucchini2016hidden}.
\end{enumerate}

\subsection{Simulation models}
We generated artificial data from four simulation models.
Model 1 is a 2-state HMM (see Figure \ref{fig:sim1})  considered in \citet{langrock2015nonparametric} with emissions:
\begin{displaymath}
	y_{t}|x_{t}=1 \sim \mathcal{N}(-15,11^{2}), 
\end{displaymath}
\begin{displaymath}
	y_{t}|x_{t}=2 \sim 0.35\mathcal{N}(-5,9^{2})+0.65\mathcal{N}(30,10^{2}),
\end{displaymath}
where the states of the underlying Markov chain were generated from $\bm{\delta}=(1/2,1.2)$ and $\gamma_{12}=\gamma_{21}=0.1$ and $n=800$. For Model 2, we consider a 3-state HMM with a unimodal positively skewed emission distribution in state 1, a bimodal distribution in state 2 and a unimodal negatively skewed distribution in state 3 (see Figure \ref{fig:sim2}). We use B-splines to construct these densities and the details of the spline parameters are omitted here. The states were generated using $\bm{\delta}=(1/3,1/3,1/3)$, $\gamma_{11}=\gamma_{22}=\gamma_{33}=0.85$, $\gamma_{2,1}=\gamma_{2,3}=0.075$, $\gamma_{1,2}=\gamma_{3,2}=0.1$ and $\gamma_{1,3}=\gamma_{3,1}=0.05$,
from which $n=1500$ observations were simulated from the corresponding emission distribution using the inverse transform sampling scheme \citep{devroye1986sample}. 
Model 3 is motivated and modified from the \textit{bimod} model considered in \citet{yau2011bayesian}. We construct the emissions using a mixture of a Laplace and a generalized Student's t distribution (see Figure \ref{fig:sim3}): 
\begin{displaymath}
	y_{t}|x_{t}=1 \sim 0.5\mathbf{Laplace}(-1,0.2)+0.5\mathbf{t}_{3}(2,2),
\end{displaymath}
\begin{displaymath}
	y_{t}|x_{t}=2 \sim 0.5\mathbf{Laplace}(0.5,0.2)+0.5\mathbf{t}_{3}(3.5,2),
\end{displaymath}
where $\mathbf{Laplace}(\mu,\sigma)$ denotes a Laplace distribution with location parameter $\mu$ and scale parameter $\sigma$ and $\mathbf{t}_{\nu}(\mu,\sigma)$ denotes a generalised t distribution with $\nu$ degrees of freedom (assume $\nu>2$), mean $\mu$ and standard deviation $\sigma$. By construction, one of the emission can be obtained by translating the other one horizontally, and the emissions have varying degree of smoothness across the domain, which can be expected to pose challenges to P-spline based inference methods. For this model we set $\bm{\delta}=(0.5,0.5)$, $\gamma_{12}=\gamma_{21}=0.05$ and $n=2000$. Finally, Model 4 is a 2-state HMM considered in \citet{yau2011bayesian} (the \textit{trimod} case) with emissions specified as a mixture of three well-separated normal distributions (see Figure \ref{fig:sim4}): 
\begin{displaymath}
	y_{t}|x_{t}=1 \sim \frac{1}{3}\mathcal{N}(-4,1)+\frac{1}{3}\mathcal{N}(0,1)+\frac{1}{3}\mathcal{N}(8,1),
\end{displaymath}
\begin{displaymath}
	y_{t}|x_{t}=2 \sim \frac{1}{3}\mathcal{N}(-3,1)+\frac{1}{3}\mathcal{N}(1,1)+\frac{1}{3}\mathcal{N}(9,1),
\end{displaymath}
and the same Markov chain parameters as in Model 3. For this model we consider a relatively large data set of length $n=5000$. 
Note that Models 3 and 4 pose the most serious  challenges even when the correct number of states is assumed to be known as in \citet{yau2011bayesian}.

\subsection{Setup and estimation for the Bayesian P-spline-based HMM}
Motivated by \citet{lang2004bayesian}, we set up the spatially adaptive Bayesian P-spline model for the emissions as follows. 
We consider a second-order random walk prior for the reparameterized spline coefficients for state $i$, $\tilde{a}_{i}=(\tilde{a}_{i,1},\ldots,\tilde{a}_{i,K+4})$:
\begin{equation}\label{eq:bspprior}
	\tilde{a}_{i,j}|\tilde{a}_{i,j-1},\tilde{a}_{i,j-2},\tilde{\tau}_{i},(\lambda_{i,j})_{j=3,\ldots,K+4}\sim \mathcal{N}(2\tilde{a}_{i,j-1}-\tilde{a}_{i,j-2},\frac{1}{\tilde{\tau}_{i}\lambda_{i,j}}), \quad j=3,\ldots,K+4,
\end{equation}
where $\tilde{\tau}_{i}$ and $\lambda_{i,j}$ may be understood as state-specific ``global" and ``local" smoothing parameters, respectively. We additionally assign noninformative normal priors $\mathcal{N}(0,100^{2})$ to $\tilde{a}_{i,1}$ and $\tilde{a}_{i,2}$ to ensure that the joint prior for $\tilde{a}_{i}$ is proper. 
For $\lambda_{i,j}$, we follow \citet{lang2004bayesian} to use independent Gamma priors, i.e. $\lambda_{i,j}\sim\mathbf{Gamma}(\frac{\alpha_{\lambda}}{2},\frac{\alpha_{\lambda}}{2})$. For the global smoothing parameter, to address its impact on the smoothness of the spline fit, we adopt a robust specification by introducing the following hyperpriors \citep{jullion2007robust}
\begin{gather*} 
	\tilde{\tau}_{i}|\tilde{\tau}^{'}\sim \mathbf{Gamma}(\alpha_{\tilde{\tau}},\alpha_{\tilde{\tau}}\tilde{\tau}^{'}), \quad i=1,\ldots,N,\\
	\tilde{\tau}^{'}\sim \mathbf{Gamma}(\alpha_{\tilde{\tau}^{'}},\beta_{\tilde{\tau}^{'}}).
\end{gather*}
In our implementation we use $\alpha_{\lambda}=1$ as in \citet{lang2004bayesian} and we choose $\alpha_{\tilde{\tau}^{'}}=\beta_{\tilde{\tau}^{'}}=10^{-3}$ following the suggestion in \citet{jullion2007robust}. The choice of $\alpha_{\tilde{\tau}}$ is not influential and we set $\alpha_{\tilde{\tau}}=1$ as in \citet{bremhorst2016flexible} and \citet{maturana2021bayesian}. Details for the associated MCMC algorithm are given below. Note that other versions of spatially adaptive Bayesian P-spline models may exist, but our goal here is to compare with the state-of-the-art approach commonly used in practice.

For the Bayesian P-spline-based model the knot configuration $(K,R_{K})$ is prefixed (in our implementation we use the same knot configuration as for the fpSP method to facilitate comparison). The parameter set is $(\Tilde{A}$, $\Lambda$, $\Gamma$, $\tilde{\bm{\tau}}$, $\tilde{\tau}^{'})$, where $\Tilde{A}=(\Tilde{a}_{1},\ldots,\Tilde{a}_{N})$, $\Lambda=(\lambda_{i,j})_{i=1,\ldots,N;j=3,\ldots,K+4}$ and $\tilde{\bm{\tau}}=(\tilde{\tau}_{1},\ldots,\tilde{\tau}_{N})$. We assume that the joint posterior distribution of the state sequence and model parameters takes the form
\begin{displaymath}
	f(\mathbf{x}^{(n)},\Gamma,\Tilde{A},\Lambda,\tilde{\bm{\tau}},\tilde{\tau}^{'}|\mathbf{y}^{(n)})\propto f(\Gamma)f(\tilde{\tau}^{'})f(\tilde{\bm{\tau}}|\tilde{\tau}^{'})f(\Lambda)f(\Tilde{A}|\tilde{\bm{\tau}},\Lambda)f(\mathbf{y}^{(n)},\mathbf{x}^{(n)}|\Tilde{A},\Gamma),
\end{displaymath}
where $f(\tilde{\bm{\tau}}|\tilde{\tau}^{'})=\prod_{i=1}^{N}f(\tilde{\tau}_{i}|\tilde{\tau}^{'})$, $f(\Lambda)=\prod_{i=1}^{N}\prod_{j=3}^{K+4}f(\lambda_{i,j})$, $f(\Tilde{A}|\tilde{\bm{\tau}},\Lambda)=\prod_{i=1}^{N}f(\Tilde{a}_{i}|\tilde{\tau}_{i},(\lambda_{i,j}))$ and the complete likelihood $f(\mathbf{y}^{(n)},\mathbf{x}^{(n)}|\cdot)$ is given by \eqref{eq:completellk} with spline coefficients derived from $\Tilde{A}$. We use the same prior distribution for $\Gamma$ as for our proposed spline-based HMM and the priors $f(\tilde{\tau}^{'})$, $f(\tilde{\tau}_{i}|\tilde{\tau}^{'})$, $f(\lambda_{i,j})$ and $f(\Tilde{a}_{i}|\tilde{\tau}_{i},(\lambda_{i,j})_{j=3,\ldots,K+4})$ are specified as described above. Posterior simulation for the resulting model can be achieved using a Metropolis-within-Gibbs sampler as outlined in Algorithm 2.

\setcounter{algocf}{1}
\begin{algorithm}[h!]
	\label{alg:bpSP}
	\SetAlgoLined
	Initialize $\Tilde{A}$, $\Lambda$, $\Gamma$, $\tilde{\bm{\tau}}$, $\tilde{\tau}^{'}$;
	\For{i=1, \ldots, T}{
		(a) update the hidden state sequence $\mathbf{x}^{(n)}$\;
		(b) update the transition probability matrix $\Gamma$\;
		(c) update the set of reparametrized B-spline coefficients $\Tilde{A}$\;
		(d) update the local smoothness parameters $\lambda_{i,j}$\;
		(e) update the global smoothness parameters $\tilde{\bm{\tau}}$\;	
		(f) update the hyperparameter $\tilde{\tau}^{'}$
	}
	\caption{MCMC algorithm for Bayesian P-spline-based HMMs}
\end{algorithm}
Steps (a) and (b) can be performed exactly as in our RJMCMC algorithm (see steps (a) and (b) of Algorithm 1), and the details are omitted here. For step (c), we update the vector of state-specific spline coefficients $\Tilde{a}_{i}$ one-at-a-time via a random walk MH step as used in step (d) of Algorithm 1. A joint blockwise update can be challenging due to the potential high-dimensionality of $\Tilde{a}_{i}$. For each coefficient, a separate variance parameter is used for the MH update and they are tuned adaptively as described in Section \ref{sec:AP_RJMCMC}. For step (d), we can derive that the full conditional distribution of $\lambda_{i,j}$ is
\begin{displaymath}
	f(\lambda_{i,j}|rest)\propto \lambda_{i,j}^{\frac{\alpha_{\lambda}}{2}-1}\exp(-\frac{\alpha_{\lambda}\lambda_{i,j}}{2})\lambda_{i,j}^{\frac{1}{2}}\exp(-\frac{\tilde{\tau}_{i}u_{i,j}^{2}\lambda_{i,j}}{2}), \quad i=1,\ldots,N, \quad j=3,\ldots,K+4,
\end{displaymath}
where $u_{i,j}=\tilde{a}_{i,j}-2\tilde{a}_{i,j-1}+\tilde{a}_{i,j-2}$. This is the kernel of a Gamma distribution and we can update $\lambda_{i,j}$ via
\begin{displaymath}
	\lambda_{i,j}|rest\sim \mathbf{Gamma}(\frac{\alpha_{\lambda}+1}{2},\frac{\alpha_{\lambda}+\tilde{\tau}_{i}u_{i,j}^{2}}{2}), \quad i=1,\ldots,N, \quad j=3,\ldots,K+4.
\end{displaymath}
For step (e), note that the full conditional distribution of $\tilde{\tau}_{i}$ is
\begin{displaymath}
	f(\tilde{\tau}_{i}|rest)\propto \tilde{\tau}_{i}^{\alpha_{\tilde{\tau}}-1}\exp(-\alpha_{\tilde{\tau}}\tilde{\tau}^{'}\tilde{\tau}_{i})\tilde{\tau}_{i}^{\frac{K+2}{2}}\exp(-\frac{\tilde{\tau}_{i}}{2}\sum_{j=3}^{K+4}\lambda_{i,j}u_{i,j}^{2}), \quad i=1,\ldots,N.
\end{displaymath}
Therefore we update $\tilde{\tau}_{i}$ by drawing from
\begin{displaymath}
	\tilde{\tau}_{i}|rest\sim \mathbf{Gamma}(\frac{K+2}{2}+\alpha_{\tilde{\tau}},\alpha_{\tilde{\tau}}\tilde{\tau}^{'}+\frac{1}{2}\sum_{j=3}^{K+4}\lambda_{i,j}u_{i,j}^{2}), \quad i=1,\ldots,N.
\end{displaymath}
Step (f) is also a Gibbs step. The full conditional distribution of $\tilde{\tau}^{'}$ is
\begin{displaymath}
	f(\tilde{\tau}^{'}|rest)\propto (\tilde{\tau}^{'})^{\alpha_{\tilde{\tau}^{'}}-1}\exp(-\beta_{\tilde{\tau}^{'}}\tilde{\tau}^{'})\prod_{i=1}^{N}(\tilde{\tau}^{'})^{\alpha_{\tilde{\tau}}}\exp(-\alpha_{\tilde{\tau}}\tilde{\tau}_{i}\tilde{\tau}^{'}),
\end{displaymath}
and thus we draw:
\begin{displaymath}
	\tilde{\tau}^{'}|rest\sim \mathbf{Gamma}(N\alpha_{\tilde{\tau}}+\alpha_{\tilde{\tau}^{'}},\alpha_{\tilde{\tau}}\sum_{i=1}^{N}\tilde{\tau}_{i}+\beta_{\tilde{\tau}^{'}}).
\end{displaymath}
For the same reason stated before, here MCMC inference is subject to the label switching problem, which can be tackled using the Kullback-Leibler relabelling algorithm described above.

\subsection{Setup and estimation for the GMM-based HMM}
For the GMM-based HMM, the emission distributions are given by
\begin{displaymath}
	f_{i}(y_{t})=\sum_{j=1}^{K_{i}}w_{i,j}f_{N}(y_{t}|\mu_{ij},\sigma^{2}_{ij}), \quad i=1,\ldots,N,
\end{displaymath}
where $f_{N}(\cdot|\mu,\sigma^{2})$ denotes the Gaussian density with mean $\mu$ and variance $\sigma^{2}$, $w_{i,j}>0$ for $j=1,\ldots,K_{i}$ with $\sum_{j=1}^{K_{i}}w_{i,j}=1$ are mixing weights for state $i$ and $\mathbf{K}=(K_{1},\ldots,K_{N})$ specifies the number of Gaussian mixture components for each state. As for GMMs, estimation of the resulting GMM-based HMM can be performed in a maximum likelihood framework, where there are two key issues to be tackled.
The first arises from the fact that the ordinary maximum likelihood estimate of the GMM is not well-defined as the likelihood is unbounded with $\sigma_{ij}\to 0$ \citep{kiefer1956consistency}. Therefore a naive implementation may lead to degenerate solutions. To overcome this issue we follow popular suggestions in the literature (see e.g. \citet{ciuperca2003penalized}) to introduce a penalty term to the log-likelihood function, which can be regarded as a Bayesian regularization approach, leading to a penalized log-likelihood 
\begin{displaymath}
	\log L_{P}=\log f(\mathbf{y}^{(n)}|\cdot)+\sum_{i,j}\log(p_{\lambda}(\sigma_{ij})),
\end{displaymath}
where $f(\mathbf{y}^{(n)}|\cdot)$ is given by \eqref{eq:obsllk} and $p_{\lambda}()$ is a suitable penalty function with hyperparameters $\lambda$. Here, inspired by \citet{ciuperca2003penalized} and \citet{baudry2015mixtures}, we choose $p_{\lambda}()$ to be the density of an inverse gamma distribution, and we set both of the shape and scale parameters to be $0.001$, which coincides with the commonly used noninformative prior distribution for Gaussian variance.
Note that we used ``non-informative" penalization parameters for the variance parameters to reflect the fact that, in reality, the true nature of the emissions would usually be unknown. 
The second key issue is to select the number of mixture components $K_{i}$, which is a long standing problem for GMMs.
When GMM is used for density estimation, the BIC has been shown to be theoretically and practically adequate \citep{roeder1997practical}. We thus follow \citet{sun2009large} and \citet{shokoohi2019capturing} to use the BIC to choose $\mathbf{K}$ from a pre-defined grid such that each $K_{i}$ varies from 1 to 7, and the ``optimal" values selected by this procedure are reported in Table \ref{tab:GMMconstants}. We note that this may not be the best approach but we found it worked well for our purpose. 
\begin{table}
	\caption{``Optimal" numbers of Gaussian mixture components selected by BIC for the GMM method. \label{tab:GMMconstants}}
	\centering
	\begin{tabular}{cllll}
		\hline
		Parameters & Model 1 & Model 2 & Model 3 & Model 4  \\ \hline
		$\mathbf{K}$    & $(1,2)$ & $(2,2,3)$   &  $(4,4)$ & $(3,3)$  \\  \hline
	\end{tabular}
\end{table}

\subsection{Results}
To take care of the variability in the simulated data set, for each of the four simulation models $60$ random replicates of the data were generated and the four methods were applied to each set of replicates.
For our proposed algorithm the unspecified constants are set to $a=\min(\mathbf{y}^{(n)})-10$, $b=\max(\mathbf{y}^{(n)})+10$ and $\alpha=0.65$ in all scenarios (experiments suggest that our results are not very sensitive to these specific choices).
For the fpSP method we select the number of equidistant knots $K$, and the state-specific smoothing parameters $\mathbf{\lambda}=(\lambda_{1},\ldots,\lambda_{N})$ either based on the original choices in \citet{langrock2015nonparametric} (for Model 1), or on pre-experiments of our own (for Models 2-4), and the details are summarized in Table \ref{tab:lgconstants}.
Settings for the bpSP and GMM methods are as described in the previous sections. For Bayesian MCMC methods (adSP and bpSP), we ran the corresponding MCMC samplers for a total of $120$k iterations. We discarded the first $60$k iterations as burn-in, which was found to be sufficient to obtain reliable results in all four scenarios. Our results are then based on post-thinning samples with a thinning interval of $10$. To obtain results for the frequentist methods (fpSP and GMM), we tested 50 different initializations and selected the final fitted model with the highest penalized model likelihood.
All computations were performed on the Cambridge Service for Data Driven Discovery (CSD3) High-Performance Computing (HPC) system using the Ice Lake CPUs.
\begin{table}
	\caption{Number of knot points and the smoothing parameter used for implementing the fpSP method. \label{tab:lgconstants}}
	\centering
	\begin{tabular}{cllll}
		\hline
		Parameters & Model 1 & Model 2 & Model 3 & Model 4  \\ \hline
		$K$    & $27$ & $43$   &  $71$ & $51$   \\ 
		$\mathbf{\lambda}$   & $(2048,1024)$   & $(2800,1600,2400)$  & $(1,1)$  & $(600,600)$ \\ \hline
	\end{tabular}
\end{table}

\subsubsection*{Model selection}
For each simulation model, we first tested the marginal likelihood based approach described in Section \ref{sec:model selection} of the main paper for selecting the number of states, implemented in conjunction with the Bayesian MCMC based approaches (adSP and bpSP).  We place a uniform prior on $N$ over the candidate set $\{2,3,4,5\}$. Throughout we set $\beta=0.2$ and $\xi=0.01$ and experiments suggest that the results are robust to these choices provided that they are chosen to be relatively small as specified here. For the adSP method, the correct number of states is correctly identified as posterior mode in all replicates for all four simulation models, with averaged posterior probability of the correct model equal to one (rounded to 3 decimal places). 
In contrast, with the bpSP method, the correct number of states is successfully selected in all repetitions only for simulation Models 1 and 3. For Model 4, it fails to pick N=2 in 1 out of 60 repetitions. The accuracy is notably poorer for Model 2, where $N=3$ is selected in only 2 out of the 60 repetitions, with $N=2$ being favoured in the majority of the remaining repetitions.
Note that the tendency of the bpSP method to underestimate the number of states is not entirely surprising. The underlying model for bpSP is highly parameterized due to the use of a large number of basis functions and the introduction of basis-specific local smoothing parameters. On the other hand, the marginal likelihood penalizes more complex models, and thus the increased complexity from adding a new state may outweigh the improvement in the fit.

\subsubsection*{Comparison with fixed $N$}
\begin{figure}[h!]
	\centering
	\includegraphics[width=0.98\textwidth]{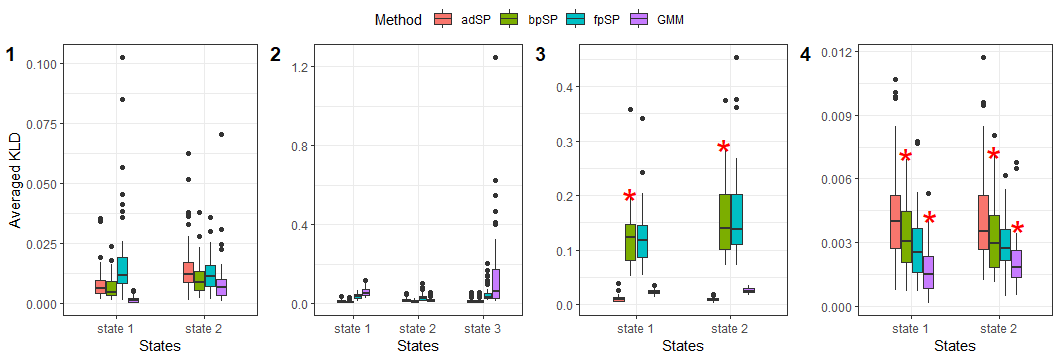}
	\includegraphics[width=0.98\textwidth]{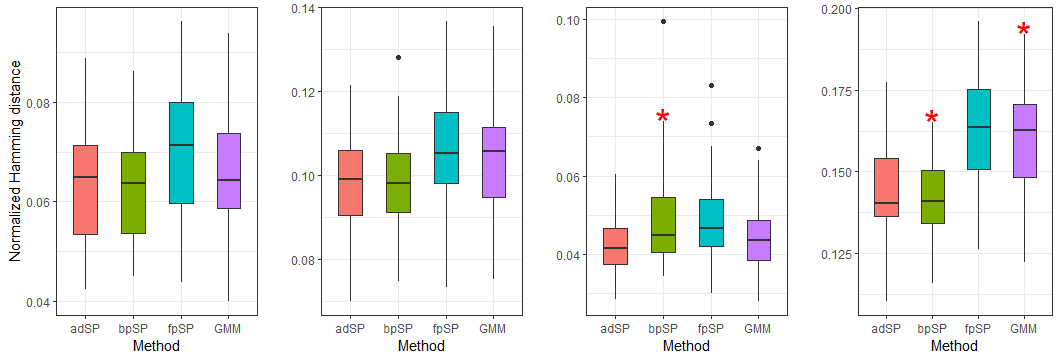}
	\includegraphics[width=0.98\textwidth]{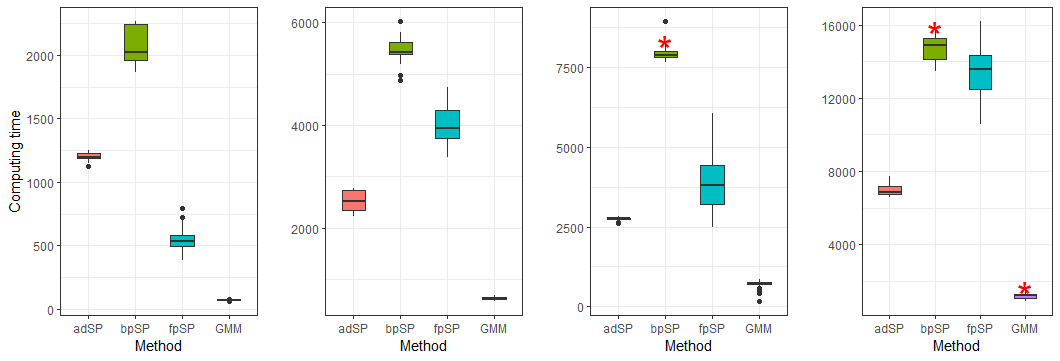}
	\caption[Summary of the estimation results]{\label{fig:simresults} 
		Boxplot summaries of the simulation results based on the average KLD (top panel), decoding error (middle panel), and computing time (bottom panel) for Models 1 (column 1), 2 (column 2), 3 (column 3), and 4 (column 4).  The upper whisker of the boxplot is marked with a red asterisk to indicate that it is based only on the convergent cases for that method. 
 }
\end{figure}

\begin{figure}[h!]
	\centering
	\includegraphics[width=0.35\textwidth]{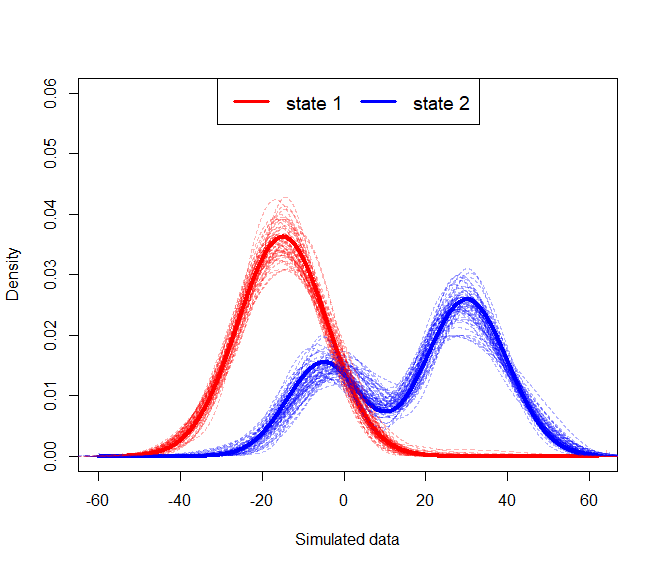}\hspace{-0.05in}\includegraphics[width=0.35\textwidth]{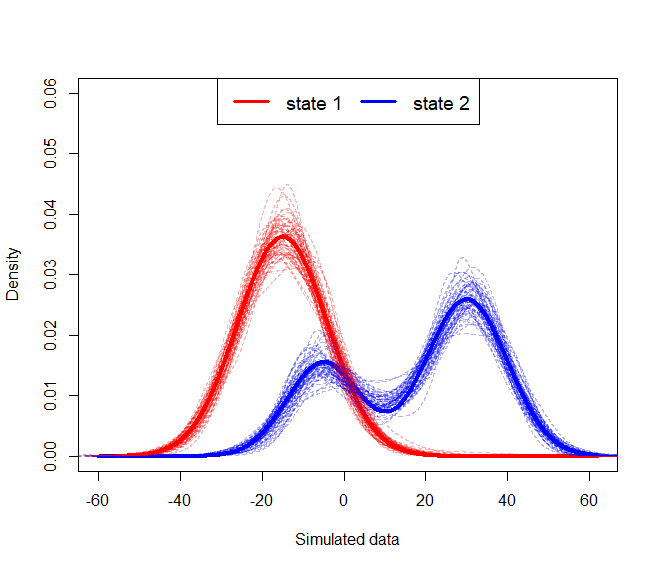}\\
	\vspace{-0.3in}
	\includegraphics[width=0.35\textwidth]{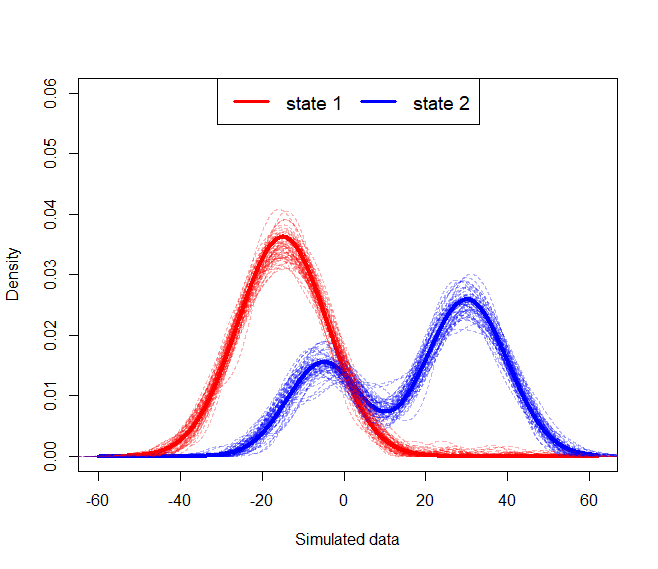}\hspace{-0.05in}\includegraphics[width=0.35\textwidth]{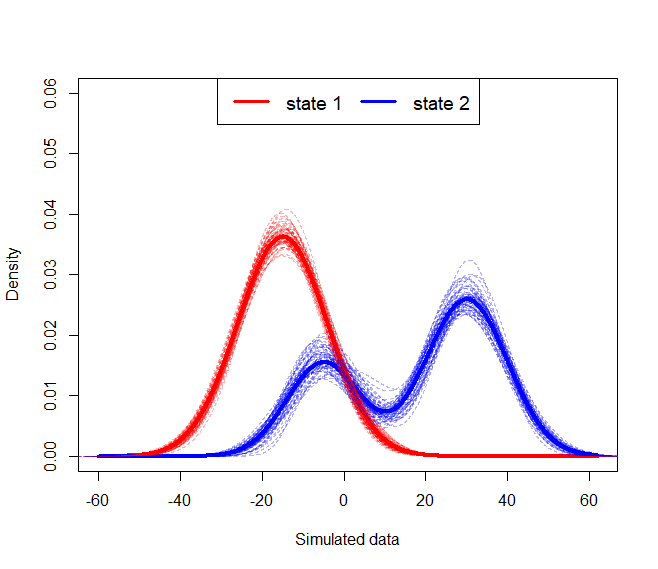}
	\caption[Estimation results for 60 simulations of Model 1]{\label{fig:sim1} Estimation results for 60 simulations of Model 1. Top left, top right, bottom left and bottom right panels show the true (solid curves) and estimated (dashed curves) densities of the emission distributions obtained in each replication using the adSP, bpSP, fpSP and GMM methods, respectively. }
\end{figure}

\begin{figure}[h!]
	\centering
	\includegraphics[width=0.35\textwidth]{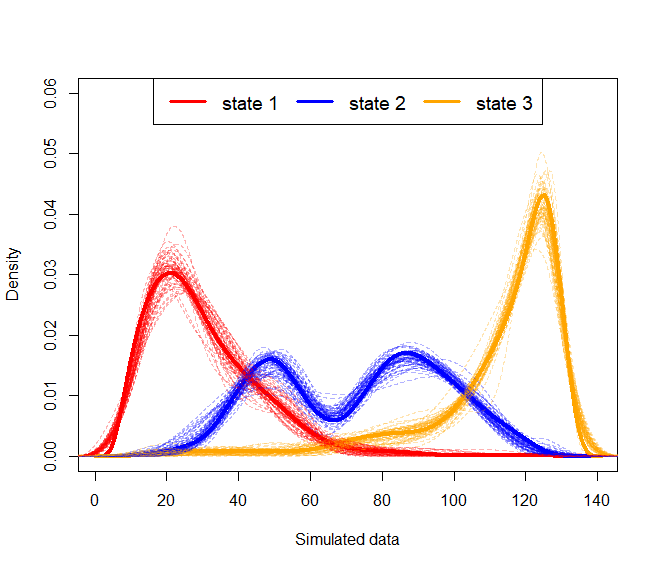}\hspace{-0.05in}\includegraphics[width=0.35\textwidth]{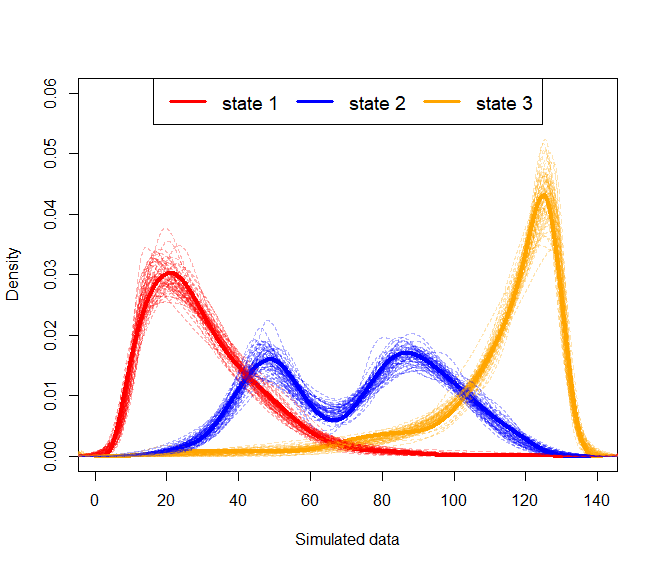}\\
	\vspace{-0.3in}
	\includegraphics[width=0.35\textwidth]{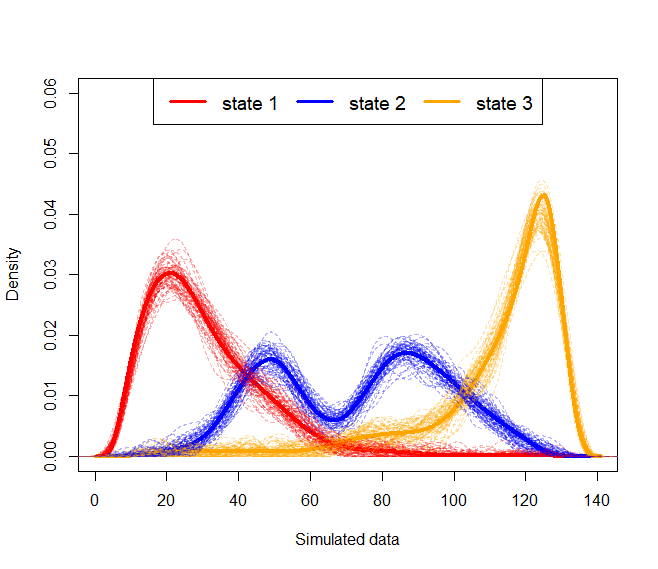}\hspace{-0.05in}\includegraphics[width=0.35\textwidth]{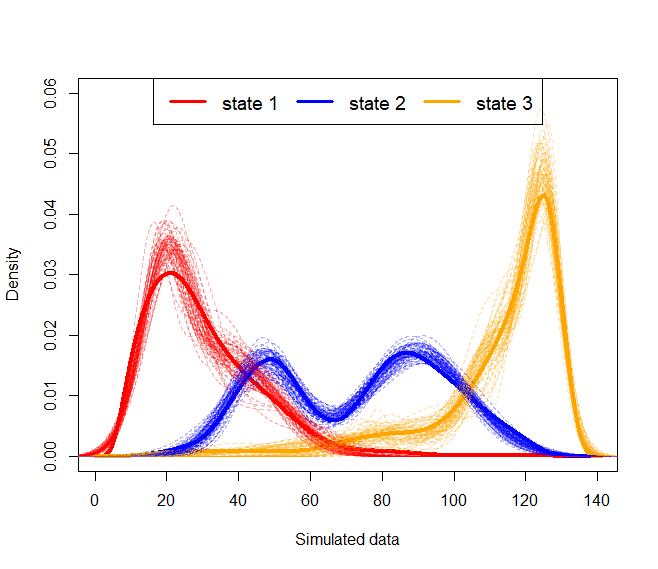}
	\caption[Estimation results for 60 simulations of Model 2]{\label{fig:sim2} Estimation results for 60 simulations of Model 2. The settings are the same as in Figure \ref{fig:sim1}.}
\end{figure}

\begin{figure}[h!]
	\centering
	\includegraphics[width=0.35\textwidth]{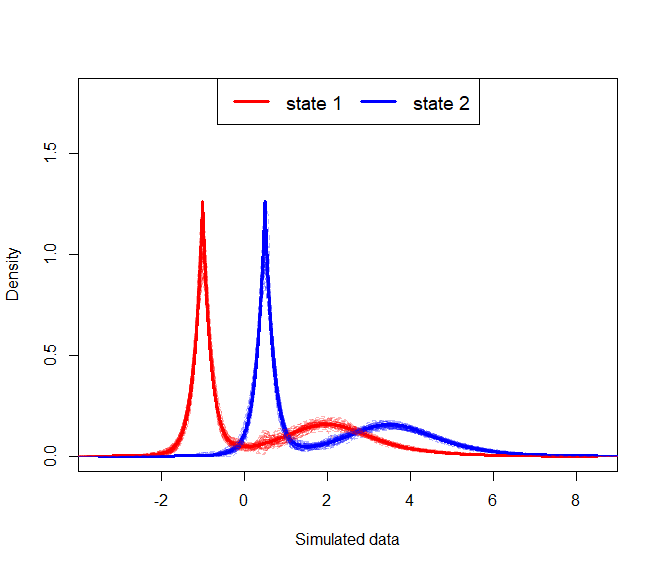}\hspace{-0.05in}\includegraphics[width=0.35\textwidth]{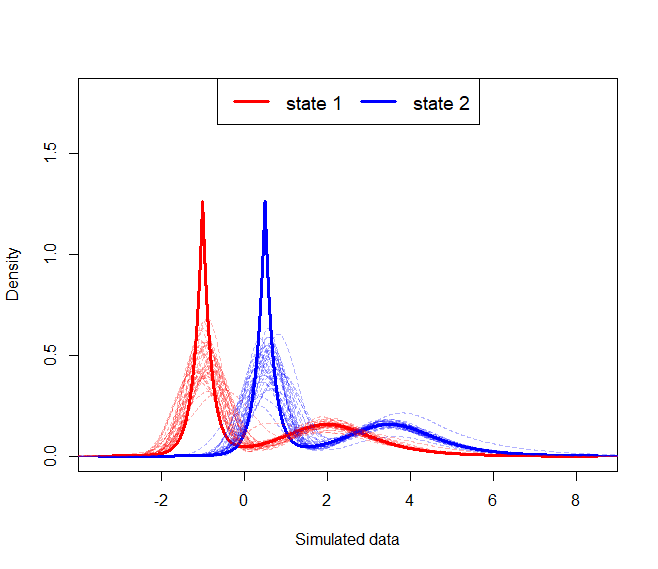}\\
	\vspace{-0.3in}
	\includegraphics[width=0.35\textwidth]{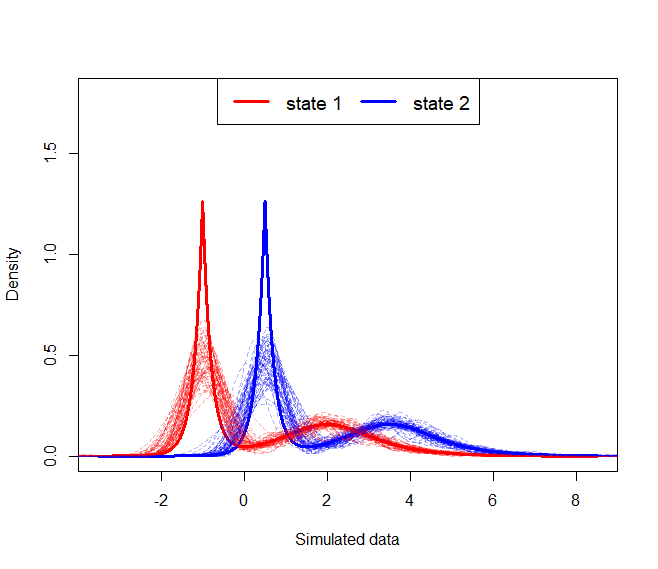}\hspace{-0.05in}\includegraphics[width=0.35\textwidth]{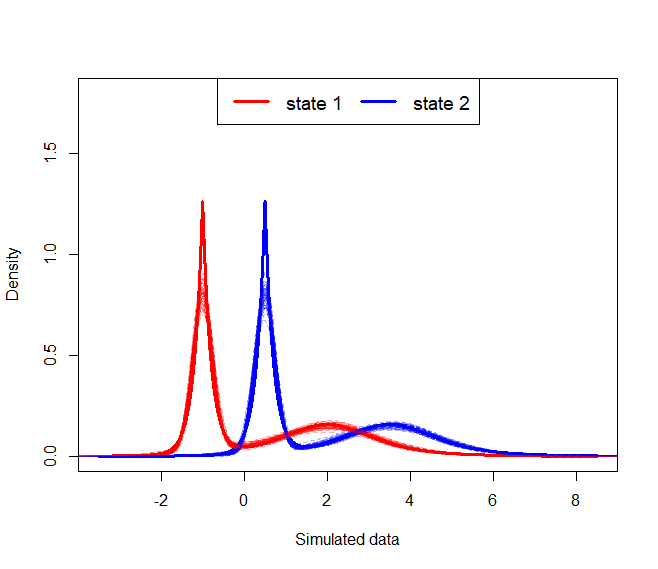}
	\caption[Estimation results for 60 simulations of Model 3]{\label{fig:sim3} Estimation results for 60 simulations of Model 3. The settings are the same as in Figure \ref{fig:sim1} (for the bpSP method only the 34 convergent cases are included).}
\end{figure}

\begin{figure}[h!]
	\centering
	\includegraphics[width=0.35\textwidth]{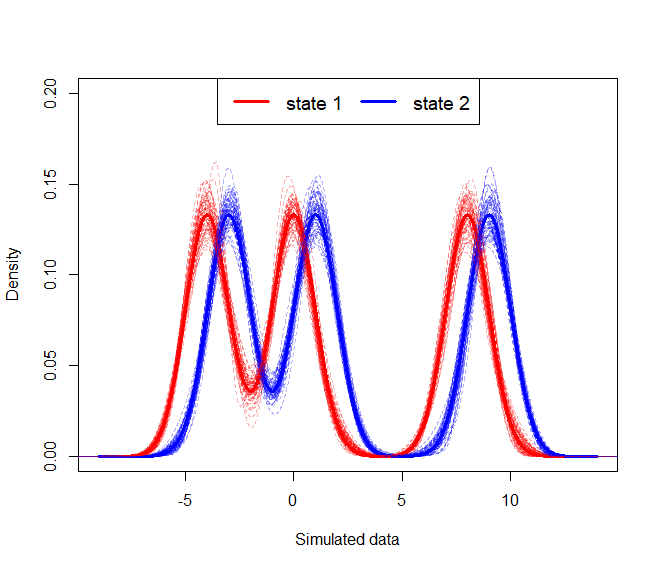}\hspace{-0.05in}\includegraphics[width=0.35\textwidth]{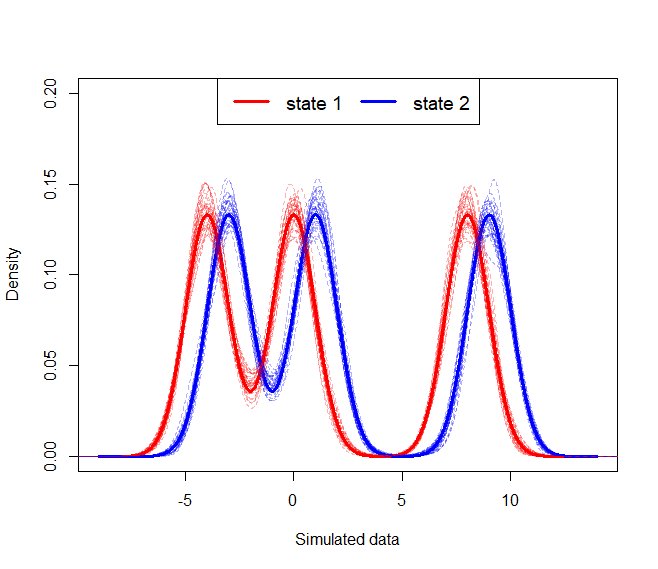}\\
	\vspace{-0.3in}
	\includegraphics[width=0.35\textwidth]{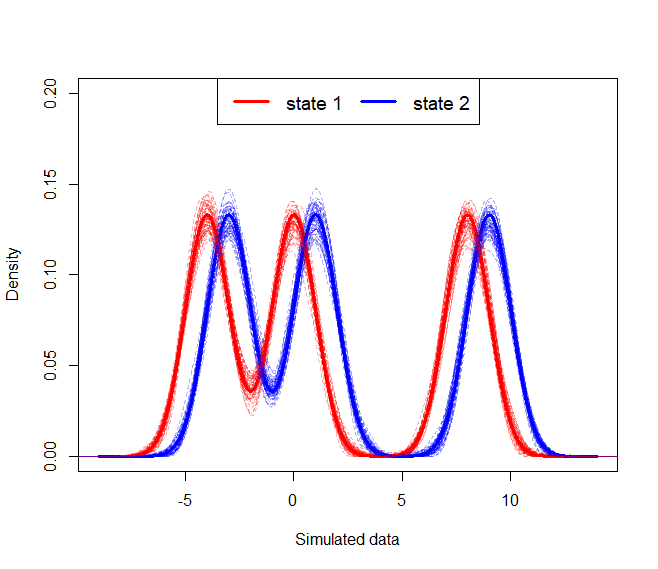}\hspace{-0.05in}\includegraphics[width=0.35\textwidth]{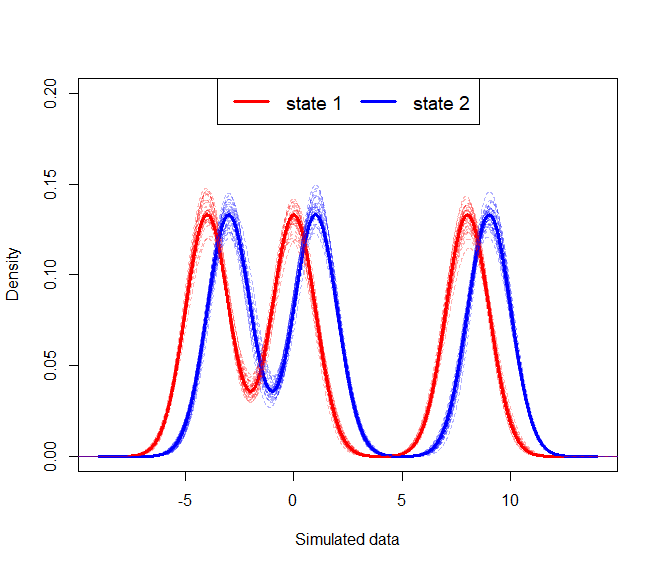}
	\caption[Estimation results for 60 simulations of Model 4]{\label{fig:sim4} Estimation results for 60 simulations of Model 4. The settings are the same as in Figure \ref{fig:sim1}. For the bpSP and GMM methods only 35 and 32 convergent cases are included.}
\end{figure}

Suppose that   $N$ is fixed at the true value. We compare the performance of the four competing methods across the four simulation models in Figure  \ref{fig:simresults}.
Additional Figures, namely Figures \ref{fig:sim1}, \ref{fig:sim2}, \ref{fig:sim3}, and \ref{fig:sim4}, show the estimated emission density obtained with each method for simulation models 1, 2, 3 and 4, respectively.
For Model 1, the adSP method indicated 6 or 7 as the posterior mode for $K$ with frequencies of $38.3\%$ and $50\%$, respectively. In contrast, the bpSP and fpSP methods used considerably more parameters with 31 basis elements for the estimation, yet the adSP  method  achieved comparable density fits in terms of the average KLD. The GMM approach achieved slightly better density fits which is to be  expected given that the true emissions are indeed Gaussian mixtures. In terms of decoding, adSP and bpSP demonstrate slightly better performance than the fpSP and GMM methods, with mean accuracies of $93.6\%$, $93.7\%$, $92.9\%$ and $93.5\%$, respectively, averaged over the $60$ repetitions. For  Model 2, the adSP method suggested $K=9$ as the posterior mode in $36/60$ of the simulation runs, while due to inefficient knot placements, the two P-spline-based approaches use a much larger $K=47$. Despite this, the fpSP method gave generally poorer fits. The GMM method performed the worst in this scenario, struggling to estimate the emissions of states 1 and 3 which possess excessive skewness, see also Figure \ref{fig:sim2}. The decoding accuracy of Bayesian spline-based methods outperformed that of fpSP and GMM, with average decoding accuracies of $90.2\%$, $90.2\%$, $89.3\%$ and $89.6\%$ for the adSP, bpSP, fpSP and GMM methods, respectively. In addition, we note that the GMM-based HMM tended to underestimate the probabilities of staying in states 1 and 3 (i.e. $\gamma_{1,1}$ and $\gamma_{3,3}$), possibly due to the lack of fit of the corresponding emissions.

Models 3 and 4 were designed to be more challenging, and our adSP method converged well in both cases while the other methods experienced varying degrees of numerical or convergence issues. To facilitate illustration, here we based our evaluation metrics summaries only on the convergent cases for the competing methods, though this may not provide a completely "fair" comparison to our method. For model 3, $K=14$ or 15 is suggested by adSP with frequencies of $46.7\%$ and $26.7\%$, respectively. The fpSP and bpSP methods, by contrast, require over 70 basis functions to capture the peak and smooth parts of the emissions. Despite careful selection of the initial parameter values, bpSP failed to converge to the correct posterior in 26 out of the 60 simulation runs. Furthermore, even with the use of much more complex spline models, both bpSP and fpSP methods still produced significantly larger average KLDs for the emissions and failed to adequately model the densities over the domain (see Figure \ref{fig:sim3}). While the GMM method achieved better density estimates than fpSP and bpSP, it still slightly underperformed compared to adSP. In terms of decoding, the adSP method demonstrated slightly better performance than the other methods, with averaged accuracies of $95.8\%$, $95.2\%$, $95.1\%$ and $95.6\%$ for adSP, bpSP, fpSP and GMM, respectively. For model 4, our adSP method successfully identified the tri-modal nature of the emissions in all 60 replicates despite the vague prior information we use. In $25/60$ of the simulation runs, our algorithm suggested using $K=16$, while in $22/60$ of the runs, $K=15$ was suggested. The bpSP and GMM methods, however, failed to converge (or converged to sub-optimal solutions) in 25 and 28 of the repetitions, respectively. When considering only the cases where convergence was achieved, bpSP, fpSP and GMM methods obtained slightly better density estimates than the adSP method. This result is understandable given the smooth nature of the true emissions and the fact that a much larger number of basis elements ($K=51$) are employed in the P-spline-based approaches, and the fact that the ground truth is a Gaussian mixture. However, when looking at decoding accuracy, our proposed adSP method outperformed the frequentist counterparts with an average accuracy of $85.6\%$, compared to $83.7\%$ and $84\%$ for the fpSP and GMM methods, respectively. The results demonstrate again the better decoding ability of the adSP over the frequentist counterparts, despite potentially slightly worse density fits in some cases. Our adSP method  also compares favourably with the Bayesian nonparametric approach of \citet{yau2011bayesian} on retrieving the transition dynamics of the hidden Markov chain, which is one of the main foci in their method. The average posterior means ($\pm1$ standard deviation) of the transition probabilities obtained from our method are $\gamma_{1,2}=0.056\ (\pm0.0096)$ and $\gamma_{2,1}=0.056\ (\pm0.0097)$, which is consistent with the true value $\gamma_{1,2}=\gamma_{2,1}=0.05$, and is comparable to those reported in \citet{yau2011bayesian} (Table 2, case $T=5000$). However, we note that  in contrast to \citet{yau2011bayesian} we did not need to assume a known translation nature of the emission  as our adSP method was able to identify this from the data.

In the bottom panel of Figure \ref{fig:simresults}, we compare the computational time required for each method in each of the 4 scenarios. For MCMC-based methods, this includes the total sampling time, including the burn-in period, while for frequentist approaches, this includes the total time for performing the penalized likelihood optimization based on the 50 different initializations. We can see that within the spline-based methods, our adSP is the most efficient method (except for Model 1), whereas bpSP is the most computationally intensive. Among the 4 methods, GMM required the least computing time, which is reflective of its simpler model structure. However, it is important to note that for fpSP and GMM, the time required for selecting the smoothing parameters or the number of mixture components is not taken into account here. Additionally, significant additional computational effort is still needed for uncertainty quantification of the parameters, especially for fpSP.

\subsection{Performance details of the proposed RJMCMC algorithm}
\begin{figure}[H]
	\centering
	\includegraphics[width=0.35\textwidth]{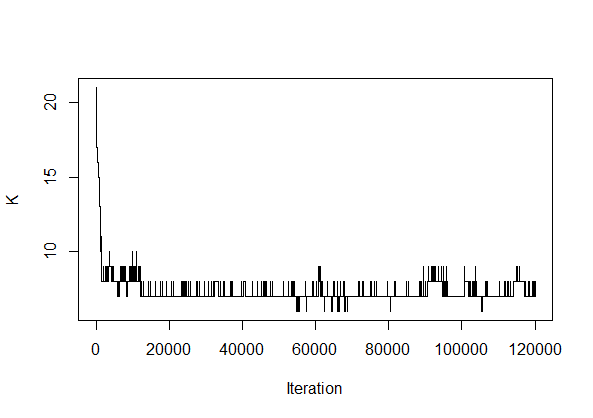}\includegraphics[width=0.35\textwidth]{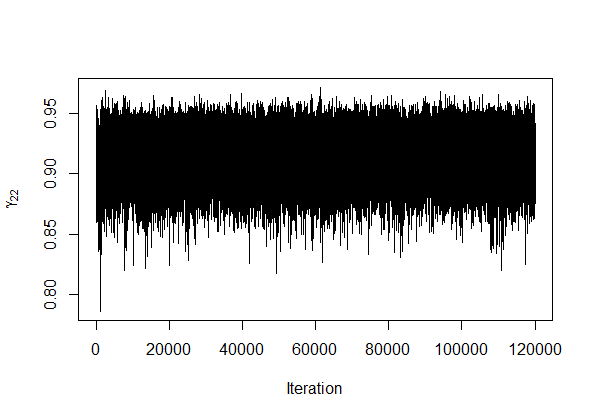}\\
	\vspace{-1em} 
	\includegraphics[width=0.35\textwidth]{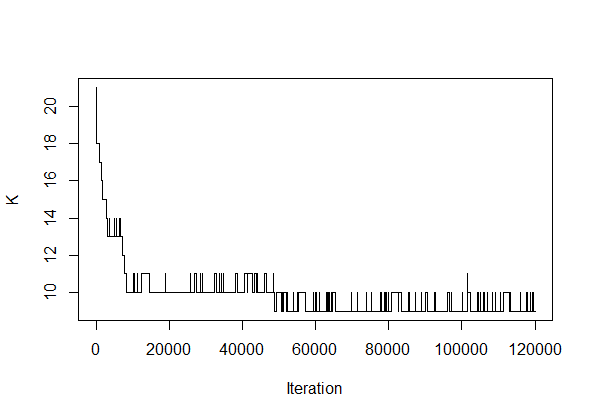}\includegraphics[width=0.35\textwidth]{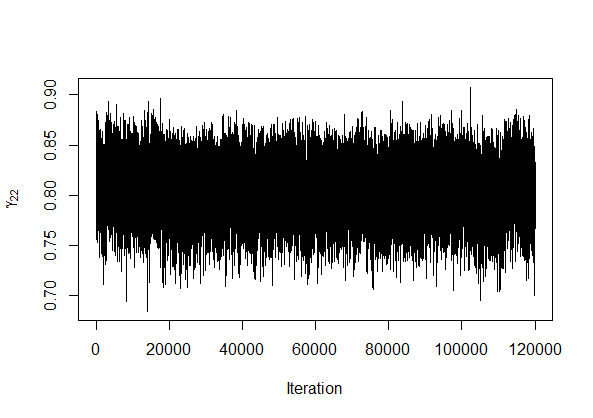}\\
	\vspace{-1em} 
	\includegraphics[width=0.35\textwidth]{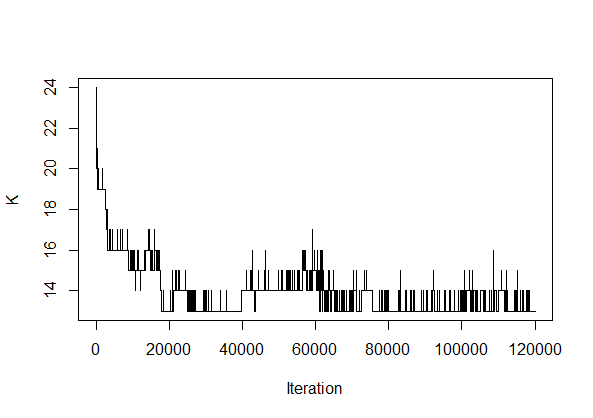}\includegraphics[width=0.35\textwidth]{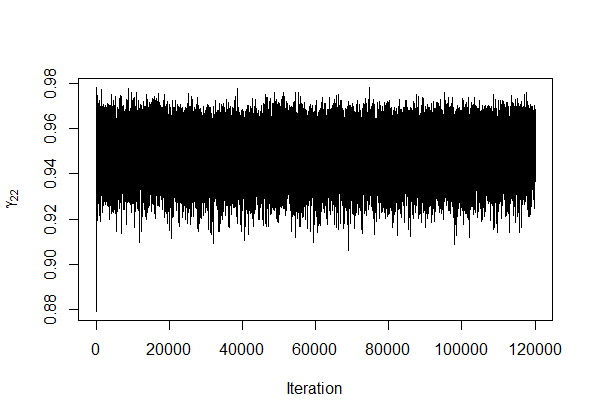}\\
	\vspace{-1em} 
	\includegraphics[width=0.35\textwidth]{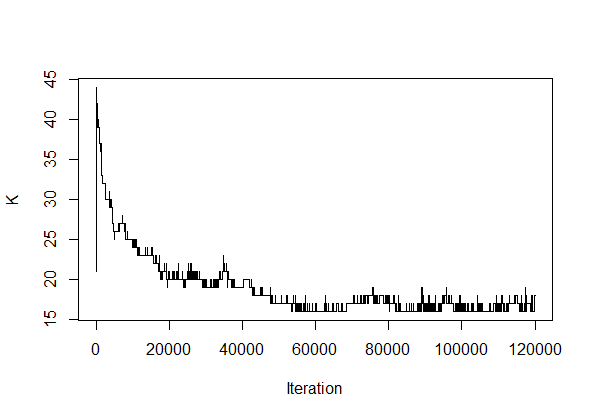}\includegraphics[width=0.35\textwidth]{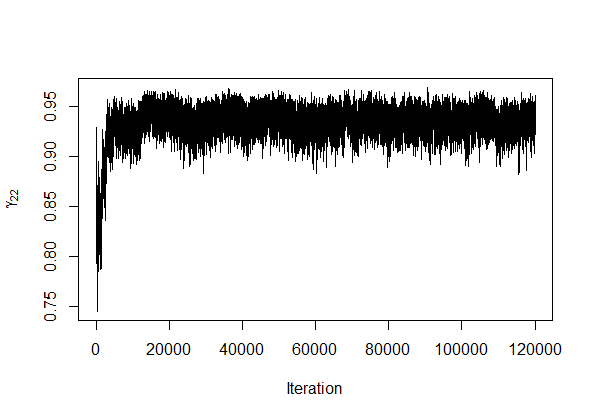}
	\caption[Convergence diagnostics]{\label{fig:simdiagnostics} Selected MCMC outputs for a single data set simulated from models 1 (top panel), 2 (middle-top panel), 3 (middle-bottom panel), and 4 (bottom panel), conditioned on the true value of $N$. The left panel shows the trace of $K$, and the right panel shows the trace of the transition probability $\gamma_{2,2}$ for the complete MCMC run, including burn-in. }.
\end{figure}

\begin{figure}[ht]
	\centering
	\includegraphics[width=0.35\textwidth]{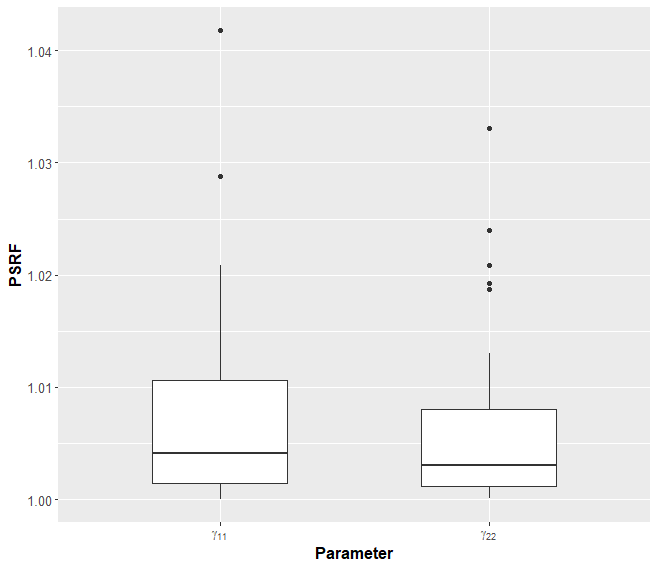}\hspace{-0in}\includegraphics[width=0.35\textwidth]{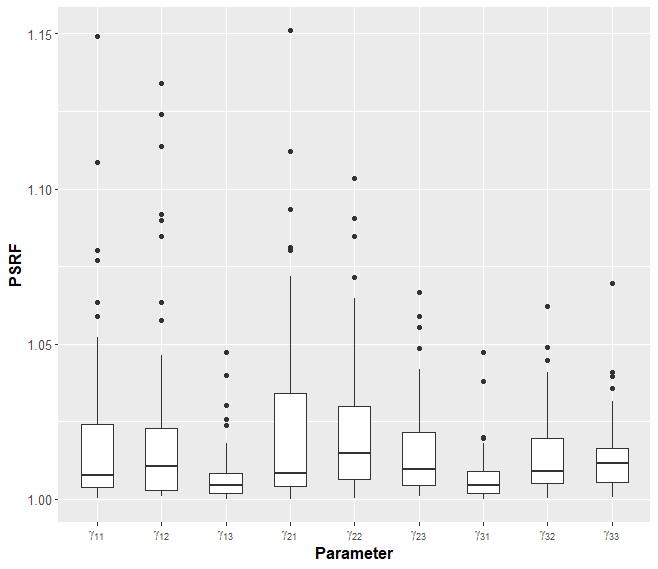}
	\includegraphics[width=0.35\textwidth]{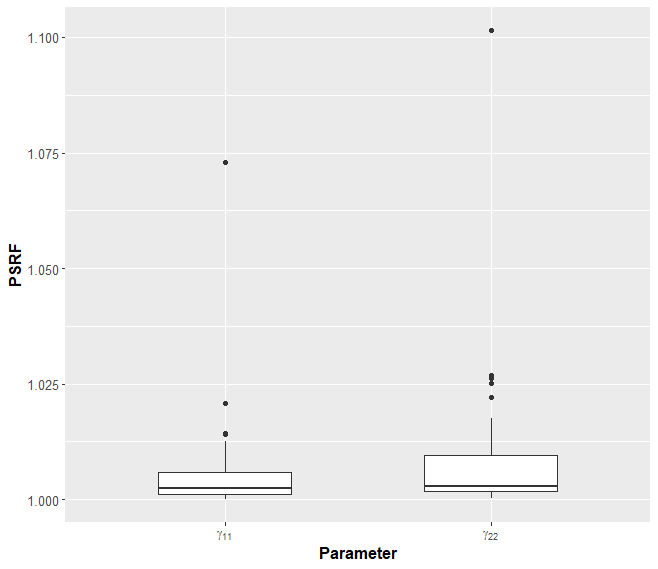}\hspace{-0in}\includegraphics[width=0.35\textwidth]{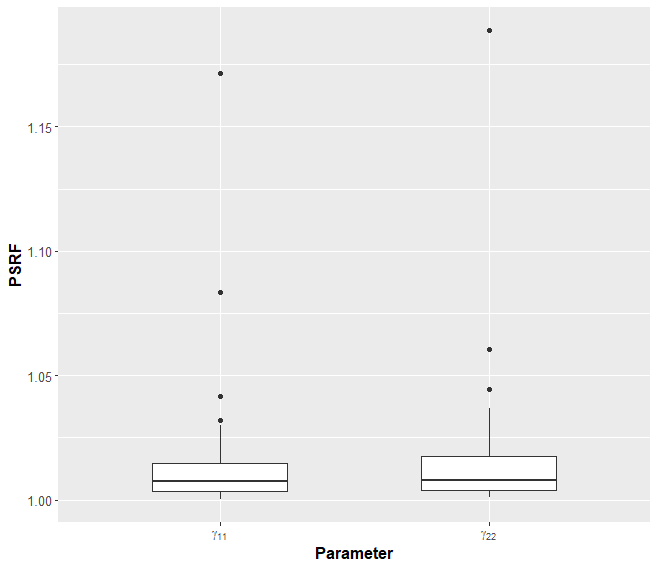}
	\caption[PSRF]{\label{fig:psrf} Boxplot summary of potential scale reduction factors (PSRF) for parameters in the transition matrix for simulation models 1 (top left), 2 (top right), 3 (bottom left) and 4 (bottom right). Note that for models 1, 3 and 4, the PSRF values for $\gamma_{12}$ and $\gamma_{21}$ are identical to those for $\gamma_{11}$ and $\gamma_{22}$, respectively, so they are omitted here. }
\end{figure}

For adaptive MCMC steps (steps (c)-(e) of Algorithm 1), the empirical acceptance rates are closed to the pre-specified desired levels. We examined the trace plots and running averages of selected parameters, including the tuning variance parameters, across MCMC iterations and found acceptable mixing patterns in most cases. However, step (c) might mix more slowly compared to other MH steps due to the relatively high dimensionality of the spline coefficients vector. 
For the dimension changing moves, the averaged acceptance rates for models 1-4 are $0.27\%$, $0.15\%$, $0.55\%$ and $0.41\%$, respectively. While these rates are lower than desired in our simulation cases, we did not detect any apparent convergence issues from our diagnostic trace plots (see Figure \ref{fig:simdiagnostics}). To further assess the convergence of the Markov chain, we performed 4 independent MCMC runs with different initializations for each of the 60 replicates generated for each of the 4 simulation models considered. We employed the Gelman-Rubin convergence diagnostic \citep{gelman1992inference}, with a focus on parameters in the fix-dimensional transition matrix. The obtained potential scale reduction factors (PSRF) are summarized in Figure \ref{fig:psrf}, which are generally very close to 1, supporting our visual diagnostic on the trace plots. 

We investigated various modifications of the proposed algorithm. One possible modification is to replace the random walk MH with a Metropolis adjusted Langevin algorithm (MALA) \citep{roberts1996exponential} for updating the spline coefficients. MALA is a specific class of MH algorithms which exploits the local gradient information of the target distribution to propose new state, making it highly effective for sampling complex distributions. In our context, the proposal takes the form
\begin{displaymath}
	q(\Tilde{A}_{K}^{'}|\Tilde{A}_{K})=\mathcal{N}_{K+4}(\Tilde{A}_{K}+\frac{h}{2}\Sigma\nabla_{\Tilde{A}_{K}}\log f, h\Sigma),
\end{displaymath}
where $h$ is a positive real number, $\Sigma$ is a symmetric positive definite matrix and \\
$\nabla_{\Tilde{A}_{K}}\log f=(\frac{\partial}{\partial {\Tilde{a}_{i,j}}}\log f)_{i=1,\ldots.N;j=1,\ldots,K+4}$,
with 
\[
\frac{\partial}{\partial {\Tilde{a}_{i,j}}}\log f=\sum_{t:x_{t}=i}\frac{\partial}{\partial {\Tilde{a}_{i,j}}}\log f_{i}(y_{t})+\frac{\partial}{\partial {\Tilde{a}_{i,j}}}\log f_{LG}(\Tilde{a}_{i,j}|\zeta,1).
\]
Both $h$ and $\Sigma$ are tuning parameters that need to be selected using pilot runs or tuned on-the-fly using adaptive MALA techniques, see, e.g. \citet{atchade2006adaptive,christensen2005scaling}. However, the potential gain in sampling efficiency from using MALA comes at the cost of a much higher computational cost at each iteration due to the need to evaluate gradients. In our simulation scenarios, we found the random walk MH to be sufficient. However, in more challenging cases, one may consider combing MALA and random walk MH updates. For instance, MALA could be used for the first few thousand iterations to speed up convergence to stationarity, and then random walk MH can be used for the remaining iterations. 

For the reversible jump move, the acceptance rate may be mildly affected by the standard deviation $\tau$ of the truncated normal distribution used to generate the new knot or the proposal distribution for $u_{i}$ used in the birth move. 
In our experiments, with the functional form of $\tau$ fixed, the results are not very sensitive to the value of $\alpha$, provided that it is chosen in a reasonable way. For instance we prefer to set $0<\alpha<1$ when the averaged distance between knots is much larger than 1 and vice versa. The use of other potential proposal distributions for $u_{i}$ within the Beta family was also investigated, but no clear evidence was found in terms of superiority of different choices over the noninformative choice $U(0,1)$. A potentially promising strategy is to generate the random variables $u_{i}$ used in equation \eqref{eq:birth} from a more informative proposal distribution such as a truncated normal distribution centred at its maximum likelihood estimate (approximated via some numerical optimization routines). However, the computational cost of the algorithm will increase dramatically, which may prevent a successful application of the algorithm in some settings. We also compared our algorithm with the one that integrates out the latent state sequence $\mathbf{x}^{(n)}$, using the forward algorithm in \eqref{eq:obsllk}, and subsequently draws $\mathbf{x}^{(n)}$ at each iteration using the FFBS algorithm, conditional on the simulated model parameters. Although this strategy yielded a slightly higher acceptance rate, which is not surprising given the reduced dimensionality of the parameter space, there is no noticeable gain in terms of overall computational cost or estimation accuracy.

\section{Additional simulation studies}
\label{sec:AP_simulation2}
\subsection{Systems with more complex transition dynamics}
\begin{figure}[h!]
	\centering
	\includegraphics[width=0.8\textwidth]{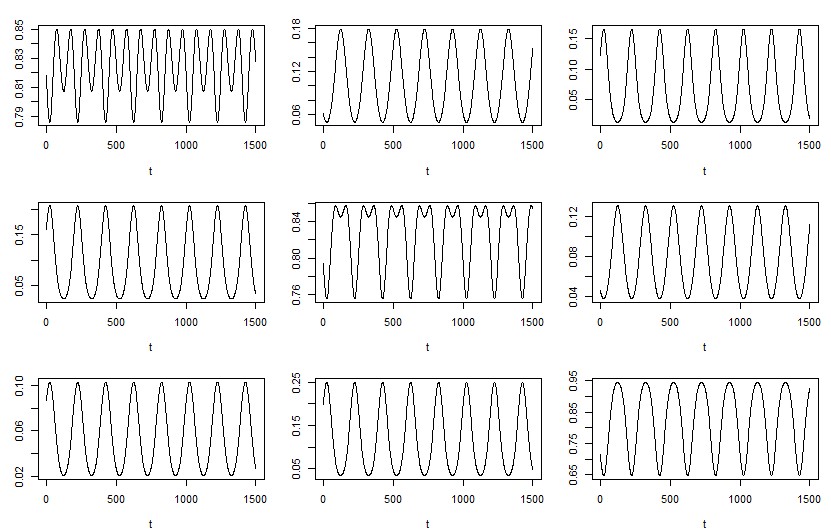}
	\caption[Nonhomogeneous transitions]{\label{fig:nhtrans} Transition probabilities as a function of time. The $(i,j)$-th panel corresponds to the $(i,j)$-th entry in the transition matrix. }
\end{figure}

In reply to an anonymous referee we modify Model 2 of Section \ref{sec:appsimulation} to further assess the robustness of our proposed method against misspecification of the transition model, i.e., when the true generating process of the latent state sequence $\{x_{t}\}$ does not follow a homogeneous Markov chain as assumed before. We considered two possible deviations in this context, using the same set of emission distributions as in Model 2. In the first case, we allow $x_{t}$ to follow a semi-Markov process where the associated sojourn time in each state is explicitly modelled via a separate distribution, rather than the Markov implied geometric distribution \citep{yu2010hidden}. More specifically, let $\tilde{\mathbf{x}}=\{\tilde{x}_{t}\}$ denotes the ``super-states" defined on $\{1,\ldots,N\}$ following a Markov process with transition matrix given by $\Omega=(\Omega_{i,j})_{i,j=1,\ldots,N}$, with $\Omega_{i,j}=P(\tilde{x}_{t}=j|\tilde{x}_{t-1}=i)$ and $\Omega_{i,i}=0$, $i=1,\ldots,N$. Therefore no consecutive realizations of $\tilde{x}_{t}$ can be equal. 
Conditional on $\tilde{x}_{t}$, a dwell time $d_{t}$ is generated according to a distribution $g_{\tilde{x}_{t}}$ parametrized by some state-specific hyperparameters, indicating how long we stay at the current super state $\tilde{x}_{t}$.
Such runs of ``$d_{t}$-values" of $\tilde{x}_{t}$, stacked together, form a semi-Markov chain $\{x_{t}\}$.
The generative procedure of $\{x_{t}\}$ may thus be summarized as
\begin{displaymath}
	\tilde{x}_{t}|\tilde{x}_{t-1} \sim \Omega_{\tilde{x}_{t-1}},
\end{displaymath}
\begin{displaymath}
	d_{t}|\tilde{x}_{t} \sim g_{\tilde{x}_{t}}, \quad t=1,2,\ldots,T,
\end{displaymath}
where $\Omega_{i}$ denotes the $i$-th row of $\Omega$. Here $N=3$ and we set $\Omega_{1,2}=\Omega_{2,1}=\Omega_{1,3}=\Omega_{3,1}=\Omega_{2,3}=\Omega_{3,2}=0.5$ and $T$ to be the smallest value such that the sample size passes $n=1500$. The sojourn time distributions $g_{\tilde{x}_{t}}$ for states 1, 2 and 3 (cf. Model 2 of Section \ref{sec:appsimulation}) are given by $NB(3,0.45)$, $NB(2,0.3)$ and $Po(6)+1$, respectively, where $NB(k,p)$ denotes the negative binomial distribution with density $P(d|k,p)=\big(\begin{smallmatrix}
	d-1 \\
	k-1 
\end{smallmatrix}\big)p^{k}(1-p)^{d-k}$
and $Po(\lambda)$ denotes a Poisson distribution with rate $\lambda$.
Note that these hyperparameters were chosen such that the mean dwell time for each state is roughly the same as those implied by the Markov transition model used in Model 2. However, in this construction, the process no longer has the memoryless property. In the second scenario we consider a time inhomogeneous Markov chain model for the state process such that the transition probabilities between states exhibit oscillatory patterns over time. Motivated by the settings of the circadian HMM as developed in \citet{huang2018hidden}, we specify the entries of $\Gamma$ as  
\begin{displaymath}
	P(x_{t}=k|x_{t-1}=j,z_{t})=\frac{\exp(\beta^{0}_{j,k}+\beta^{1}_{j,k}z_{t})}{\sum_{l=1}^{N}\exp(\beta^{0}_{j,l}+\beta^{1}_{j,l}z_{t})}, \quad k,j=1,\ldots,N,
\end{displaymath}
where $x_{t}=sin(2\pi t/T)+cos(2\pi t/T)$ which is an oscillator with period $T$.
In our experiments we set $n=1500$ as in Model 2, $T=200$ and the coefficients are chosen as $\beta^{0}_{1,2}=\beta^{0}_{3,2}=\log(2/17)$, $\beta^{0}_{1,3}=\beta^{0}_{3,1}=\log(1/17)$, $\beta^{0}_{2,1}=\beta^{0}_{2,3}=\log(3/34)$ and $\beta^{1}_{1,2}=-0.45$, $\beta^{1}_{1,3}=0.9$, $\beta^{1}_{2,1}=0.8$, $\beta^{1}_{2,3}=-0.4$, $\beta^{1}_{3,1}=0.7$, $\beta^{1}_{3,2}=0.84$. 
$\beta^{0}_{i,i}$ and $\beta^{1}_{i,i}$, $i=1,\ldots,N$, are fixed to 0.
Figure \ref{fig:nhtrans} shows the resulting transition probabilities as a function of time.

For each of the two scenarios considered, 60 random replicates of the data were generated and the proposed method was implemented for each replicate under the same setting as described in Section \ref{sec:appsimulation}. 
We started by performing model selection. With a uniform prior on $N$ over the candidate set $\{2,3,4,5\}$, the correct value $N=3$ was identified by the proposed marginal likelihood estimator in all repetitions of both simulation scenarios, with averaged posterior probability of the correct model equal
to one (rounded to 3 decimal places).
Conditional on $N=3$, we examined estimation accuracy of our method in these two cases in terms of density estimation and state decoding. 
The average KLD (as defined in Section \ref{sec:appsimulation}) for each state obtained by averaging over the 60 replications is $(0.014,0.033,0.022)$ (state 1, state 2 and state 3) and $(0.023,0.024,0.009)$ for the semi-Markov and time-inhomogeneous cases, respectively. 
Thus our density fits are only slightly worse than those obtained for the correctly specified Model 2 for which the averaged KLDs for the corresponding states are $(0.013,0.018,0.015)$ (cf. upper panel of Figure \ref{fig:simresults}).
Our robust performance is also reflected in the decoding accuracy (averaged over the 60 repetitions), with $90.2\%$ and $89.2\%$ for the former and latter scenarios (cf. middle panel of Figure \ref{fig:simresults}), respectively.

\subsection{Systems with larger number of regimes}
\begin{figure}[H]
	\centering
	\includegraphics[width=0.4\textwidth]{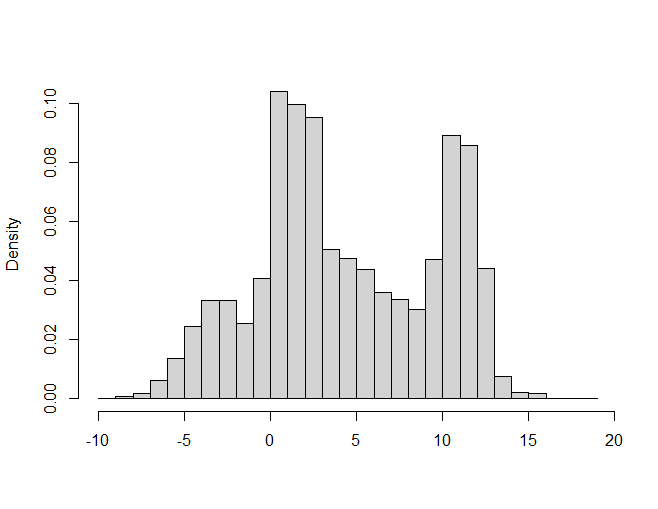}\hspace{-0in}\includegraphics[width=0.4\textwidth]{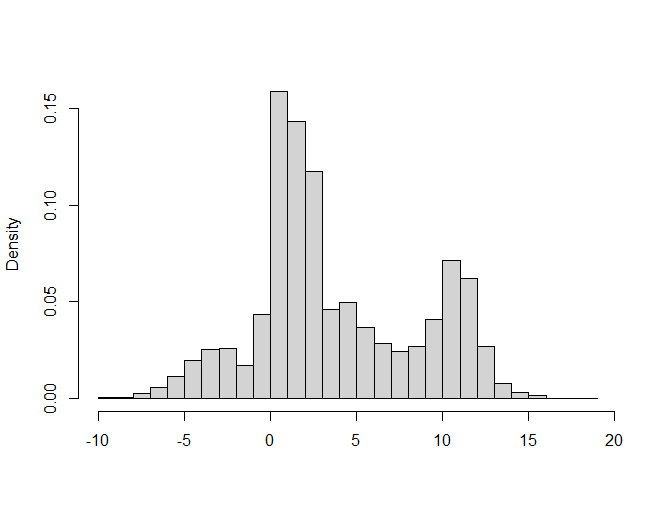}
	\includegraphics[width=0.4\textwidth]{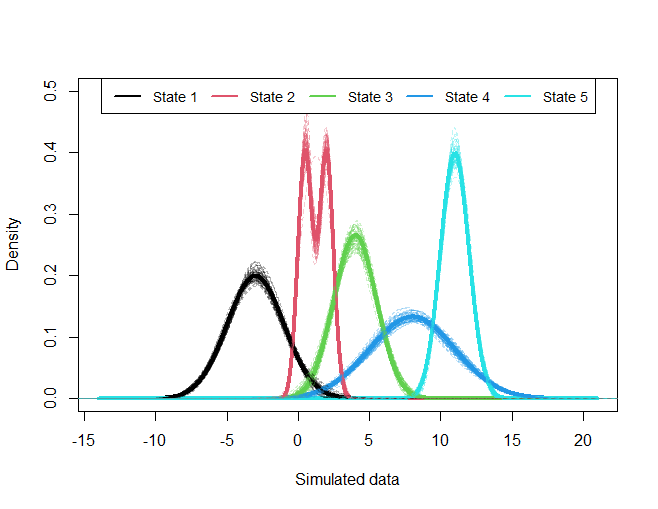}\hspace{-0in}\includegraphics[width=0.4\textwidth]{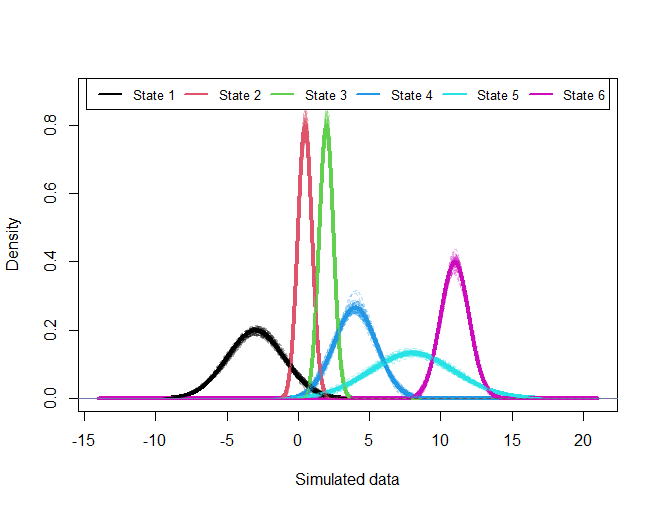}
	\includegraphics[width=0.4\textwidth]{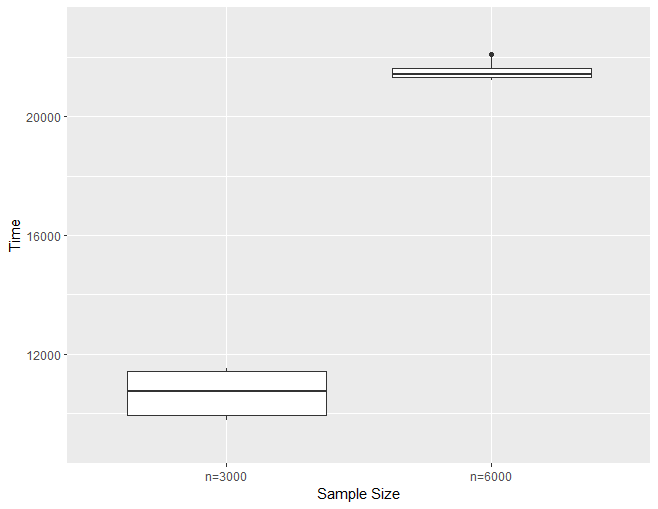}\hspace{-0in}\includegraphics[width=0.4\textwidth]{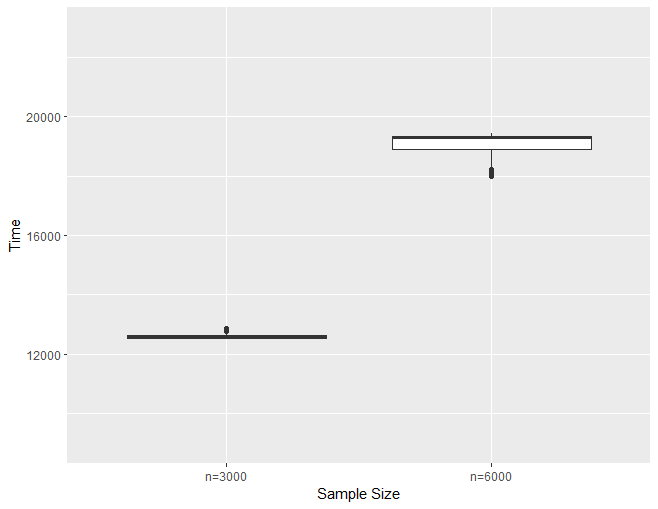}
	\caption[largerN]{\label{fig:GaussianN5} Results for the 5-state (left panel) and 6-state (right panel) models. Top row: marginal histograms of a single data set simulated from each true model; middle row: true (solid curves) and estimated (dashed curves) emission densities obtained in each replication using the proposed algorithm, conditional on the true number of states; bottom panel: total sampling time including burn-in (in seconds) required for the 5 and 6-state models with sample sizes of $n=3000$ and $n=6000$. }
\end{figure}

To show the feasibility of the extended algorithm with state-specific knots (as described in Section \ref{sec:AP_statespec_adsp}) in situations with a relatively large number of states, as requested by an anonymous referee, we considered two additional simulation models with 5 and 6 hidden states.
For the 5-state model, the emissions are specified as $y_{t}|x_{t}=i \sim N(\mu_{i},\sigma^{2}_{i})$, for $i=1,3,4,5$, and $y_{t}|x_{t}=2 \sim 0.5N(\mu_{21},\sigma^{2}_{21})+0.5N(\mu_{22},\sigma^{2}_{22})$. The parameter values are set as follows: $\mu_{1}=-3$, $\mu_{21}=0.5$, $\mu_{22}=2$, $\mu_{3}=4$, $\mu_{4}=8$, $\mu_{5}=11$, $\sigma_{1}=2$, $\sigma_{21}=\sigma_{22}=0.5$, $\sigma_{3}=1.5$, $\sigma_{4}=3$ and $\sigma_{5}=1$.
The states were generated using a uniform initial distribution and the transition matrix
\begin{displaymath}
	\Gamma=\begin{pmatrix} 
		0.75 & 0.1 & 0.05 & 0.05 & 0.05  \\
		0.05 & 0.8 & 0.05  & 0.05 & 0.05\\
		0.05 & 0.1 & 0.75 & 0.05 & 0.05\\
		0.05 & 0.05 & 0.05  & 0.8  & 0.05 \\
		0.05 & 0.05 & 0.05 & 0.05 & 0.8
	\end{pmatrix}.
\end{displaymath}
The 6-state model is derived from the 5-state model by splitting state 2, which has a Gaussian mixture emission, into 2 separate states. The transition matrix for this model is specified as 
\begin{displaymath}
	\Gamma=\begin{pmatrix} 
		0.75 & 0.09 & 0.04 & 0.04 & 0.04 & 0.04\\
		0.04 & 0.80 & 0.04 & 0.04 & 0.04 & 0.04\\
		0.04 & 0.04 & 0.80 & 0.04 & 0.04 & 0.04\\
		0.04 & 0.04 & 0.09 & 0.75 & 0.04 & 0.04\\
		0.04 & 0.04 & 0.04 & 0.04 & 0.80 & 0.04\\
		0.04 & 0.04 & 0.04 & 0.04 & 0.04 & 0.80
	\end{pmatrix}.
\end{displaymath}
For each model, we examined two different sample sizes: $n=3000$ and $n=6000$. For each sample size, we generated 60 random replicates of the data. Note that in both cases, neither the number nor the structure of the emissions can be directly identified by visually inspecting the empirical marginal distributions (see top panel of Figure \ref{fig:GaussianN5}).

\begin{figure}[H]
	\centering
	\includegraphics[width=0.8\textwidth]{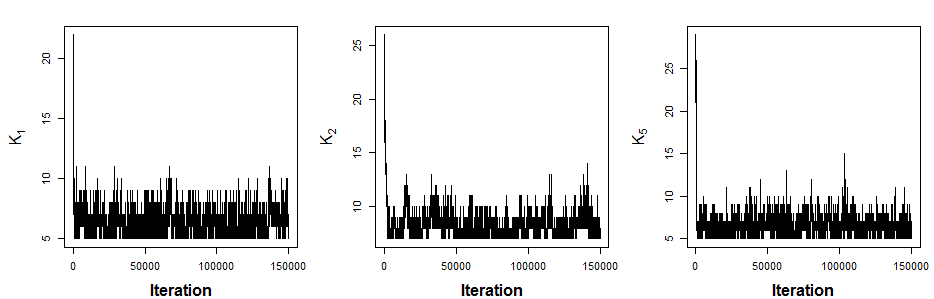}\\
	\includegraphics[width=0.8\textwidth]{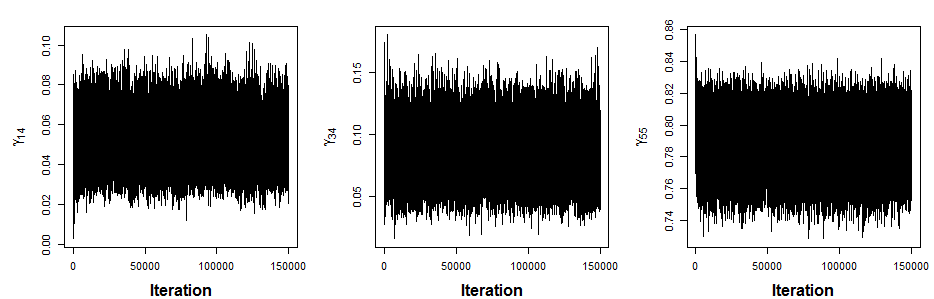}
	\caption[largerN]{\label{fig:GaussianN5diag} Selected MCMC trace plots for a single data set simulated from the 5-state model ($n=6000$), with burn-in included. Upper panel: traces of $K_{1}$, $K_{2}$ and $K_{5}$; lower panel: traces of transition probabilities $\gamma_{1,4}$, $\gamma_{3,4}$ and $\gamma_{5,5}$. MCMC run was conditioned on $N=5$. }
\end{figure}

\begin{figure}[H]
	\centering
	\includegraphics[width=0.8\textwidth]{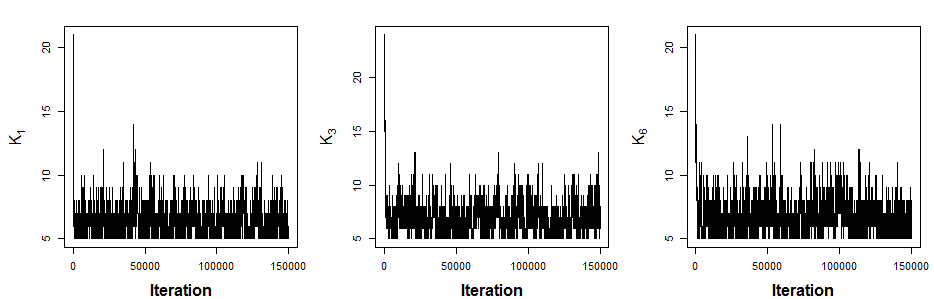}\\
	\includegraphics[width=0.8\textwidth]{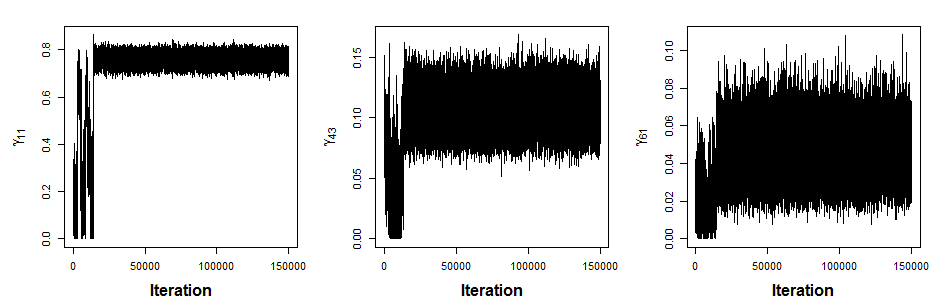}
	\caption[largerN]{\label{fig:GaussianN6diag} Selected MCMC trace plots for a single data set simulated from the 6-state model ($n=6000$), with burn-in included. Upper panel: traces of $K_{1}$, $K_{3}$ and $K_{6}$; lower panel: traces of transition probabilities $\gamma_{1,1}$, $\gamma_{4,3}$ and $\gamma_{6,1}$. MCMC run was conditioned on $N=6$. }
\end{figure}

We first performed model selection on $N$ from the candidate set $\{3,4,5,6,7\}$ using the marginal likelihood based approach with $\beta=0.2$ and $\xi=0.01$ as before, assuming a uniform prior on $N$. For each model and each $N$ considered, our algorithm was run for total of $150$k iterations, with the first $100$k iterations discarded as burn-in. Our results were then based on post-thinning samples with a thinning interval of $10$. Figures \ref{fig:GaussianN5diag} and \ref{fig:GaussianN6diag} show selected trace plots for the state-specific number of knot points and entries in the transition matrix for a randomly chosen replication of the data for each model with $n=6000$, conditioned on the correct number of states. The plots exhibit reasonably good mixing patterns.

When $n=3000$, the correct number of states was identified as the posterior mode in in 57 out of 60 repetitions for the 5-state model, and 58 out of 60 repetitions for the 6-state model. Conditional on replicates where $N$ was correctly selected, the algorithm achieved an average decoding accuracy of $91.2\%$ and $91.8\%$ for the 5 and 6-state cases, respectively. When increasing the sample size to $n=6000$, we were able to identify the correct model with increasingly strong support and achieve higher estimation and decoding accuracy. For both models, the correct number of states was selected in all 60 replications, with an average (over $60$ replications) posterior probability of the true model equal to 1 (rounded to 3 decimal places). The decoding accuracy improved as sample size increased, with average values of $91.3\%$ and $92.2$ for the 5 and 6-state cases, respectively. Conditional on the true value of $N$, Figure \ref{fig:GaussianN5} (middle panel) shows the true and estimated emission densities obtained in the 60 repetitions for the 5 and 6-state models, and we can see that our algorithm provided good fits and revealed the true nature of the emissions in almost all repetitions.
The bottom panel of Figure \ref{fig:GaussianN5} shows the total sampling time (including burn-in) required for the 5 and 6-state models with the two different sample sizes. The cost of the algorithm scales roughly linearly with $n$. However, we did not see a dramatic increase in computing time when moving from $N=5$ to $N=6$. In particular, with $n=6000$, the average computing time are approximately 5.97 and 5.3 hours for the 5 and 6 state models, respectively. This could be due to the simpler structure of the 6-state model compared to the 5-state model. Overall, the algorithm is computationally feasible in all scenarios.

Though not presented here, we also investigated the estimation performance of the proposed algorithm (using shared knots or state-specific knots) with other simulation settings where the transition model is correctly specified or misspecified. Our preliminary findings indicate that the performance generally improves as serial correlation and/or sample size increase, while it declines as the number of states and/or the overlap of the emission distributions increase. These patterns are commonly observed in parametric HMM inference. It is also important to note that, as with any estimator, the performance of our method depends on the settings of the true simulation model. One can imagine that it would be very challenging or even impossible to accurately estimate the number of states and/or achieve highly precise parameter estimates when the data size is too ``small" and/or the temporal dependency in the signal is weak. Obtaining an exact quantification of such dependency and estimation errors for the proposed method requires further theoretical research.

\section{Further details of the empirical application of Section 4.1}
\label{sec:AP_shark}
\begin{figure}[h!]
	\centering
	\includegraphics[width=0.5\textwidth]{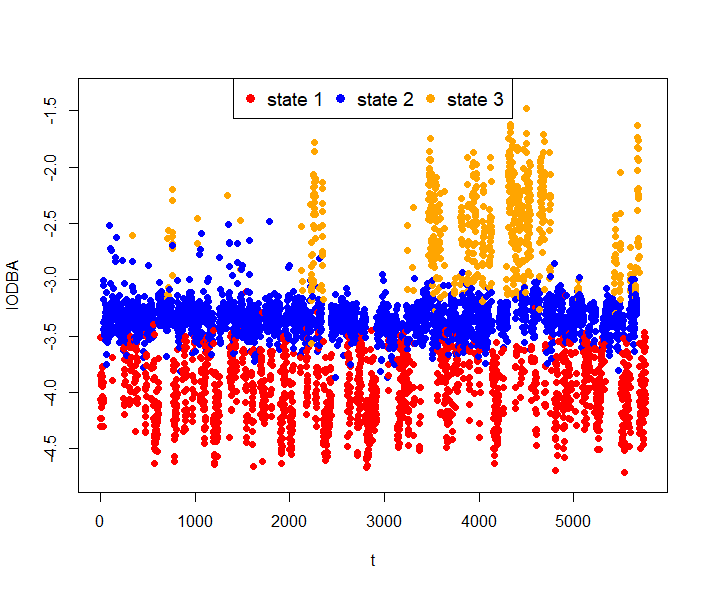}\includegraphics[width=0.5\textwidth]{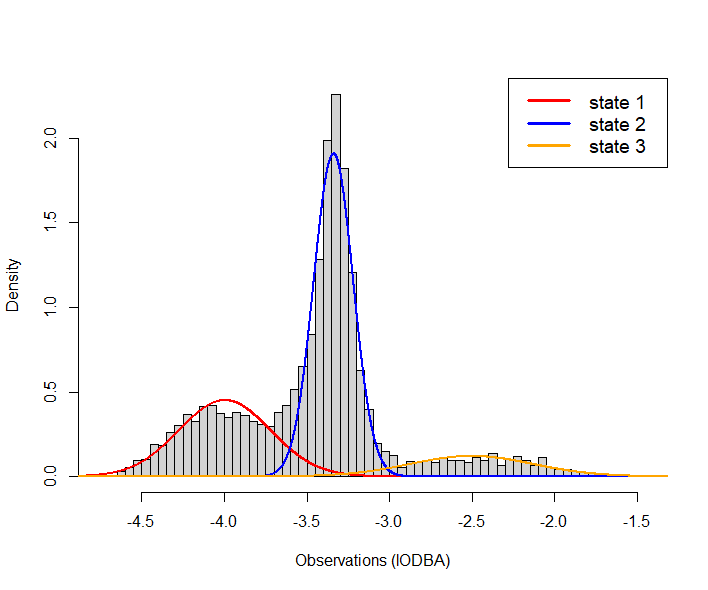}
	\caption[Estimation results for lODBA data with $N=3$ states]{\label{fig:sharkN3_2} Left panel: 15s-averaged lODBA series, where colour indicates the locally decoded state at each time obtained with our method (state labels are sorted according to their mean lODBA levels); right panel: Histogram of 15s-averaged lODBA values along with the estimated emission densities (weighted according to their proportion in the stationary distribution of the estimated Markov chain) obtained from the Gaussian HMM.}
\end{figure}

Here we provide additional details for the 3 and 9-state HMMs for the shark activity data. For the 3-state model, the algorithm's constant were set to $a=-5.5$, $b=-1$ and $\alpha=2$. Posterior inference was performed based on 50k iterations of the Algorithm 1 after a burn-in of 70k iterations, with a thinning interval of 10. The entire sampling process took 
about 2.32 hours in R (based on a laptop having Intel(R) Core(TM) i7-1185 CPU, at 3.0
GHz and 16 GB RAM). We performed convergence diagnostic by running 4 independent MCMC chains with different initializations, and the PSRF values for the nine transition probabilities were calculated to be $(1.00, 1.01, 1.03, 1.01, 1.02, 1.03, 1.01, 1.14, 1.05)$. These values are all fairly close to one, which suggests that the chains have likely converged satisfactorily. Our posterior summaries for the entries of the transition probability matrix are
\begin{displaymath}
	\hat{\Gamma}=\begin{pmatrix} 
		0.944_{(0.006)} & 0.055_{(0.006)} & 0.001_{(0.0008)} \\
		0.029_{(0.003)} & 0.96_{(0.004)} & 0.012_{(0.002)} \\
		0.009_{(0.004)} & 0.052_{(0.01)} & 0.939_{(0.01)}
	\end{pmatrix},
\end{displaymath}
where the point estimates are the posterior means and the associated standard deviations are shown in brackets. In particular, the diagonal entries of $\hat{\Gamma}$ are all close to 1, highlighting the serial dependence. The estimated state sequence (indicated in colours) obtained via local decoding is shown in the left panel of Figure \ref{fig:sharkN3_2}. Note that, as was pointed out in \citet{langrock2018spline}, there could be a potential lack of fit if one chooses to model this type of data by some common parametric HMMs. To illustrate this, we fitted a Gaussian HMM with $N=3$ to the lODBA data using the standard MLE approach and the resulting estimates of the emissions are shown in the right panel of Figure \ref{fig:sharkN3_2}, which exhibits a certain level of underfitting (e.g. fails to accurately capture the spikiness and the slight right tail of the emission of the middle state). The diagonal entries of the transition matrix were estimated as $\hat{\gamma}_{1,1}=0.941$, $\hat{\gamma}_{2,2}=0.952$ and $\hat{\gamma}_{3,3}=0.915$, suggesting that the state persistency was underestimated in all three states, especially for state 3. 

\begin{figure}[h!]
	\centering
	\includegraphics[width=0.51\textwidth]{Figures/sharkN9density}\includegraphics[width=0.51\textwidth]{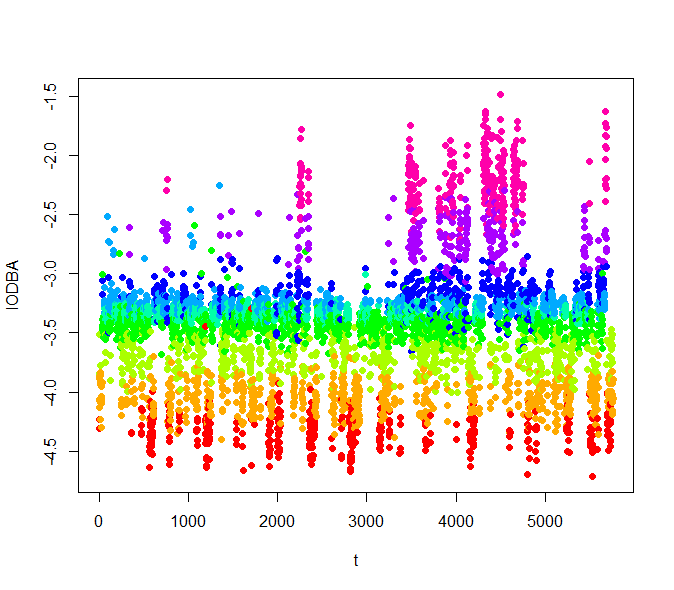}
	\caption[Estimation results for lODBA data with $N=9$ states]{\label{fig:sharkN9} Left panel: histogram of 15s-averaged lODBA values along with the nine estimated emission densities (weighted according to their proportion in the stationary distribution of the estimated Markov chain) obtained from our method (using state-specific knots); right panel: the corresponding locally decoded time series of the 9-state model. Here the state labels are sorted according to their mean lODBA levels.}
\end{figure}

Using our model selection method we found that the data support a model with $N=9$ states, which points to the potential rich information contained in this high-resolution signal and may not be surprising given the expected complexity of shark’s behaviour in reality \citep{bres1993behaviour}.
We obtained posterior estimates for this model by generating $50$k samples after discarding the initial $200$k iterations as burn-in, with a thinning interval of 10. To assess convergence, we estimated the PSRF focusing on parameters in the transition matrix by executing 4 independent MCMC runs with different initializations. Among the 81 parameters, the PSRF ranges from 1 to 1.02, indicating satisfactory convergence.
Figure \ref{fig:sharkN9} displays the estimated emission densities (left panel) and the corresponding decoded times series (right panel). We can see that the estimated hidden states roughly correspond to 9 different levels of activity which interestingly, resolves the multimodality seen in the 3-state model (e.g. state 1) into a mixture of unimodal emission densities. The posterior means of the transition probabilities are
\begin{displaymath}
	\hat{\Gamma}_{9}=\begin{pmatrix}
		0.843 & 0.131 & 0.01 & 0.003 & 0.002 & 0.003 & 0.004 & 0.002 & 0.002 \\
		0.073 & 0.747 & 0.151 & 0.008 & 0.004 & 0.004 & 0.01 & 0.001 & 0.001 \\
		0.018 & 0.14 & 0.642 & 0.158 & 0.011 & 0.011 & 0.015 & 0.002 & 0.001 \\
		0.001 & 0.004 & 0.077 & 0.721 & 0.125 & 0.059 & 0.01 & 0.002 & 0.001 \\
		0.001 & 0.003 & 0.008 & 0.11 & 0.778 & 0.084 & 0.009 & 0.006 & 0.001 \\
		0.003 & 0.005 & 0.014 & 0.08 & 0.069 & 0.761 & 0.06 & 0.007 & 0.001 \\
		0.003 & 0.031 & 0.048 & 0.045 & 0.016 & 0.067 & 0.701 & 0.086 & 0.003 \\
		0.004 & 0.004 & 0.007 & 0.01 & 0.011 & 0.013 & 0.18 & 0.614 & 0.157 \\
		0.003 & 0.003 & 0.003 & 0.003 & 0.003 & 0.003 & 0.007 & 0.112 & 0.863 \\
	\end{pmatrix}
\end{displaymath}
Almost all of the estimated states are persistent in the sense that there usually is a large probability of staying in the current state (see diagonal entries of $\hat{\Gamma}_{9}$). In comparison to the 3-state model, the diagonal entries are naturally lower as the shark's movement is now sub-divided into more states. 
The fitted model with 9 states indicates, for example, that states 1-3, 4-7 and 8-9 or 7-9 tend to have a higher probability of transitioning within the group of states than to other states. Furthermore, these three ``clusters" roughly correspond to the lower, middle and higher activity mode of the 3-state model. To assign biologically meaningful interpretations to these states, however, additional ecological information and investigation are required and we have not pursued this further here. Nevertheless, our analysis demonstrates the ability of our method to deal with model selection in an HMM with a relatively large number of states, while Langrock et al.'s approach can be expected to be severely challenged with increasing $N$. Our flexible HMM approach could also be used in an explorative way to identify a suitable parametric model. We can see from Figure \ref{fig:sharkN9} that most of the emission densities of the 9-state model have a relatively regular shape. An interesting further question would be whether the 3-state spline model can be replaced by a 9-state parametric approach, assuming for instance a mixture of Gaussian emission densities, which can capture a higher degree of multimodality in the marginal distribution.

\section{Further details of the empirical application of Section 4.2}
\label{sec:AP_condHMM}
In this section we present additional implementation details and estimation results for applying the proposed conditional HMM approach on the example subjects.
\subsection{Further details of the prior setting and the MCMC algorithm}
The priors for parameters in the sub-HMM are specified as follows. We used the same priors for the number of knot points and spline coefficients as described in Section \ref{sec:The model} of the main paper. For the knot location vector, $R_{K}$, we used the same uniform prior as in Section \ref{sec:The model}, but with an additional constraint that ensures the minimum distance between two adjacent knots is greater than a threshold, set to 0.6 here. This constraint is introduced for the sub-HMM as it directly models the 30-s PA counts data, helping to prevent numerical issues that may arise from the data's high level of discreteness.
For the transition matrix of sub-HMM there is no conjugate prior available as the associated hidden state process is not simulated, and a MH update is required instead.
Following the reparametrization scheme used in Section \ref{sec:The model}, we reparameterize each row of the transition probability matrix as $\gamma_{i,j}=\Tilde{\gamma}_{i,j}/\sum_{l=1}^{N}\Tilde{\gamma}_{i,l}$, $\Tilde{\gamma}_{i,j}>0$, and place a vague gamma prior on the $\Tilde{\gamma}_{i,j}$, i.e. $f(\Tilde{\gamma}_{i,j})$=gamma$(1,1)$, resulting in a Dir($1,\ldots,1$) distribution on $(\gamma_{i,1},\ldots,\gamma_{i,N})$.
For the weights involved in the emissions we apply the same reparametrization strategy:
$w_{i,j}=\Tilde{w}_{i,j}/(\Tilde{w}_{i,1}+\Tilde{w}_{i,2})$, $\Tilde{w}_{i,j}>0$, $i=1,\ldots N_{S}$, $j=1,2$., with a vague gamma$(1,1)$ prior on the $\Tilde{w}_{i,j}$, leading to a Dir($1,1$) distribution on $(w_{i,1},w_{i,2})$. 
Using the notation of Section \ref{sec:The model}, we assume the following factorization of the joint distribution in \eqref{subhmmcondpost} with all the reparametrizations
\begin{equation}\label{subhmmjoint}
	f(\Tilde{\bm{\theta}}^{S}|\mathbf{y}^{(n)},\mathbf{x}^{(n)})\propto f(\zeta)f(K)f(\Tilde{W})f(\Tilde{\Gamma})f(R_{K}|K)f(\Tilde{A}_{K}|K,\zeta)f(\mathbf{y}^{(n)}|\bm{\theta}^{S},\mathbf{x}^{(n)})
\end{equation}
where the reparameterized parameter vector $\Tilde{\bm{\theta}}^{S}=(\zeta,K,\Tilde{W},\Tilde{\Gamma},R_{K},\Tilde{A}_{K})$, $\Tilde{\Gamma}=
(\Tilde{\gamma}_{i,j})_{i,j=1,...,N_{S}}$,\\
 $\Tilde{W}=(\Tilde{w}_{i,k})_{i=1,\ldots,N_{S};\ k=1,2}$, and the $\Tilde{w}_{i,k}$ and $\Tilde{\gamma}_{i,j}$ are assumed to be a-priori independent. For the main-HMM, we use the same prior settings for the transition matrix and spline parameters as in Section \ref{sec:The model},  and the same prior specification for the weight parameters involved in the emissions as the sub-HMM.

Posterior inference for the main and sub-HMMs can be achieved by sequentially sampling from $f(\bm{\theta},\mathbf{x}^{(n)}|\mathbf{y}^{(n)})$ and $f(\bm{\theta}^{S}|\mathbf{y}^{(n)})$, 
where $\bm{\theta}$ represent the parameter set of the main-HMM with the additional weight parameters included.
The MCMC methodology described in Section \ref{sec:The algorithm} of the main paper (Algorithm 1) can be used to simulate from $f(\bm{\theta},\mathbf{x}^{(n)}|\mathbf{y}^{(n)})$, with an additional MH step to update the state-specific weight parameters and the details for this update are given below. For the sub-HMM, we simulate from $f(\bm{\theta}^{S}|\mathbf{y}^{(n)})$ according to \eqref{subhmmAS} in Section \ref{sec:condHMM} of the main paper. Conditional on each realisation of $\mathbf{x}^{(n)}$ (obtained from the posterior simulation for the main-HMM), we simulate from $f(\bm{\theta}^{S}|\mathbf{x}^{(n)},\mathbf{y}^{(n)})$ by drawing a sample for $\bm{\theta}^{S}$ from the joint density defined in \eqref{subhmmjoint} using essentially the same sampling scheme as used for the main-HMM, where the RJMCMC updates are run for several iterations and the last sample is kept.
Tuning of the MH scaling parameters was achieved via a separate pilot run using the same adaptive procedure as described in Section \ref{sec:AP_RJMCMC}, where the conditioning variable $\mathbf{x}^{(n)}$ may be fixed at a specific realisation from its posterior distribution or the local decoding result obtained from the simulation output for the main-HMM. 
To perform posterior simulation of the hidden state process associated with the sub-HMM for a given segment(s) of the time series (e.g. data points that were assigned to state 1 by the local decoding result of the main-HMM), a standard FFBS algorithm can be used by conditioning on the sampled $\bm{\theta}^{S}$.
During the forward procedure, data points that were assigned to states 2 or 3 by the the main-HMM are treated as ``missing". Mathematically this is done by replacing the emission densities associated with those time points by the constant of one.
In running the backward simulation procedure, the whole sub-state sequence associated with the given series is simulated. However, only sub-states that correspond to the ``non-missing" data points (assigned by state 1 of the main-HMM) are of interest.

To update the state-specific weight parameters in the zero-inflated emission distributions, we use a log-normal random walk for the reparameterized weights 
$\Tilde{w}_{i,j}$, $i=1,\ldots N$, $j=1,2$:
\begin{displaymath}
	\log(\Tilde{w}_{i,j}^{'})=\log(\Tilde{w}_{i,j})+\phi_{i,j},
\end{displaymath}
where $\phi_{i,j}\sim \mathcal{N}(0,\tau_{w}^{2})$. 
The acceptance probabilities of this move for the main and sub-HMM are 
\begin{displaymath}	\min\bigg(1,\frac{f(\mathbf{y}^{(n)},\mathbf{x}^{(n)}|\bm{\theta}^{'})f(\Tilde{W}^{'})}{f(\mathbf{y}^{(n)},\mathbf{x}^{(n)}|\bm{\theta})f(\Tilde{W})}\prod_{i=1}^{N}\prod_{j=1}^{2}\frac{\Tilde{w}_{i,j}^{'}}{\Tilde{w}_{i,j}}\bigg)
\end{displaymath}
and
\begin{displaymath}
	\min\bigg(1,\frac{f(\mathbf{y}^{(n)}|\mathbf{x}^{(n)},\bm{\theta}^{S'})f(\Tilde{W}^{'})}{f(\mathbf{y}^{(n)}|\mathbf{x}^{(n)},\bm{\theta}^{S})f(\Tilde{W})}\prod_{i=1}^{N}\prod_{j=1}^{2}\frac{\Tilde{w}_{i,j}^{'}}{\Tilde{w}_{i,j}}\bigg),
\end{displaymath}
respectively, where $\Tilde{W}^{'}$ denotes the vector of proposed $\Tilde{w}_{i,j}^{'}$ and $\bm{\theta}^{'}$ and $\bm{\theta}^{S'}$
denote the corresponding updated parameter set for the main and sub-HMM, respectively. This scheme can be easily adjusted for cases where multiple point masses are used, as in our MESA application.

\subsection{Further details of the MESA application}
\begin{figure}[h!]
	\centering
	\includegraphics[width=0.45\textwidth]{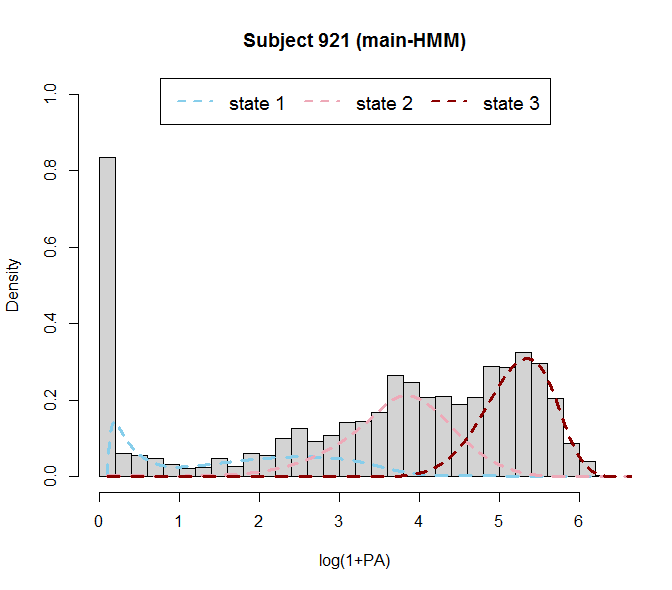}\includegraphics[width=0.45\textwidth]{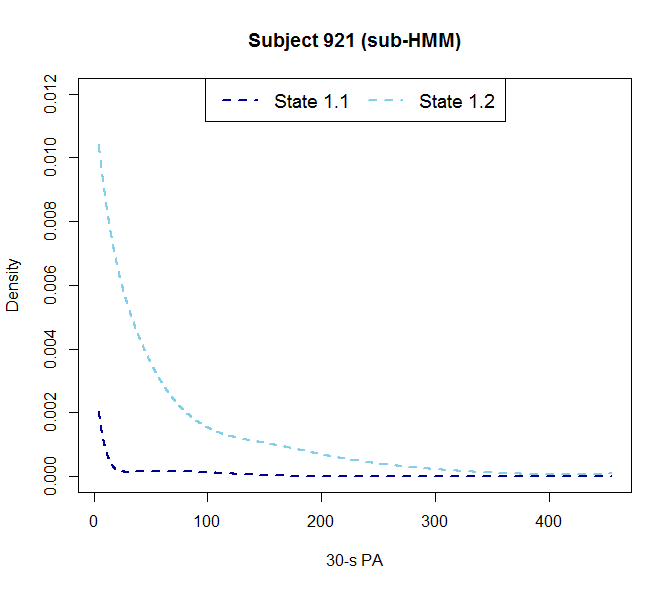}
	\includegraphics[width=0.45\textwidth]{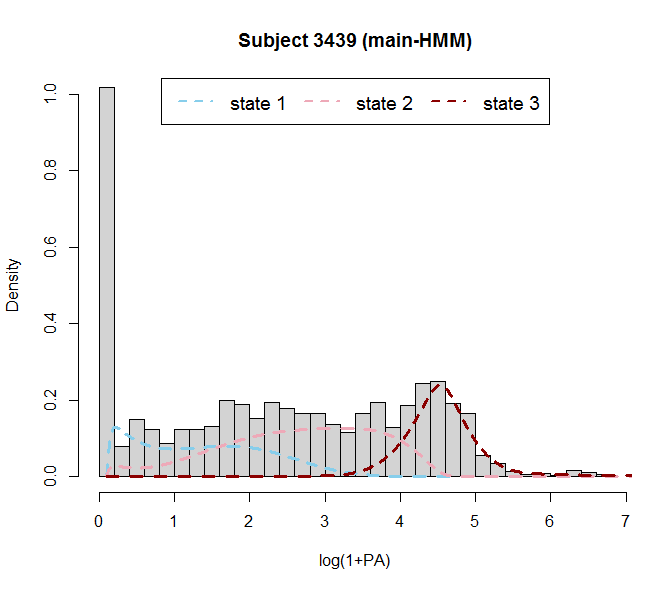}\includegraphics[width=0.45\textwidth]{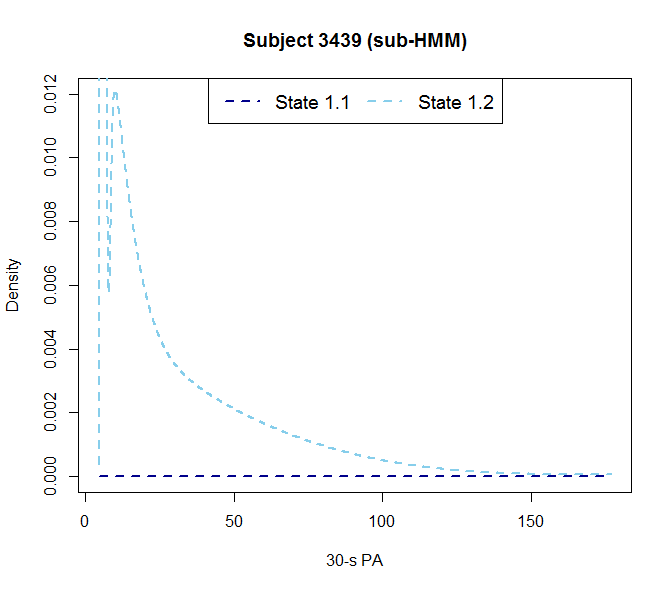}
	\caption[Estimated emission densities for the main and sub-HMMs]{\label{fig:examsubjemi} Left panel: histogram of 5-min transformed PA data along with the estimated emission densities (weighted according to their proportion in the stationary distribution of the estimated Markov chain) for the main-HMM; right panel: estimated emission densities for the sub-HMM. The weights for the point masses are not shown in the graph.
	}
\end{figure}

Here we present additional implementation details and estimation results for applying the proposed conditional HMM approach on the example subjects. 
For the main-HMM, we pre-processed the data by averaging the PA over 5-min non-overlapping intervals, followed by a log-transformation as used in \citet{li2020novel}, i.e. $\log(1+PA)$, to handle the high variability observed in the MESA data. For the emission model, we included two point masses, one at 0 and another at log(1.1) (the second smallest possible value for 5-min averaged PA after the log-transformation) to handle the high occurrence of both values.
For the algorithm's constants, we have chosen $a=0.1$, $b=\max(\log(1+PA))+3$ and $\alpha=2$. The main-HMM results were based on 25k iterations of the proposed algorithm, after discarding the first 50k as burn-in.
To assess the MCMC convergence, we estimated the PSRF by running 4 independent chains focusing on the transition matrix as before. The obtained PSRF values were close to one for both subjects, with values ranging between $(1,1.04)$ for subject 921 (subject A) and $(1,1.02)$ for subject 3439 (subject B), respectively. Figure \ref{fig:examsubjemi} (left panel) displays the estimated emission densities for the main-HMM (for the continuous part) obtained by averaging over the emissions generated across MCMC iterations for subjects A and B. The posterior modal number of knots is 8 for A and 7 for B. Table \ref{tab:mainHMMweights} shows the posterior means for the state-specific weights of the point masses at 0 ($w_{i,1}$) and $\log(1.1)$ ($w_{i,2}$). 

Inference for the sub-HMM is based on the raw 30-s PA data. To specify the emissions for the sub-states, we introduced point masses at the first four lowest values including 0, which exhibit a high frequency of occurrence. We set $\alpha=0.65$, $a=4.5$ and $b=10$ plus the maximum of the 30-s PA data corresponding to state 1 of the main-HMM. 
We ran our proposed algorithm for 25k updates (based on the 25k posterior samples of $\mathbf{x}^{(n)}$ obtained from the main-HMM analysis), discarding the first 15k as burn-in. 
The obtained PSRF values were 1.11 (for $\gamma_{1.1,1.1}$) and 1.03 (for $\gamma_{1.2,1.2}$) for subject A, and 1.04 (for $\gamma_{1.1,1.1}$) and 1.004 (for $\gamma_{1.2,1.2}$) for B.
The estimated emission densities (for the continuous part) for the two subjects are displayed in the right panel of Figure \ref{fig:examsubjemi}. The posterior modal number of knots is 3 for subject A and 4 for subject B.

\begin{table}
	\caption{Posterior means for the state-specific weights of the point masses at 0 ($w_{i,1}$) and $\log(1.1)$ ($w_{i,2}$)
		\label{tab:mainHMMweights}}
	\centering
	\renewcommand{\arraystretch}{0.85}
	\begin{tabular}{lcccccc}
		\hline
		Subject & $w_{1,1}$  & $w_{1,2}$ & $w_{2,1}$ & $w_{2,2}$ & $w_{3,1}$ & $w_{3,2}$ \\ \hline
		Subject 921   & 0.312 & 0.156 & 0.002 & 0.002 & 0.001 & 0.002   \\ 
		Subject 3439 & 0.315 & 0.132 & 0.008 & 0.012 & 0.002 & 0.002 \\  \hline
	\end{tabular}
\end{table}

\newpage
\bibliographystyle{apalike}

\bibliography{Bibliography-MM-MC}